\documentclass[11pt]{article}
\pdfoutput=1

\usepackage[textwidth=15.5cm,textheight=22.5cm]{geometry}
\usepackage{bm}
\usepackage{amsmath,amssymb,color,graphicx,amscd,amsfonts}
\usepackage{latexsym}
\usepackage{graphicx}
\usepackage{cite}
\usepackage{subfig}
\usepackage{bbm}
\usepackage{ulem}
\usepackage{longtable}
\usepackage[usenames,dvipsnames,svgnames,table]{xcolor}

\usepackage{array}
\usepackage[utf8]{inputenc}
\usepackage{empheq}

\usepackage{datetime}

\newcommand{\adj}{\ensuremath{{}_\text{adj}}}
\newcommand{\free}{\ensuremath{\mathcal{F}}}
\newcommand{\improved}[1]{\ensuremath{#1^\text{(imp.)}}}

\begin{document}

\thispagestyle{empty}

\setcounter{page}{0}

\begin{flushright} 

YITP-18-01\\
RBRC-1267\\
%LLNL-JRNL-****
\end{flushright} 

\vspace{0.1cm}

\begin{center}
{\LARGE

Gauged And Ungauged: 
A Nonperturbative Test

\rule{0pt}{20pt}  }
\end{center}

\vspace*{0.2cm}

\renewcommand{\thefootnote}{\alph{footnote}}

\begin{center}
    Evan B{\sc erkowitz}$^a$,
    Masanori H{\sc anada}$^{bcd}$,
    Enrico R{\sc inaldi}$^{ef}$
    and 
    Pavlos V{\sc ranas}$^{gf}$

\vspace{0.3cm}

$^a$ {\it Institut f\"{u}r Kernphysik and Institute for Advanced Simulation, \\
Forschungszentrum J\"{u}lich, 52425 J\"{u}lich, Germany}

$^b${\it Yukawa Institute for Theoretical Physics, Kyoto University,\\
Kitashirakawa Oiwakecho, Sakyo-ku, Kyoto 606-8502, Japan},

$^c${\it The Hakubi Center for Advanced Research, Kyoto University,\\
Yoshida Ushinomiyacho, Sakyo-ku, Kyoto 606-8501, Japan}

$^d${\it Department of Physics, University of Colorado, Boulder, Colorado 80309, USA}

$^e$
{\it RIKEN-BNL Research Center, Brookhaven National Laboratory, Upton, NY 11973, USA}

$^f$
{\it Nuclear Science Division, Lawrence Berkeley National Laboratory, \\
Berkeley, CA 94720, USA}

$^g$ {\it Nuclear and Chemical Sciences Division, Lawrence Livermore National Laboratory, \\
Livermore CA 94550, USA}

\vspace{0.5cm}
e.berkowitz@fz-juelich.de, hanada@yukawa.kyoto-u.ac.jp, \\
erinaldi@bnl.gov, vranas2@llnl.gov

\end{center}

\vspace{1cm}

\begin{abstract}
We study the thermodynamics of the `ungauged' D0-brane matrix model by Monte Carlo simulation.  
Our results appear to be consistent with the conjecture by Maldacena and Milekhin.

\end{abstract}

\newpage
%%%%%%%%%%%%%%%%%%%%%%%%%%%%%%%%%%%%%%
%%%%%%%%%%%%%%%%%%%%%%%%%%%%%%%%%%%%%%
%%%%%%%%%%%%%%%%%%%%%%%%%%%%%%%%%%%%%%
%\tableofcontents
%%%%%%%%%%%%%%%
%%%%%%%%%%%%%%%
\section{Introduction}
\hspace{0.25in}
%%%%%%%%%%%%%%%
%%%%%%%%%%%%%%%
The meaning of gauge symmetry in holography is not immediately clear. 
Although the most well understood version is indeed {\it gauge}/gravity duality \cite{Maldacena:1997re}, the dual field theory may not have to be a gauge theory.

Gauge symmetry is a source of various headaches at the technical level as well.
For example:
\begin{itemize}
\item
In the Hamiltonian formulation of quantum field theory, the algebra of gauge invariant operators has a complicated structure, and even state counting is difficult.  
This is one of the reasons that the Hamiltonian formulation of lattice gauge theory \cite{Kogut:1974ag} is not very practical. 

\item
The temporal component of the gauge field is often related to the sign problem.

\item
It is not easy to keep both gauge symmetry and supersymmetry on lattice. 

\item
It is not straightforward to define the entanglement entropy (see e.g. \cite{Ghosh:2015iwa,Aoki:2015bsa}). 

\end{itemize}
For $(0+1)$-dimensional theories, the {\it ungauged} counterparts do not seem pathological. 
In the Hamiltonian language, the gauge singlet condition (Gauss's law) is omitted.
In the path-integral language, the gauge field $A_t$ is turned off. 
It does not lead to possible pathologies in higher dimensions---the breakdown of Lorentz symmetry, for example---either. 
If such an ``ungauged theory'' makes sense, it would avoid the problems listed above. 

A priori, however, it is not clear how much the gauged and ungauged theories differ. 
It is often said that the gauge symmetry is not important in the deconfining phase.\footnote{
The explanation is that ``color degrees of freedom can be seen directly''.
This is somewhat misleading, because physical states are still gauge invariant. 
See e~.g.~\cite{Hanada:2014noa} regarding this point.  
}
While it is qualitatively true, it is more subtle at the quantitative level;
at least at high temperature, where the coupling constant is small, free energies of gauged and ungauged theories are different.\footnote{
In the case of the D0-brane matrix model which we will study in this paper, $E\sim 6N^2T$ and $E\sim 6.75N^2T$ for the gauged and ungauged theories, respectively. See Appendix~\ref{sec:high-T}. 
}
For example, nothing is known for the strongly coupled region near the transition to the confined phase;
the two theories may or may not resemble one another.
Below the transition temperature, the two theories are clearly different, in that the ungauged theory cannot be confining by definition. 
Still, the contribution from the non-singlet sector might be so small that the physics in the singlet sector is practically unaffected. 

Recently Maldacena and Milekhin \cite{ungauged} considered this problem by taking the D0-brane matrix model\cite{Witten:1995im,deWit:1988ig,Banks:1996vh,Itzhaki:1998dd} as a concrete example.
In terms of string theory, the gauge singlet sector describes closed strings.
Gauge non-singlets naturally appear when open strings are allowed.
As a string theory, such a setup seems to be fine.
Then it would make sense to consider the duality between the ungauged theory and string theory. 
They proposed a reasonable dual gravity prescription, and made a striking conjecture: 
the difference between gauged and ungauged theories is exponentially suppressed at low temperature. 
For example the difference of the energy should scale as $d\adj N^2 C\adj e^{-C\adj/T}$, where $d\adj$ is a positive integer and 
$C\adj$  is an order one positive number which corresponds to the energy of the adjoint excitation. 

In this work, in order to test the Maldacena-Milekhin conjecture, we perform Monte Carlo calculations for the ungauged matrix model at small temperatures. 
First, we introduce the gauged and ungauged D0-brane matrix models in Sec.~\ref{sec:model}.  
The dual gravity descriptions are reviewed in Sec.~\ref{sec:gravity_dual}.  
The lattice regularization used for the simulations is explained in Sec.~\ref{sec:lattice_action}. 
In Sec.~\ref{sec:bosonic}, we study the bosonic analogue of the ungauged D0-brane matrix model numerically.
Although the bosonic models do not admit dual gravity descriptions, they illuminate the numerical approach we have adopted.
Sec.~\ref{sec:fermionic} is the main part of this paper, which tests the Maldacena-Milekhin conjecture.

%%%%%%%%%%%%%%%
%%%%%%%%%%%%%%%
\section{Gauged and ungauged D0-brane matrix model}\label{sec:model}
\hspace{0.25in}
%%%%%%%%%%%%%%%
%%%%%%%%%%%%%%%

The Euclidean action of the original, `gauged' D0-brane matrix model \cite{Witten:1995im,deWit:1988ig,Banks:1996vh,Itzhaki:1998dd}
 is given by 
\begin{eqnarray}\label{eq:original-bfss}
S_{\rm gauged}
&=&
\frac{1}{2g_{YM}^2}\int_0^\beta dt {\rm Tr}\left\{
(D_t X_M)^2 
-
\frac{1}{2}[X_M,X_{M'}]^2 
+
\bar{\psi}^\alpha D_t\psi_\alpha
-
\bar{\psi}^\alpha\gamma^M_{\alpha\beta}[X_M,\psi^\beta] 
\right\},  
\nonumber\\
\end{eqnarray}
where $X_M$ $(M=1,2,\cdots,9)$ are $N\times N$ Hermitian matrices and $D_tX_M$ is the covariant derivative given by $(D_tX_M)=\partial_t X_M-i[A_t,X_M]$ and $A_t$ is the $U(N)$ gauge field. 
We impose the thermal boundary conditions, $A_t(t+\beta)=A_t(t)$, $X_M(t+\beta)=X_M(t)$, $\psi(t+\beta)=-\psi(t)$. 
By doing so, the circumference of the Euclidean circle $\beta$ is the inverse temperature: $\beta=1/T$. 
The gamma matrices $\gamma^M_{\alpha\beta}$ $(M=1,2,\cdots,9)$ are the $16\times 16$ left-handed part of the gamma matrices in ($9+1$)-dimensions. 
$\psi_\alpha$ $(\alpha=1,2,\cdots,16)$ are $N\times N$ real fermionic matrices.
This theory is the dimensional reduction of 4D ${\cal N}=4$ super Yang-Mills theory to ($0+1$)-dimensions. 
We often set the 't Hooft coupling $\lambda=g_{YM}^2N$ to one, without losing generality.
Equivalently, all dimensionful quantities are measured in units of the 't Hooft coupling; 
for example the temperature $T$ actually refers to the dimensionless combination $\tilde{T}\equiv\lambda^{-1/3}T$. 
It also means the energy scale is related to the strength of the interaction: low temperature (small $T$) and strong coupling (large $\lambda$) are equivalent, 
in the sense that $\tilde{T}$ is small.
In the same manner, long distance is strong coupling. 

The action and partition function are given by 
\begin{eqnarray}
Z_{\rm gauged}=
\int [dA_t][dX][d\psi]
e^{-S_{\rm gauged}}.  
\end{eqnarray}

The `ungauged' theory is defined simply by dropping the gauge field $A_t$, as  
\begin{eqnarray}
Z_{\rm ungauged}=
\int [dX][d\psi]
e^{-S_{\rm ungauged}}, 
\end{eqnarray}
\begin{eqnarray}
S_{\rm ungauged}
&=&
\frac{1}{2g_{YM}^2}\int dt {\rm Tr}\Bigg\{
(\partial_t X_M)^2 
-
[X_M,X_{M'}]^2 
+
\bar{\psi}^\alpha \partial_t\psi_\alpha
-
\bar{\psi}^\alpha\gamma^M_{\alpha\beta}[X_M,\psi^\beta] 
\Bigg\}. \label{eq:action-ungauged}
\end{eqnarray}
In the Hamiltonian language, the ungauging procedure we just described is equivalent to removing the gauge singlet constraint.

%%%%%%%%%%%%%%%
%%%%%%%%%%%%%%%
\subsection{Dual gravity descriptions}\label{sec:gravity_dual}
\hspace{0.25in}
%%%%%%%%%%%%%%%
%%%%%%%%%%%%%%%
The dual gravity description of the gauged D0-brane matrix model near the 't Hooft large-$N$ limit was proposed in Ref.~\cite{Itzhaki:1998dd}.\footnote{
It has also been proposed that this model can describe M-theory, in a region with much stronger coupling \cite{deWit:1988ig,Banks:1996vh,Itzhaki:1998dd}. 
}
The dual is a black zero-brane consisting of $N$ D0-branes and open strings connecting them. 
The internal energy $E=\frac{\partial }{\partial \beta}(\beta\free)$, where $\beta\free=-\log Z(\beta)$, is identified with the energy of the black hole above extremality. 
At low temperature (strong coupling), the dual gravity calculation\footnote{reviewed, for example, in Ref.~\cite{Hanada:2013rga}} provides us with  
\begin{eqnarray}\label{eq:sugra-energy}
\tilde{E}\equiv
\lambda^{-1/3}E=AN^2\tilde{T}^{14/5}, 
\end{eqnarray}
where $A\simeq 7.41$ is an analytically calculable number, up to the $\alpha'$ and $g_s$ corrections (higher order in $\tilde{T}$ and $1/N^2$).
This conjectured duality has been confirmed numerically with good precision, by comparing direct numerical Monte Carlo results of the gauged D0-brane matrix model and the dual gravity calculations, including stringy corrections.\footnote{
The large-$N$, continuum result is available in Ref.~\cite{Berkowitz:2016jlq}. 
Numerical studies by several independent groups \cite{Anagnostopoulos:2007fw,Catterall:2008yz,Hanada:2008ez,Kadoh:2012bg,Hanada:2013rga,Kadoh:2015mka,Filev:2015cmz,Hanada:2016zxj} 
obtained consistent results. 
} 
The duality has been tested for other quantities as well: 
the supersymmetric Polyakov loop \cite{Hanada:2008gy} agrees with dual gravity calculations based on the minimal surface prescription \cite{Maldacena:1998im}, 
and correlation functions \cite{Hanada:2011fq} agree with the calculation based on the generalized conformal symmetry \cite{Sekino:1999av} analogous to the GKPW relation \cite{Gubser:1998bc,Witten:1998qj}. 

In the Hamiltonian formulation, the Hilbert space of the gauged theory consists of gauge singlets, which are obtained by acting traces of products of scalars on the vacuum, such as ${\rm Tr}(\hat{X}_{M_1}\hat{X}_{M_2}\hat{X}_{M_3}\hat{X}_{M_4})|{\rm \it{Vac}}\rangle$. 
Such single trace operators are similar to Wilson loops in QCD, which are identified with QCD flux strings, 
and naturally  leads to an intuitive interpretation as closed strings. %; see e.~g.~\cite{Asplund:2008xd,Hanada:2016pwv} for more detail. 
It motivates us to identify microstates of black zero-brane with closed string states, as was speculated before the duality had been found \cite{Susskind:1993ws,Horowitz:1996nw}. 

\begin{figure}
\begin{center}
\scalebox{0.4}{
\includegraphics{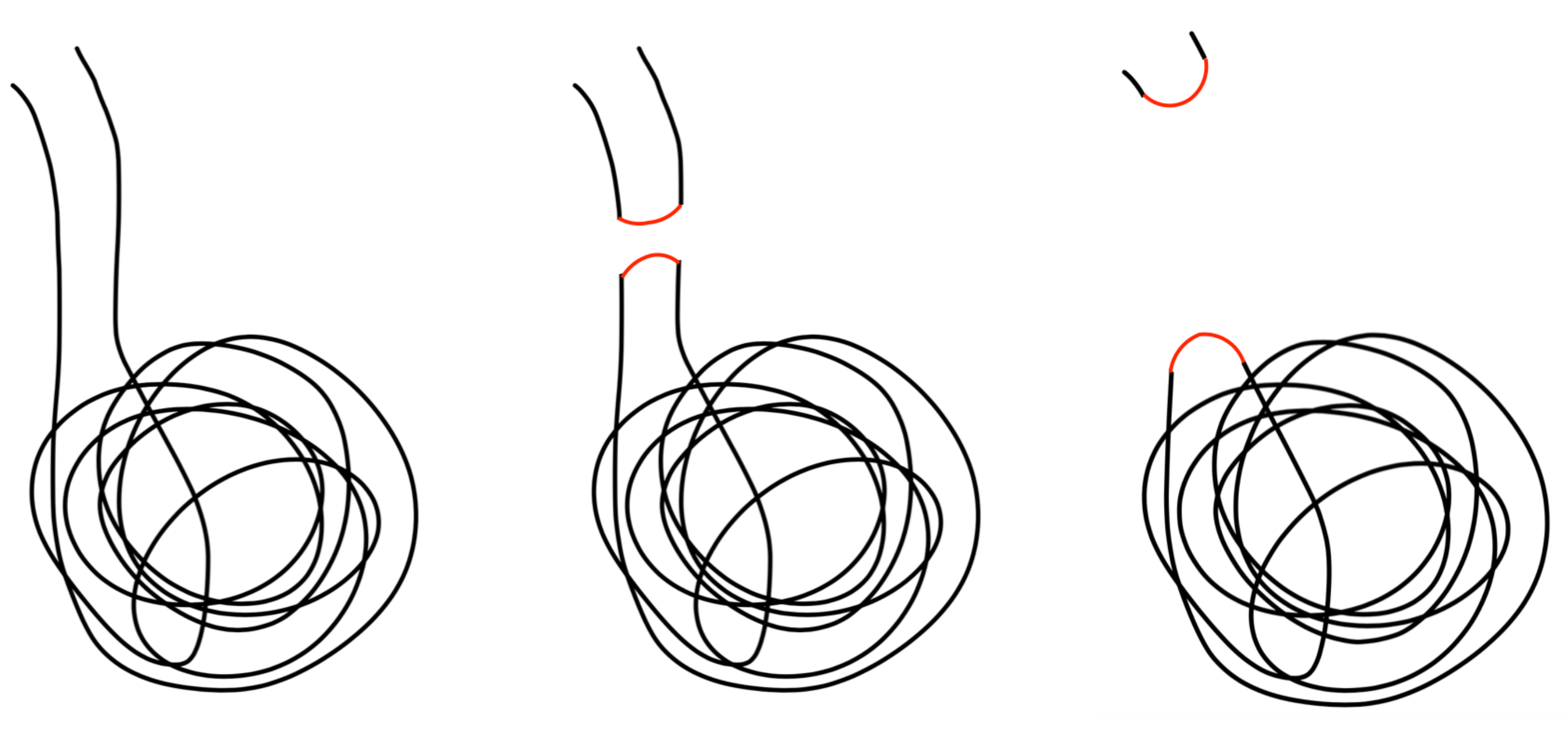}}
\end{center}
\caption{
Dual gravity description of the ungauged D0-brane matrix model proposed by Maldacena and Milekhin. 
A long open string (left) can split to closed and open strings (middle). 
Then the open string can shrink toward the boundary, while the bight of the closed string can shrink toward the center (right).
}\label{fig:ungauged-dual-string-picture}
\end{figure}

Gauge non-singlets admit products of scalars without trace, introducing open strings. 
For example, $(\hat{X}_{M_1}\hat{X}_{M_2}\hat{X}_{M_3}\cdots \hat{X}_{M_L})|{\rm \it{Vac}}\rangle$ (no trace!) can describe an open string consisting of $L$ bits.  
In the dual gravity description, because there is no special point in the bulk where open strings can end, 
it is natural to assume the open strings ends at the boundary of the space, as the left picture in Fig.~\ref{fig:ungauged-dual-string-picture} \cite{ungauged}. 

Such a long open string can interact with itself at self-intersecting point; on the gauge theory side, the kinetic (electric) term 
describes such an interaction. (It is essentially the same as the self-interaction of the long closed string explained in Ref.~\cite{Hanada:2014noa}.
The crucial point is that, although the interaction at each intersection is $1/N$-suppressed, the growth of intersection points overcome such suppression.) 
Correspondingly, on the gravity side, a long open string can split into a long closed string and a simpler open string, as in the middle picture in Fig.~\ref{fig:ungauged-dual-string-picture}. 
But then the open string can shrink toward the boundary, and the bight of the closed string can also shrink, like in the right picture of Fig.~\ref{fig:ungauged-dual-string-picture}, 
so that the energy (which is roughly proportional to the length of the string) is minimized.  
In this way, Maldacena and Milekhin \cite{ungauged} conjectured that the dual gravitational system is a black zero-brane plus short open strings localized near the boundary. 
They have estimated the contribution of such short open strings, and concluded that the free energy $\free$ should increase by 
\begin{eqnarray}
\beta\Delta\free=\beta\free_{\rm ungauged}-\beta\free_{\rm gauged}=-d\adj N^2e^{-C\adj/T}+\cdots, 
\label{conjecture_free_energy}
\end{eqnarray}
where $d\adj$ is a positive integer and $C\adj$ is the energy of the adjoint excitation.
The dots represent terms negligible at large $N$ and small $T$ such as those from higher representations. 
From this, $E=\frac{\partial }{\partial \beta}(\beta\free)$ should change by 
\begin{eqnarray}
\Delta E=E_{\rm ungauged}-E_{\rm gauged}=d\adj C\adj N^2e^{-C\adj/T}+\cdots, 
\end{eqnarray}
if the conjecture is correct. 
See Ref.~\cite{ungauged} for more precise arguments. 
Our goal in this paper is to test this conjecture by studying the gauge theory side. 

%%%%%%%%%%%%%%%
%%%%%%%%%%%%%%%
\subsection{Lattice regularization}\label{sec:lattice_action}
\hspace{0.25in}
%%%%%%%%%%%%%%%
%%%%%%%%%%%%%%%
We use the simulation code developed for Monte Carlo String/M-theory Collaboration, which is freely available to anybody \cite{simulation_code}. 
The ungauged version is obtained simply by turning off the gauge field (and the associated Faddeev-Popov term) from the code for the gauged theory.\footnote{
The ungauged version is also available upon request to M.~H. 
}

In order to make the lattice regularization simpler, we use a slightly different (though equivalent) form of the fermionic action,  
\begin{eqnarray}
\label{eq:sf}
S_f
=
\frac{N}{\lambda}\int_0^\beta dt\ {\rm Tr}\left\{
i\bar{\psi}\gamma^{10}D_t\psi
-
\bar{\psi}\gamma^M[X_M,\psi] 
\right\}.   
\end{eqnarray}
Here $\gamma^M$ ($M=1,\cdots,10$), which are $16\times 16$ upper-right block of the 10d gamma matrices $\Gamma^{M}$.
For the later convenience, we take\footnote{
Others are taken as follows:
\begin{align*}
\gamma^1&=\sigma_3\otimes\textbf{1}\otimes\textbf{1}\otimes\textbf{1},   &
\gamma^4&=\sigma_2\otimes\sigma_2\otimes\textbf{1}\otimes\sigma_3,       &
\gamma^7&=\sigma_2\otimes\textbf{1}\otimes\sigma_1\otimes\sigma_2,       \\
\gamma^2&=\sigma_2\otimes\sigma_2\otimes\sigma_2\otimes\sigma_2,         &
\gamma^5&=\sigma_2\otimes\sigma_1\otimes\sigma_2\otimes\textbf{1},       &
\gamma^8&=\sigma_2\otimes\textbf{1}\otimes\sigma_3\otimes\sigma_2,       \\
\gamma^3&=\sigma_2\otimes\sigma_2\otimes\textbf{1}\otimes\sigma_1,       &
\gamma^6&=\sigma_2\otimes\sigma_3\otimes\sigma_2\otimes\textbf{1},       &
\gamma^9&=-i\textbf{1}\otimes\textbf{1}\otimes\textbf{1}\otimes\textbf{1}.
\end{align*}
By taking $\Gamma^i=\sigma_1\otimes\gamma^i$ ($i=1,2,\cdots,8,10$) and 
$\Gamma^9=i\sigma_2\otimes\gamma^9$, the standard anticommutation relation holds: $\{\Gamma^M,\Gamma^N\}=2\delta^{MN}$. 
} 
\begin{eqnarray}
\gamma^{10}=\sigma_1\otimes\textbf{1}_8. 
\end{eqnarray}
 
This model is obtained by dimensionally reducing the ten-dimensional ${\cal N}=1$ super Yang-Mills theory to one dimension. 
The index $\alpha$ of the fermionic matrices $\psi_\alpha$ corresponds to the spinor index in ten dimensions, and 
$\psi_\alpha$ is Majorana-Weyl in ten-dimensional sense. 

For numerical efficiency, we take the static diagonal gauge, 
\begin{eqnarray}
A_t=&\frac{1}{\beta}\cdot{\rm diag}(\alpha_1,\cdots,\alpha_N),
\qquad
-\pi&<\alpha_i\le\pi
\end{eqnarray} 
and add the associated Faddeev-Popov term,\begin{align}
S_{F.P.}
&=
-
\sum_{i<j}2\log\left|\sin\left(\frac{\alpha_i-\alpha_j}{2}\right)\right|, 
\label{eq:Faddeev-Popov}
\end{align}
to the action.
The numerical simulations are performed in the phase-quenched limit. 
The reader can refer to Ref.~\cite{Berkowitz:2016jlq} for more details on this aspect.

%%%%%%%%%%%%%%%%%%
\subsubsection{`Naive' Regularization}
\hspace{0.51cm}
%%%%%%%%%%%%%%%%%%
We regularize the theory by introducing a lattice with $N_t$ sites. Our lattice action is 
\begin{eqnarray}
S_b
&= &
\frac{N}{2a\lambda}\sum_{t,M}{\rm Tr}\left(D_+X_M(t)\right)^2
-
\frac{Na}{4\lambda}\sum_{t,M,N}{\rm Tr}[X_M(t),X_N(t)]^2, 
\\
S_{F.P.}
&= &
-
\sum_{i<j}2\log\left|\sin\left(\frac{\alpha_i-\alpha_j}{2}\right)\right|, 
\end{eqnarray}
\begin{eqnarray}
S_f
= 
\frac{iN}{\lambda}\sum_{t}{\rm Tr}\bar{\psi}(t)
\left(
\begin{array}{cc}
0 & D_+\\
D_- & 0
\end{array}
\right)
\psi(t)
-
\frac{aN}{\lambda}\sum_{t,M}\bar{\psi}(t)\gamma^M[X_M(t),\psi(t)],  
\end{eqnarray}
where $U={\rm diag}(e^{i\alpha_1/N_t},e^{i\alpha_2/N_t}\cdots,e^{i\alpha_N/N_t})$,  
$-\pi\le \alpha_i<\pi$, and $D_\pm$ acts on $\psi$ as 
\begin{eqnarray}
D_+\psi(t)
&=&
U\psi(t+a)U^\dagger-\psi(t), 
\nonumber\\
D_-\psi(t)
&=&
\psi(t)-U^\dagger\psi(t-a)U.
\end{eqnarray} 
The action on the $X_M$ is the same: 
$D_+X_M(t)
=
UX_M(t+a)U^\dagger-X_M(t)$. 
Other than the gauge fixing, this action is the same as the one used in \cite{Catterall:2008yz}. 

%%%%%%%%%%%%%%%%%%%%%%%%%%%%%%%
\subsubsection{Tree-level Improved Lattice Regularization}\label{sec:action_lattice_improved}
\hspace{0.51cm}
%%%%%%%%%%%%%%%%%%%%%%%%%%%%%%%
The forward and backward derivatives used in the naive lattice discretization are related to the covariant derivative in the continuum theory by 
\begin{eqnarray}
D_\pm\psi(t) = aD_t\psi(t)\pm\frac{a^2}{2}D_t^2\psi(t)+O(a^3).   
\end{eqnarray}
We can reduce the discretization error by using 
\begin{eqnarray}
\improved{D}_\pm\psi(t)
\equiv
\mp\frac{1}{2}U^2\psi(t\pm 2a)U^{\dagger 2}
\pm 2U\psi(t\pm a)U^\dagger
\mp\frac{3}{2}\psi(t)
=
aD_t\psi(t) + O(a^3). 
\end{eqnarray} 
The improved action is obtained by replacing $D_\pm$ with $\improved{D}_\pm$.
The Faddeev-Popov term remains unchanged.  

%%%%%%%%%%%%%%%%%%
\subsubsection{Ungauged theory}
\hspace{0.51cm}
%%%%%%%%%%%%%%%%%%
The ungauged version is obtained by setting $\alpha_1=\cdots=\alpha_N=0$ and turning off the Faddeev-Popov term. 

%%%%%%%%%%%%%%%
%%%%%%%%%%%%%%%
\subsubsection{How to measure the internal energy}\label{sec:how-to-measure-E}
\hspace{0.25in}
%%%%%%%%%%%%%%%
%%%%%%%%%%%%%%%
In this study we calculate the internal energy
\begin{eqnarray}
E=\frac{\partial }{\partial \beta} (\beta\free)
\end{eqnarray}
in the gauge theory side for both the ungauged and the original matrix model.
This quantity can be calculated by using \cite{Catterall:2007fp}
\begin{eqnarray}
E=\frac{3}{2\beta}\left\{9(N^2N_t-1)-2\langle S_b\rangle\right\}. \label{formula-energy}
\end{eqnarray}
Note that the Faddeev-Popov term is not included in the right hand side.

The derivation of \eqref{formula-energy} is as follows. 
By rescaling the fields and time, we can take the `dimensionless coupling' to be $\lambda'=\lambda\beta^3$. In the lattice action, 
with this normalization, the $\beta$-dependence 
appears only through $\lambda'$, as an overall factor $\beta^{-3}$ in front of the action.  
From this we obtain $E=-\frac{3}{\beta}\langle S_b+S_f\rangle$, 
up to an additive constant. 
Here  $\langle S_f\rangle=\langle {\rm Tr} \frac{1}{2}{\cal D}\cdot {\cal D}^{-1}\rangle={\rm const.}$ (${\cal D}$ is the Dirac operator
defined by $S_f=\bar{\psi}{\cal D}\psi$),  
regardless of the details of the action, as long as the fermionic part is a fermion bilinear. 
So $E=-\frac{3}{\beta}\langle S_b\rangle$ up to an additive constant. 
We take the constant so that $E=0$ at $T=0$ if supersymmetry is not broken. This constant can be obtained just by counting the fermionic and gauge modes.  
The ``$-1$" in \eqref{formula-energy} is the constant mode of U(1), which is removed by hand in our calculation.  

We also use \eqref{formula-energy} for the ungauged theory. Then, if the ground state is in the singlet sector, the energy should vanish at $T=0$.

%%%%%%%%%%%%%%%
%%%%%%%%%%%%%%%
\section{Numerical exercise: bosonic theory}\label{sec:bosonic}
\hspace{0.25in}
%%%%%%%%%%%%%%%
%%%%%%%%%%%%%%%

In this section we study the bosonic analogue of the D0-brane matrix model, simply obtained by neglecting the fermionic degrees of freedom and their interactions.
The model is described by the usual matrix model once the term $S_f$ in Eq.~\eqref{eq:sf} is dropped.
The conjecture by Maldacena and Milekhin is developed in the context of gauge theories with a gravity dual.
Although the bosonic matrix model in this section has no known dual, we want to exemplify our numerical procedure in this numerically easier case, where complications due to fermionic contributions are absent.

The gauged bosonic theory has been studied as the high-temperature limit of D1-brane theory \cite{Aharony:2004ig,Aharony:2005ew,Kawahara:2007fn,Mandal:2009vz}. 
Unlike the supersymmetric theory, the gauged bosonic theory is confined at low temperature. 
For properly normalized quantities, such as $E/N^2$ or $R^2\equiv\frac{1}{N\beta}\int dt{\rm Tr}X_M^2$, the temperature independence, up to $1/N$ corrections, in the confining phase, expected from the Eguchi-Kawai equivalence\cite{Eguchi:1982nm}, has been confirmed numerically\cite{Aharony:2004ig,Aharony:2005ew,Kawahara:2007fn}.
The internal energy for this bosonic theory is related to $F^2\equiv -\frac{1}{N\beta}\int dt{\rm Tr}[X_M,X_{M'}]^2$ by $\frac{E}{N^2}=\frac{3}{4\lambda}F^2$ because $\frac{N^2}{4\lambda}F^2$ is the potential energy, and $({\rm kinetic\ energy})=2\times({\rm potential\ energy})$ holds due to the virial theorem. 

In the ungauged version of the bosonic theory, we expect an exponentially suppressed contribution from the nonsinglet sector, with Boltzmann weights $\sim e^{-C\adj/T}$, on top of the constant part coming from the ground state.
Properly accounting for these new Boltzmann weights when computing the expectation value of different observables, like the energy or $R^2$, results in an exponentially small difference.

In order to verify the presence of the contribution from the nonsinglet sector, and that it is suppressed at small temperature, we numerically calculate observables for the regularized ungauged bosonic model at nine different temperatures $0.2 \le \tilde{T} \le 0.6$ with $N=6$, 8 and 12 and $L=12$, 16, 24, 32, 48 and 64.
We take the continuum limit ($L \rightarrow \infty$) and the large-$N$ limit ($N \rightarrow \infty$).
This is the first time such limits have been studied systematically for this bosonic model and numerical details for the analysis can be found in Appendix~\ref{sec:numerics-detail}, together with summary tables.
In the following we will show the rescaled quantities measured on the lattice as $\tilde{E}/N^{2} \equiv (\lambda^{-1/3})E/N^2$ and $\tilde{R}^2 \equiv \lambda^{-2/3} R^2$.
 
\begin{figure}
\begin{center}
\scalebox{0.45}{
\includegraphics{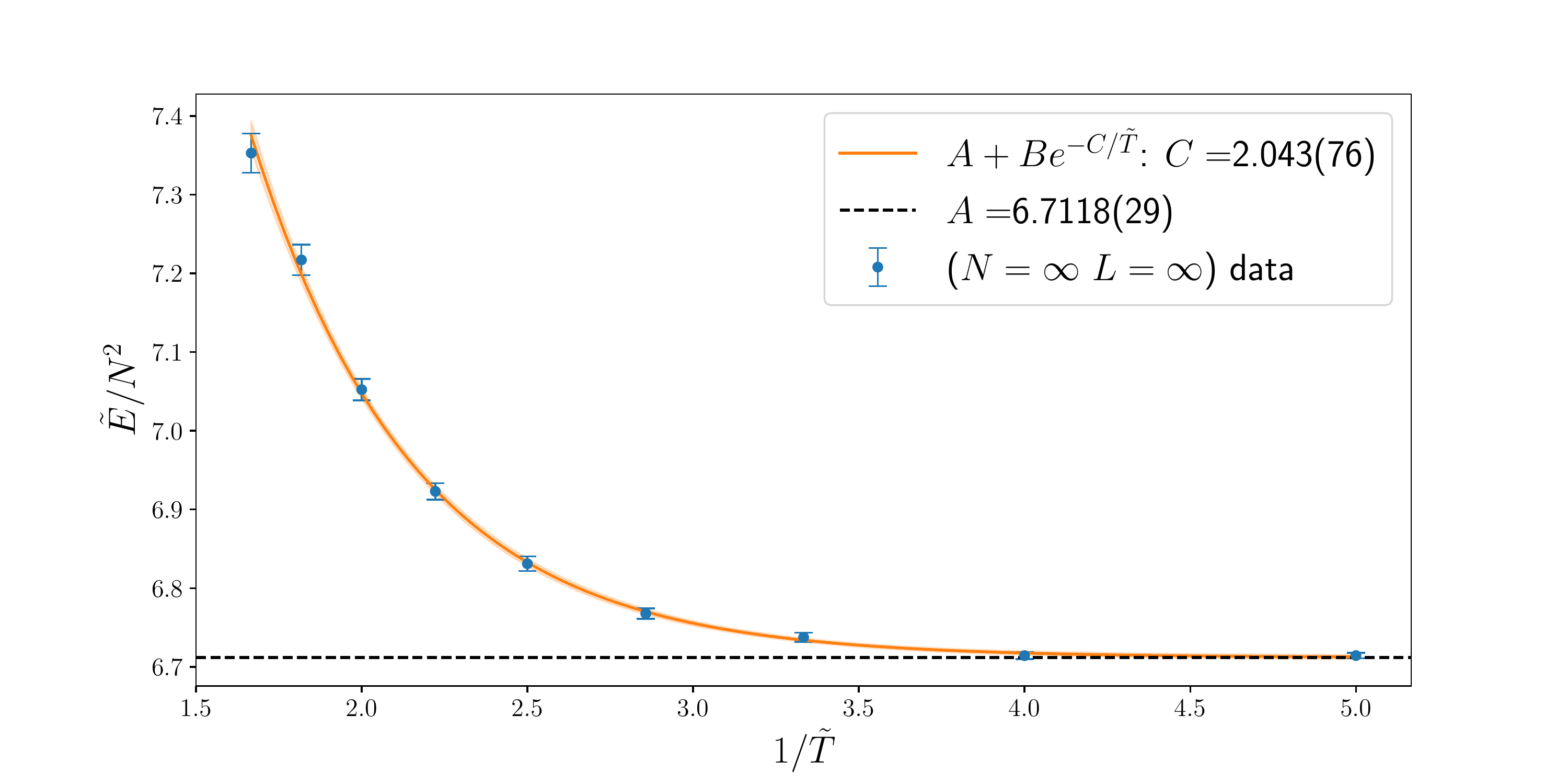}}
\scalebox{0.45}{
\includegraphics{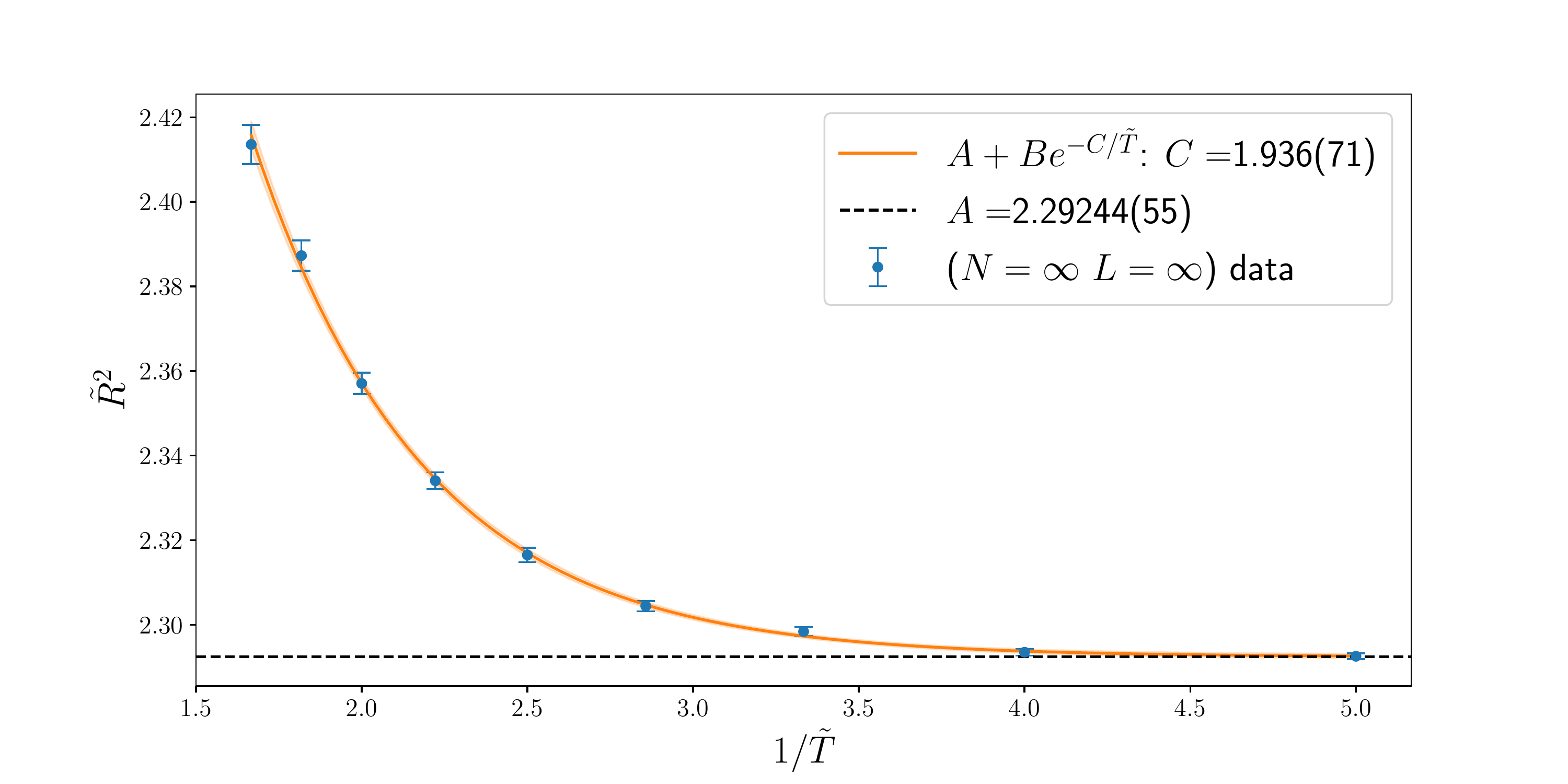}}
\end{center}
\caption{
$\tilde{E}/N^2\equiv(\lambda^{-1/3}E)/N^2$ and $\tilde{R}^2\equiv\lambda^{-2/3}R^2$ in the ungauged bosonic matrix model in the low temperature region, 
where the gauged counterpart is in the confining phase. 
Both behave as $A+Be^{-C/\tilde{T}}$, where $C\simeq 2.0$.  	
In the gauged theory, both $\tilde{E}/N^2$ and $\tilde{R}^2$ are constant in the confining phase, 
and the values obtained in Ref.~\cite{Kawahara:2007fn} agree with $A$ from the ungauged theory with good precisions. 
}\label{fig:E-and-R2-bosonic}
\end{figure}

From Ref.\cite{Kawahara:2007fn} we know that the transition temperature is $\tilde{T} \approx 0.9$ and we also know that the constant values for $\tilde{E}/N^{2}$ and $\tilde{R}^{2}$ are $\epsilon_0=6.695(5)$ and $r_0^2=2.291(1)$ from $N=32$ simulations at finite lattice spacing.
In Fig.~\ref{fig:E-and-R2-bosonic}, we plot $\tilde{E}/N^2$ and $\tilde{R}^2$ as a function of $1/\tilde{T}$ for the ungauged theory at low temperature $\tilde{T} \le 0.6$---a regime where the gauged theory is confining.
Both observables $\mathcal{O}(\tilde{T})$ are shown together with a fit of the form 
\begin{equation}
  \label{eq:bosonic-fit}
  \mathcal{O}(\tilde{T}) \; = \; A+Be^{-C/\tilde{T}} \quad ,
\end{equation}
where $C$ corresponds to $\tilde{C}\adj\equiv\lambda^{-1/3}C\adj$, while $A$, shown as a dashed black line in the plots, represents the $\tilde{T}=0$ value for the observable.
For $\tilde{E}/N^2$ we obtain $A=6.7118(29)$ which represents the internal energy of the ungauged theory at zero temperature.
For $\tilde{R}^2$ we get $A=2.29244(55)$. 
We can compare these results with the those obtained for the gauged theory in Ref.~\cite{Kawahara:2007fn}, despite the latter not being extrapolated to the continuum large-$N$ limit.
The agreement is well within the statistical accuracy of the data and it supports the conjecture by Maldacena and Milekhin that the difference of the two theories vanishes at $T=0$.

\begin{figure}
\begin{center}
\scalebox{0.45}{
\includegraphics{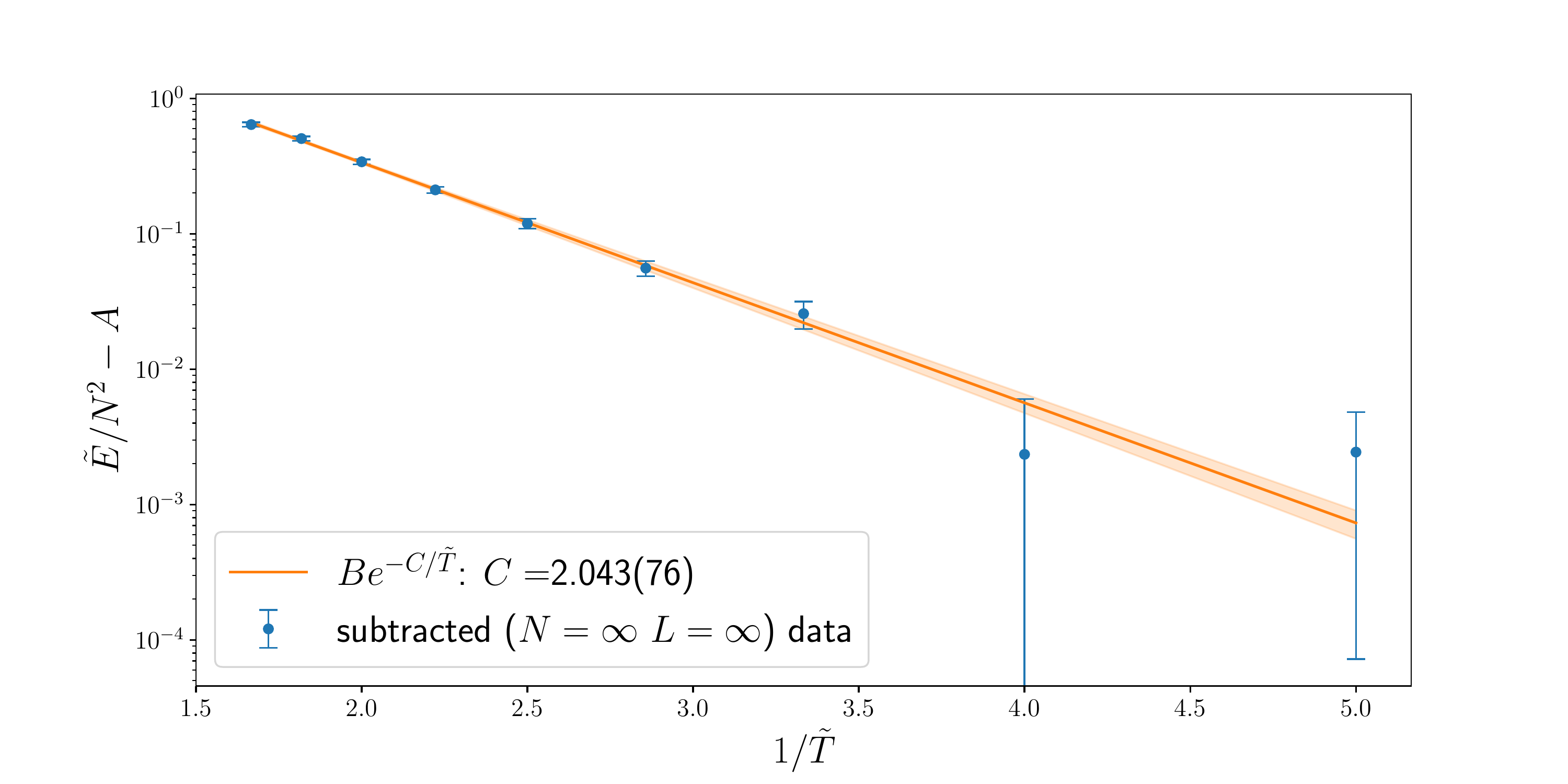}}
\scalebox{0.45}{
\includegraphics{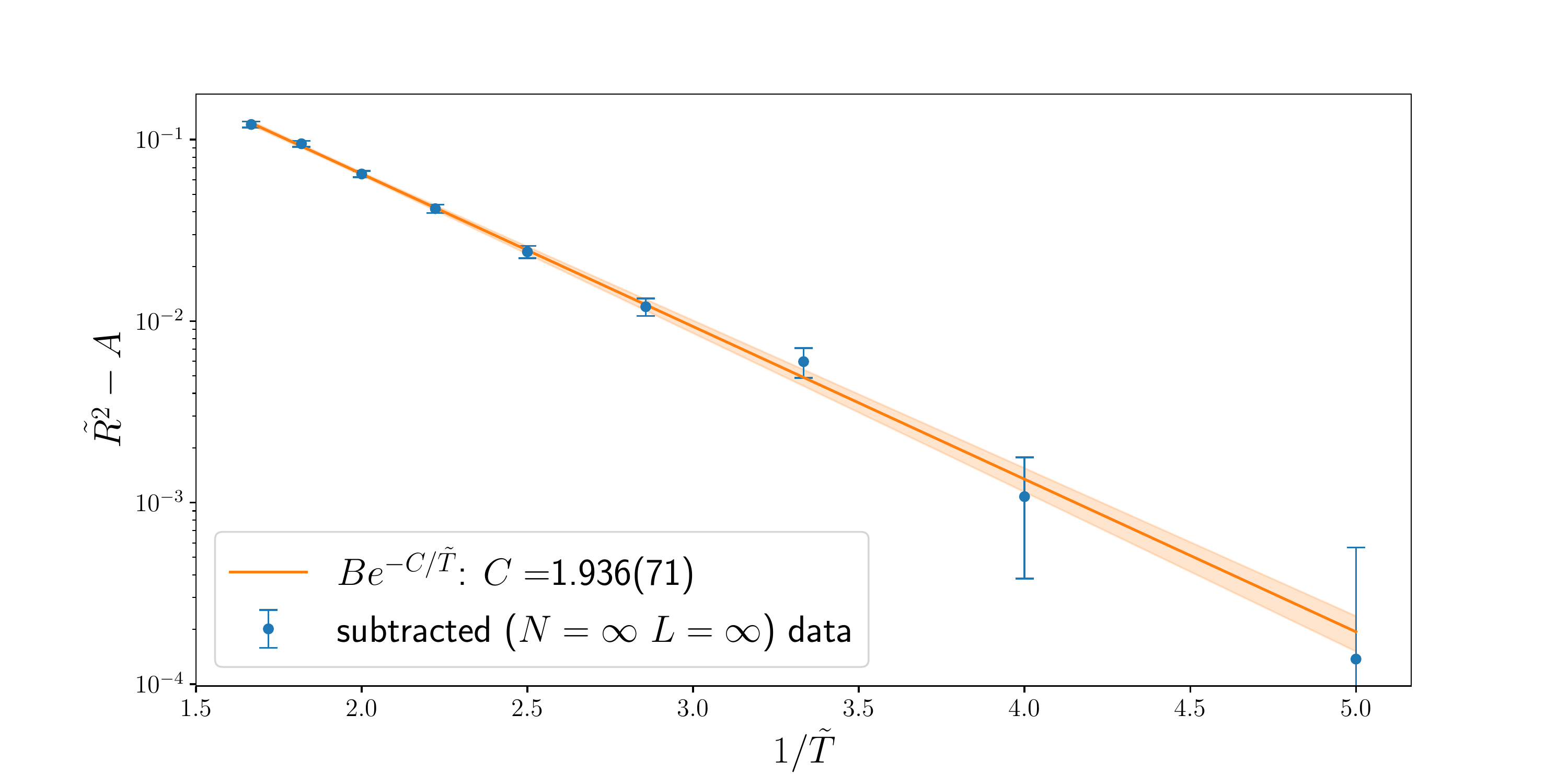}}
\end{center}
\caption{
$\tilde{E}/N^2\equiv(\lambda^{-1/3}E)/N^2$ and $\tilde{R}^2\equiv\lambda^{-2/3}R^2$ in the ungauged bosonic matrix model in the low temperature region, after subtracting the $\tilde{T}=0$ fitted value $A$ (see text and Fig.~\ref{fig:E-and-R2-bosonic}), which represents the energy of the gauged theory in the confining phase.
Both differences behave as a falling exponential $Be^{-C/\tilde{T}}$, where $C\simeq 2.0$.  	
The values of $A$ extracted from the fit agree with the values obtained in Ref.~\cite{Kawahara:2007fn} for the gauged theory.
}\label{fig:E-and-R2-bosonic-log}
\end{figure}

Furthermore, the exponent $C$ in our fit is identified with $\tilde{C}\adj$ and hence should be independent of the observable used for the fit.
We can confirm this expectation within our statistical accuracy: $C=2.043(76)$ for $\tilde{E}/N^2$ and $C=1.936(71)$ for $\tilde{R}^2$.  
The value of $B$ for $\tilde{E}/N^2$ is $20.0(2.9)$, whose error bar is large, but it could be an integer times $C$, as predicted by the conjecture.
In Fig.~\ref{fig:E-and-R2-bosonic-log} we plot the data for $\tilde{E}/N^2$ and $\tilde{R}^2$ after we have subtracted the corresponding values of $A$ obtained from the fits.
In turn, this amounts to calculate the difference between the ungauged and the gauged bosonic theory for the two aforementioned observables, and one can visually check the exponential falloff in the plot. 
The plots in Fig.~\ref{fig:E-and-R2-bosonic-log} show the same fits as the ones reported in Fig.~\ref{fig:E-and-R2-bosonic}.
Again, we remind the reader that there is no known gravity dual for the bosonic model we studied in this section, but the results are encouraging and show that a test of the conjecture with good precision is possible with lattice Monte Carlo methods.

%\clearpage
%%%%%%%%%%%%%%%
%%%%%%%%%%%%%%%
\section{Testing the Maldacena-Milekhin conjecture}\label{sec:fermionic}
\hspace{0.25in}
%%%%%%%%%%%%%%%
%%%%%%%%%%%%%%%

We now move to test the conjecture in the full D0-brane model with fermions, where it is developed.

%%%%%%%%%%%%%%%
%%%%%%%%%%%%%%%
\subsection{Taming the flat direction}
\hspace{0.25in}
%%%%%%%%%%%%%%%
%%%%%%%%%%%%%%%
The biggest obstacle in the simulation of the D0-brane matrix model is the flat directions of the eigenvalues of scalars $X_M$~\cite{deWit:1988xki,Smilga:1989ew}\cite{Anagnostopoulos:2007fw}.\footnote{
This physical feature is crucial to describe many-body states~\cite{Banks:1996vh}.
}
Namely, the bound state of eigenvalues, which is dual to a black zero-brane, is merely metastable, and the eigenvalues (D0-branes) can be emitted and fly away. 
In order to measure the energy of the black zero-brane, we have to tame the flat direction, or in other words, simulate the system in the local minimum of the action corresponding to the metastable state.
For that purpose we adopt the simplest procedure, which does not require any modification to the action: take the matrix size $N$ sufficiently large~\cite{Anagnostopoulos:2007fw,Hanada:2016zxj,Berkowitz:2016znt,Berkowitz:2016jlq,Kadoh:2015mka}. 
The emission rate of the eigenvalues decreases as $N$ becomes large and, for sufficiently large values of $N$, we can collect sufficiently many configurations of the metastable state before the onset of any instability. 

However, the severeness of the instability is regularization dependent.
In fact, bosons and fermions contribute to the attraction and the repulsion, respectively.
Therefore, if the bosonic contributions dominate due to a regularization artifact (which would disappear only in the continuum limit), the system is more stable, and the instability sets in only as one gets closer to the continuum limit.
This is the case for the regularization known as ``momentum cutoff method'' \cite{Anagnostopoulos:2007fw,Hanada:2008ez,Hanada:2013rga,Hanada:2016zxj,Hanada:2008gy,Hanada:2011fq}. 
The opposite is true for our lattice regularization: the instability becomes milder as one approaches the continuum limit.
With the regularization method we have used, we need to take the lattice size $L$ to be sufficiently large---the exact value depends on $N$.
We have also noticed, a posteriori, that the instability is more severe compared to the gauged theory, which has been studied in detail before~\cite{Berkowitz:2016jlq}.

The choices of $\tilde{T}$, $N$ and $L$ for the ungauged BFSS theory are summarized in Appendix~\ref{sec:numerics-detail}. 
We have performed a simultaneous large-$N$ and continuum extrapolation of the ungauged theory observables using different fit functions and different datasets to assess possible sources of systematic errors. 
All details are summarized in Appendix~\ref{sec:numerics-detail}.

%%%%%%%%%%%%%%%
%%%%%%%%%%%%%%%
\subsection{Comparison between gauged and ungauged theories}
\hspace{0.25in}
%%%%%%%%%%%%%%%
%%%%%%%%%%%%%%%
In this section we show the simulation results for $E/N^2$, $F^2$ and $R^2$ to test the conjecture by Maldacena and Milekhin. 
The results for $E/N^2$, $F^2$ are numerically under better control, and appear to be consistent with the conjecture. 
It is hard to make a sharp statement on $R^2$ because the $N$-dependence is large, and this observable is more subject to large fluctuations related to the flat directions.
 
%%%%%%%%%%%%%%%
%%%%%%%%%%%%%%%
\subsubsection{Energy}
\hspace{0.25in}
%%%%%%%%%%%%%%%
%%%%%%%%%%%%%%%
The difference of the energy should behave as 
\begin{eqnarray}
\Delta E=E_{\rm ungauged}-E_{\rm gauged}=d\adj C\adj N^2e^{-C\adj/T}+\cdots, 
\end{eqnarray}
where $d\adj$ is integer, if the conjecture is correct, and $C\adj$ is different from the one of the bosonic theory in Sec.~\ref{sec:bosonic}. 
The details of the analysis for the ungauged theory are summarized in Appendix~\ref{sec:numerics-detail}. 

For the ungauged theory we have data at $\tilde{T}=0.45$, 0.5, 0.6, 0.7, 0.8, 0.9 and 1.0.
At each temperature we have simulated various values of $L$ and $N$ and then performed a simultaneous large-$N$ and continuum fit, including $1/N^2$ corrections.
For the gauged theory, on the other hand, we know from the results in Ref.~\cite{Berkowitz:2016jlq} that there is a good description of the energy as a function of the temperature based on the gauge/gravity duality: $\tilde{E}(\tilde{T}) = a_0 \tilde{T}^{14/5}+a_1\tilde{T}^{23/5}+a_2\tilde{T}^{29/5}$.
We fit the parameters $a_0$, $a_1$ and $a_2$ to the continuum large-$N$ data in Ref.~\cite{Berkowitz:2016jlq} (considering only $\tilde{T}<1.0$ data) and obtain the values of $\tilde{E}_{\rm gauged} $ at the same temperatures of $\tilde{E}_{\rm ungauged}$.
The uncertainties are propagated from the full covariance matrix of the fitted parameters.
Table~\ref{tab:data-energy} summarizes the data used in the following analysis.
The energy $\tilde{E}_{\rm ungauged}$ is obtained from a large-$N$ continuum fit excluding $L=8$ points and including $1/N^2$ corrections (see Appendix~\ref{sec:numerics-detail} for more details on the fits).
\begin{table}
  \centering
  \begin{tabular}{c|c|c}
    % \hline
    $\tilde{T}$ & $\tilde{E}_{\rm ungauged}/N^2$ & $\tilde{E}_{\rm gauged}/N^2$ \\
    \hline
    0.45 & 0.848(60) & 0.593(16)\\
    0.50 & 1.18(10)  &  0.755(15\\
    0.60 & 1.500(34) & 1.126(13)\\
    0.70 & 2.029(40) & 1.544(18)\\
    0.80 & 2.588(47) & 2.010(23)\\
    0.90 & 3.164(50) & 2.557(39)\\
    1.00 & 3.805(56) & 3.26(16) \\
    % \hline
  \end{tabular}
  \caption{Data for the internal energy of the ungauged and gauged theories in the large-$N$ continuum limit.}
  \label{tab:data-energy}
\end{table}

\begin{figure}[th]
\begin{center}
\scalebox{0.45}{
\includegraphics{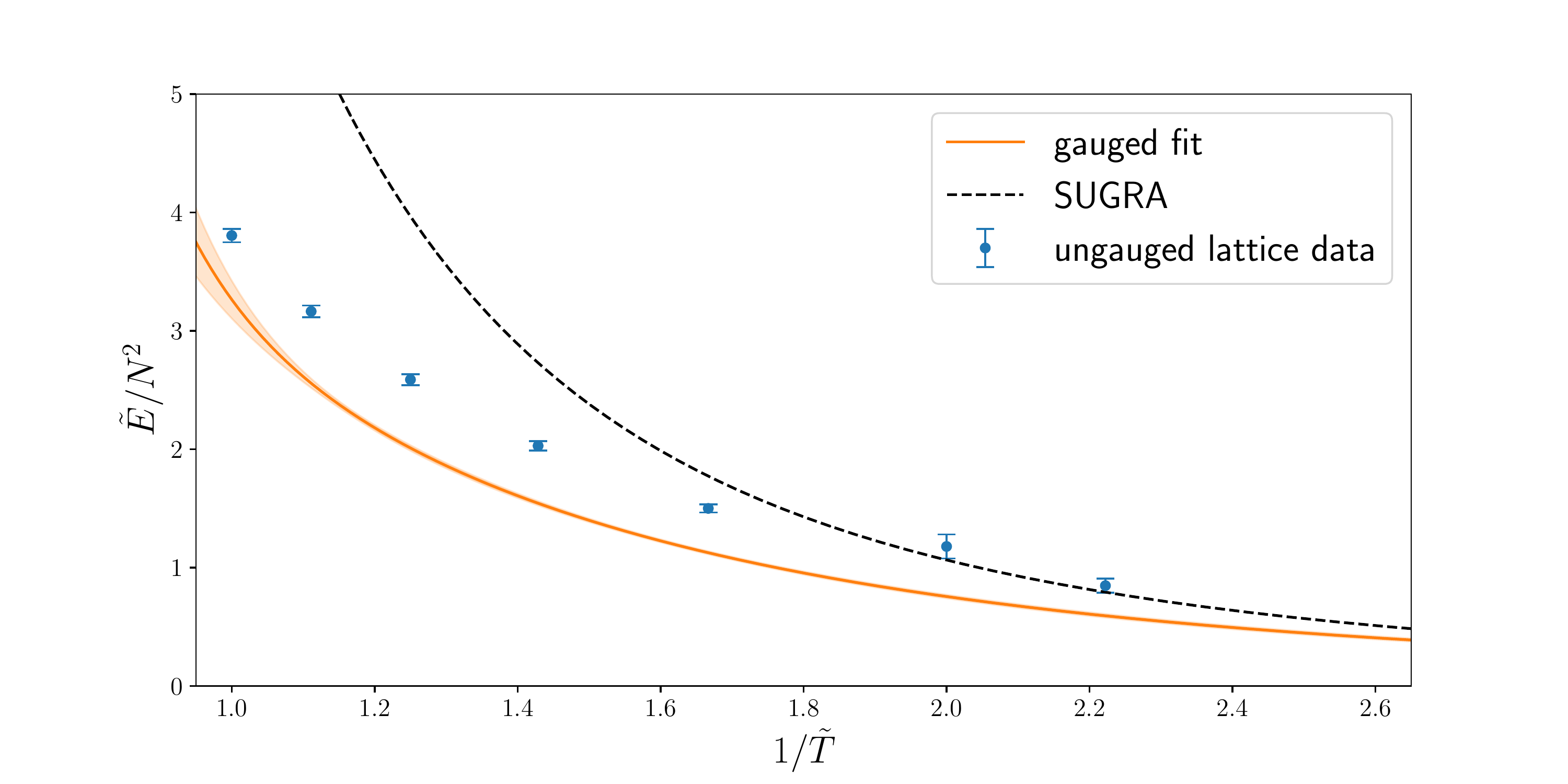}}
\scalebox{0.45}{
\includegraphics{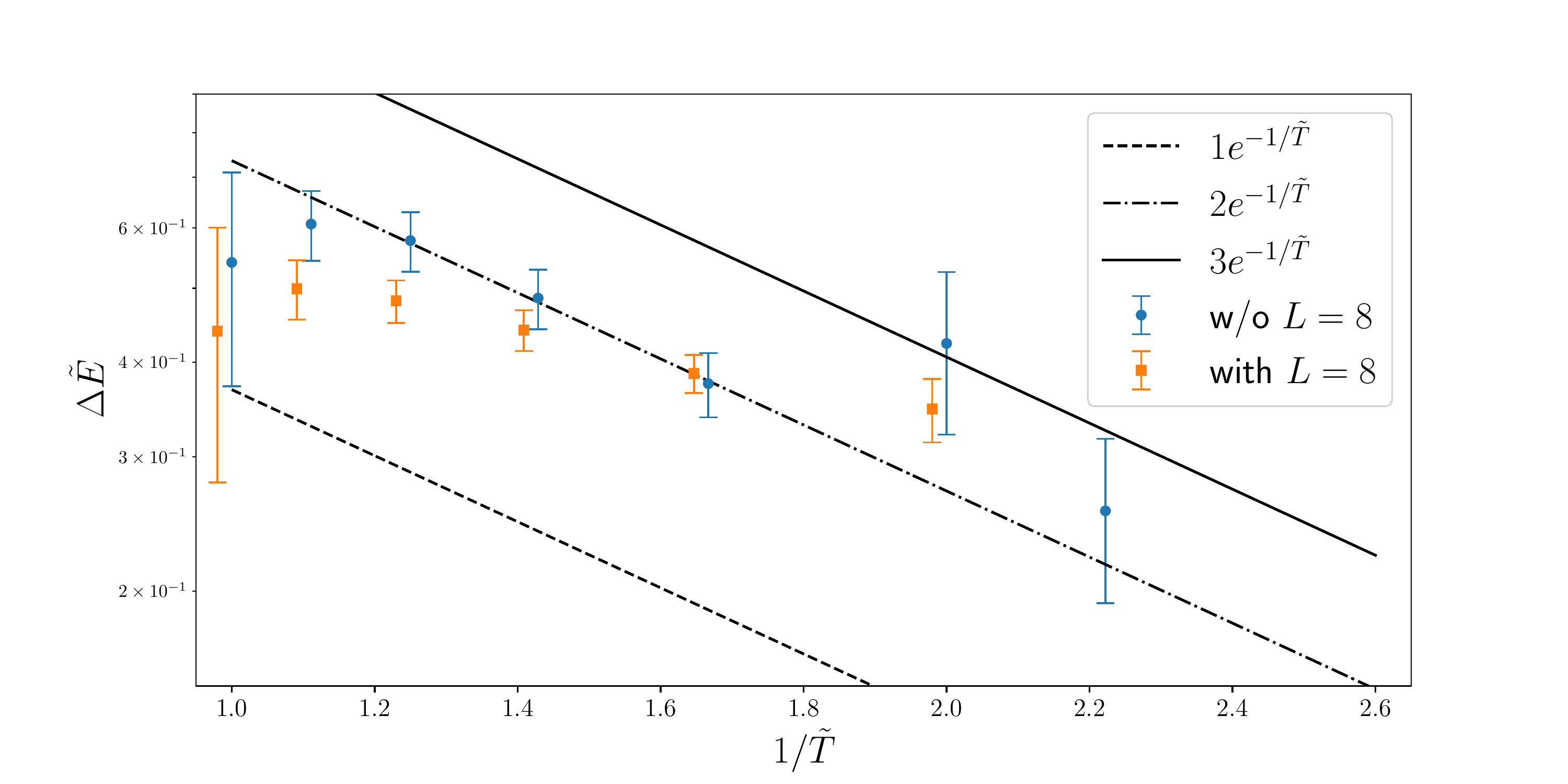}}
\end{center}
\caption{\label{fig:energy-susy-quad}
[Upper] Large-$N$ and continuum $\tilde{E}_{\rm gauged} \equiv \lambda^{-1/3}E_{\rm gauged}$ and $\tilde{E}_{\rm ungauged} \equiv \lambda^{-1/3}E_{\rm ungauged}$ as functions of $\tilde{T} \equiv \lambda^{-1/3}T$. The former is a fit to data in Ref.~\cite{Berkowitz:2016jlq}, and it is compared with the leading order supergravity solution, shown as a dashed black line.
[Lower] Large-$N$ and continuum $\Delta{\tilde{E}} \equiv \tilde{E}_{\rm ungauged} - \tilde{E}_{\rm gauged}$ vs $\tilde{T}$. Two ungauged energy are shown to asses systematic errors: one which includes $L=8$ points in the continuum extrapolation and one that does not. At $\tilde{T}=0.45$, the ungauged simulations have not been performed at $L=8$ because of numerical instabilities related to the flat direction. We also show some representative functional forms to guide the eye. 
}
\end{figure}

\begin{figure}[ht]
\begin{center}
\scalebox{0.5}{ \includegraphics{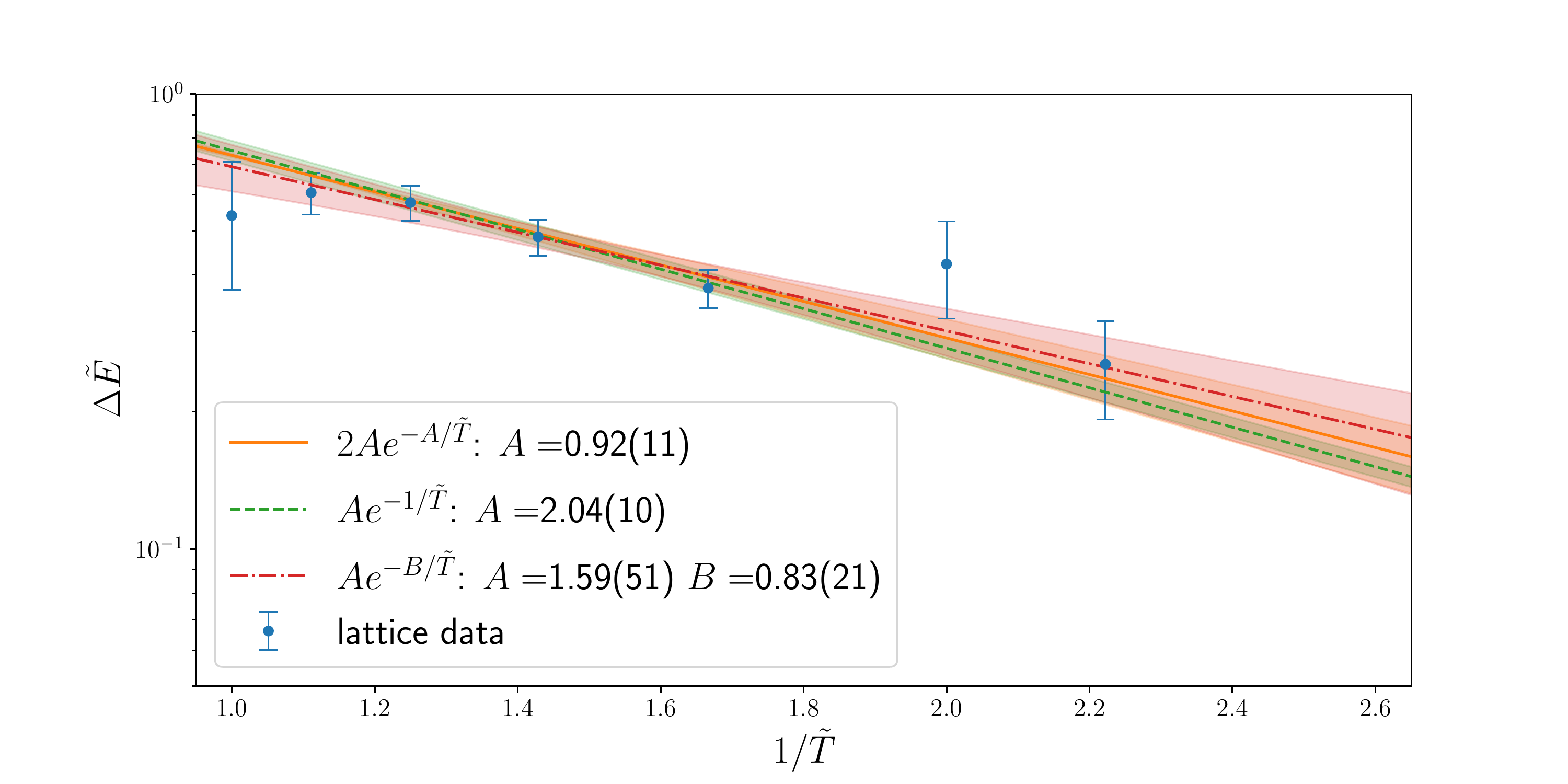}}
\end{center}
\caption{Different exponential fits are shown together with the data. Only five rightmost data points are included in the fits, corresponding to $\tilde{T} \le 0.8$. The fits are all compatible with the data and among each other, with the two-parameter fit having the largest uncertainties. }
\label{fig:dE-fit-quad}
\end{figure}

In Fig.~\ref{fig:energy-susy-quad}, we plot both the fitted functional form for $\tilde{E}_{\rm gauged}$ and raw continuum large-$N$ data for $\tilde{E}_{\rm ungauged}$ in the region $\tilde{T}\le 1.0$.
We also plot $\Delta\tilde{E} \equiv\tilde{E}_{\rm ungauged} - \tilde{E}_{\rm gauged}$ in the same Fig.~\ref{fig:energy-susy-quad}.
This energy difference is consistent with $\Delta E =d\adj C\adj e^{-C\adj/T}$, where $d\adj =2$ and $\tilde{C}\adj=\lambda^{-1/3}C\adj \simeq 1.0$.
We perform several fits on the energy difference data at $\tilde{T}\le 0.8$.
First we fit the functional form expected from the conjecture with $d\adj=2$ and obtain:
\begin{align}
  \label{eq:fit-conjecture-fixD}
  \Delta \tilde{E} \; = \; 2Ae^{-A/\tilde{T}} & \qquad A=0.92 (11) \\ \nonumber
                                              & \qquad \chi^2/{\rm dof}=0.54/4 \; ,
\end{align}
where $A$ is compatible with $\tilde{C}\adj \simeq 1.0$ and the quality of the fit is very good.
We also use the same functional form, but fix $\tilde{C}\adj = 1.0$, leading to:
\begin{align}
  \label{eq:fit-conjecture-fixE}
  \Delta \tilde{E} \; = \; Ae^{-1/\tilde{T}} & \qquad A=2.04 (10) \\ \nonumber 
                                             & \qquad \chi^2/{\rm dof}=0.62/4 \; ,
\end{align}
where we now get the amplitude compatible with $d\adj=2$.
These fits look promising, but they are not conclusive.
In fact, if we fit the same data with a fitting ansatz where the amplitude and the exponent are free parameters we obtain:
\begin{align}
  \label{eq:fit-free}
  \Delta \tilde{E} \; = \; Ae^{-B/\tilde{T}} & \qquad A=1.59 (51) \qquad B=0.83 (21) \\ \nonumber 
                                             & \qquad \chi^2/{\rm dof}=0.64/3 \; ,
\end{align}
where the best fit parameters are compatible with the previous results, but with larger uncertainties.
These fits are compared in Fig.~\ref{fig:dE-fit-quad}.

\clearpage

%%%%%%%%%%%%%%%
%%%%%%%%%%%%%%%
\subsubsection{$F^2$ and $R^2$}
\hspace{0.25in}
%%%%%%%%%%%%%%%
%%%%%%%%%%%%%%%
Let us consider other quantities like $F^2\equiv -\frac{1}{N\beta}\int dt{\rm Tr}[X_M,X_{M'}]^2$ and $R^2\equiv\frac{1}{N\beta}\int dt{\rm Tr}X_M^2$. 
The values of such quantities can be different in the singlet and nonsinglet sectors.\footnote{
The contribution from short open strings hovering at the boundary would be small but not parametrically suppressed.}
However, according to the Maldacena-Milekhin conjecture, the contribution from the nonsinglet sector is suppressed by a Boltzmann factor $e^{-C\adj/T}$ relative to the singlet sector, due to the energy gap $C\adj$.
Hence the expectation value should be dominated by the singlet sector; we expect
\begin{eqnarray}\label{eq:dF2}
\Delta \tilde{F}^2\equiv \tilde{F}^2_{\rm ungauged}-\tilde{F}^2_{\rm gauged}
=
c_{F^2}N^2e^{-C\adj/T}+\cdots,   
\end{eqnarray}
and 
\begin{eqnarray}\label{eq:dR2}
\Delta \tilde{R}^2\equiv \tilde{R}^2_{\rm ungauged}-\tilde{R}^2_{\rm gauged}
=
c_{R^2}N^2e^{-C\adj/T}+\cdots,  
\end{eqnarray}
where $\tilde{F}^2\equiv\lambda^{-4/3}F^2$ and $\tilde{R}^2\equiv\lambda^{-2/3}R^2$, respectively (In Sec.~\ref{sec:bosonic}, we have numerically observed the same phenomenon happening in the purely bosonic theory, with a different value for $C\adj$, which depends on the theory).
Unlike the case of $\Delta E$, there is no particular prediction for the overall coefficients $c_{F^2}$ and $c_{R^2}$.  

The details of the analyses for $\tilde{F}^2$ and $\tilde{R}^2$ are given in Appendix~\ref{sec:numerics-detail}.
To summarize, we choose to look at two different sets of data, reported in Tab.~\ref{tab:data-f2-r2-largeN} and Tab.~\ref{tab:data-f2-r2-largestN} for the ungauged and the gauged theories.
For the gauged theory, we use the raw data results reported in Ref.~\cite{Berkowitz:2016jlq} and we extrapolate them to the continuum and large-$N$ limit.
One dataset is extrapolated to the $N \rightarrow \infty$ limit, while the other one is not, using instead the largest $N$ at each temperature as a proxy to estimate what systematic errors we would make by neglecting finite-$N$ corrections.

We have extrapolated results in both theories for six temperatures $\tilde{T}\ge 0.5$\footnote{
Unlike $E/N^2$, we do not know reasonable fit ans\"atze for $F^2$ and $R^2$. Therefore we do not try to obtain the values at $\tilde{T}=0.45$ by fitting the temperature dependence from other values of $\tilde{T}$.  We estimate the difference only at $\tilde{T}\ge 0.5$.}
at which we compute the differences that enter Eq.~\eqref{eq:dF2} and Eq.~\eqref{eq:dR2}.

\begin{table}[ht]
  \centering
  \begin{tabular}{c|c|c||c|c}
    % \hline
    $\tilde{T}$ & $\tilde{F}^{2}_{\rm ungauged}$ & $\tilde{F}^{2}_{\rm gauged}$ & $\tilde{R}^{2}_{\rm ungauged}$ & $\tilde{R}^{2}_{\rm gauged}$ \\
    \hline
    0.5 & 19.60(15) & 18.998(29) & 3.707(33) & 3.6349(65) \\
    0.6 & 19.830(51)& 19.268(37) & 3.714(12) & 3.6509(74)\\
    0.7 & 20.111(56)& 19.488(42) & 3.729(11) & 3.6571(73)\\
    0.8 & 20.430(67)& 19.654(50) & 3.752(12) & 3.6610(87)\\
    0.9 & 20.788(79)& 19.884(58) & 3.775(13) & 3.6719(91)\\
    1.0 & 21.219(84)& 20.266(54) & 3.811(13) & 3.6940(88)\\
    % \hline
  \end{tabular}
  \caption{Data for the potential energy term $\tilde{F}^2$ and the extent of space $\tilde{R}^2$ of the ungauged and gauged theories in the large-$N$ continuum limit.}
  \label{tab:data-f2-r2-largeN}
\end{table}

\begin{table}[ht]
  \centering
  \begin{tabular}{c|c|c||c|c}
    % \hline
    $\tilde{T}$ & $\tilde{F}^{2}_{\rm ungauged}$ & $\tilde{F}^{2}_{\rm gauged}$ & $\tilde{R}^{2}_{\rm ungauged}$ & $\tilde{R}^{2}_{\rm gauged}$ \\
    \hline
    0.5 & 19.390(54) & 18.947(50) & 3.714(11) & 3.6337(97)\\
    0.6 & 19.758(55) & 19.245(50) & 3.728(10) & 3.681(12) \\
    0.7 & 20.136(59) & 19.471(56) & 3.753(10) & 3.671(12) \\
    0.8 & 20.506(74) & 19.629(68) & 3.778(12) & 3.675(12) \\
    0.9 & 20.892(74) & 19.871(74) & 3.804(11) & 3.682(14) \\
    1.0 & 21.299(80) & 20.138(83) & 3.834(12) & 3.7192(99)\\ 
  \end{tabular}
  \caption{Data for the potential energy term $\tilde{F}^2$ and the extent of space $\tilde{R}^2$ of the ungauged and gauged theories at the largest $N$ available for each $\tilde{T}$ and in the continuum limit. $N=32$ for all temperatures in the ungauged theory, and for the lowest temperature in the gauged theory, but $N=24$ for $\tilde{T} \ge 0.6$.}
  \label{tab:data-f2-r2-largestN}
\end{table}

\begin{figure}[ht]
\begin{center}
\scalebox{0.45}{
\includegraphics{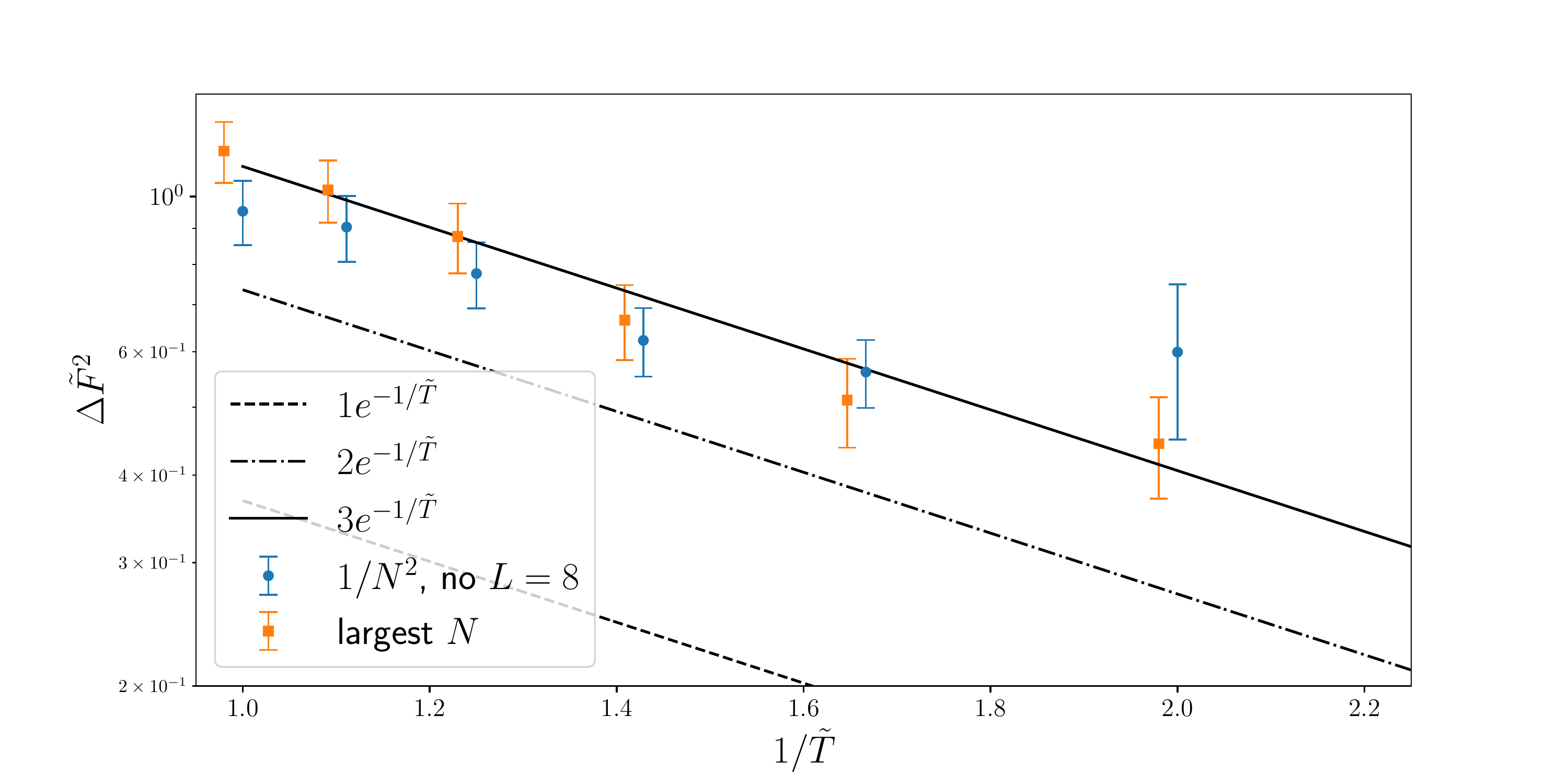}}
\scalebox{0.45}{
\includegraphics{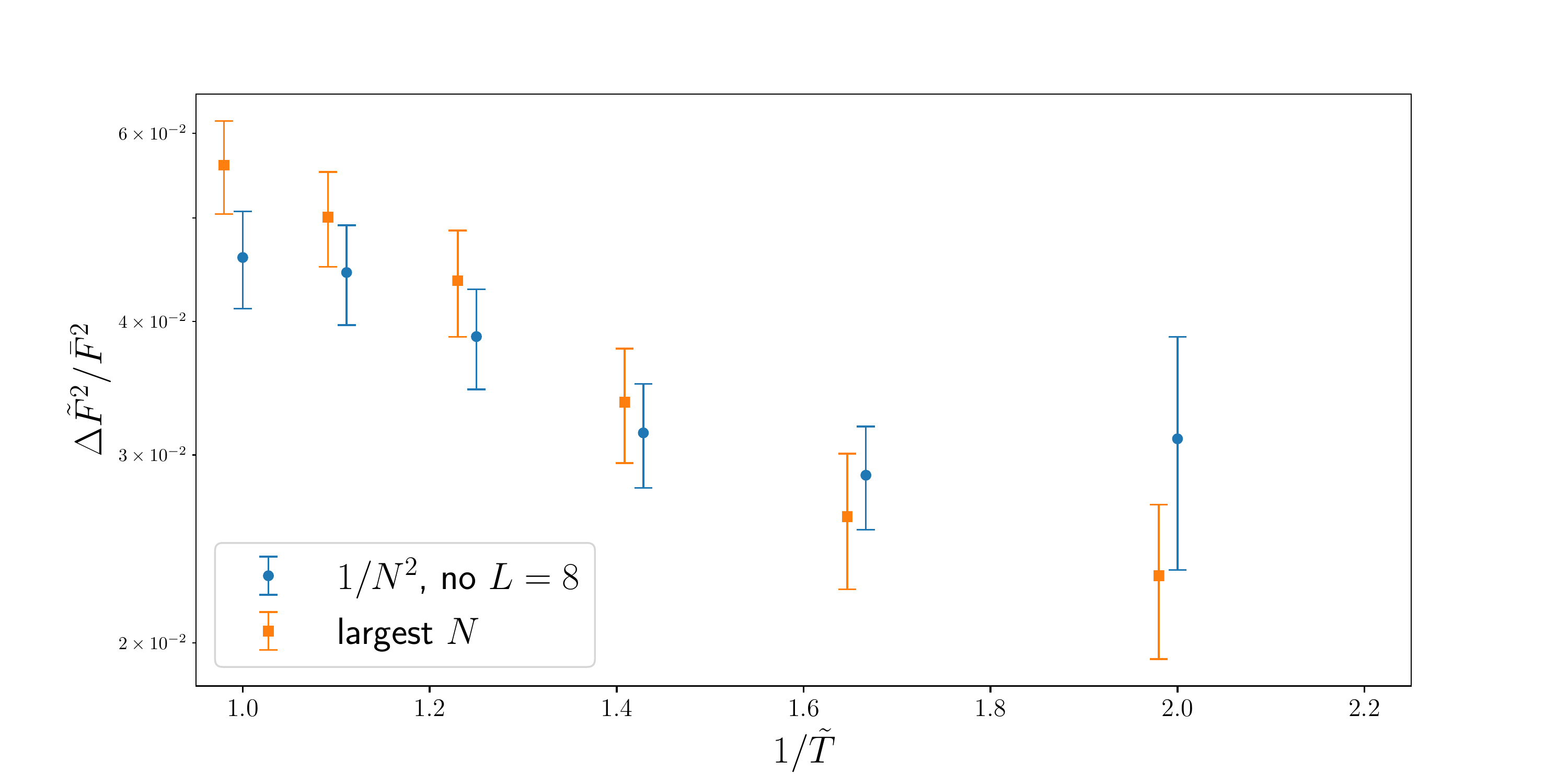}}
\end{center}
\caption{[Upper] $\Delta\tilde{F}^2$ as functions of $1/\tilde{T}$ together with some representative functional forms to guide the eye (as done for $\Delta\tilde{E}$). The datasets shown are in the large-$N$ continuum limit or at our largest $N$ value. 
[Lower] $\Delta \tilde{F}^2/\bar{F}^2$ as functions of $1/\tilde{T}$, where $\bar{F}^2$ is the average value between the ungauged and the gauged theory. There is a clear decreasing trend as the temperature decreases.
\label{fig:dF2-summary}}
\end{figure}

\begin{figure}[ht]
\begin{center}
\scalebox{0.45}{
\includegraphics{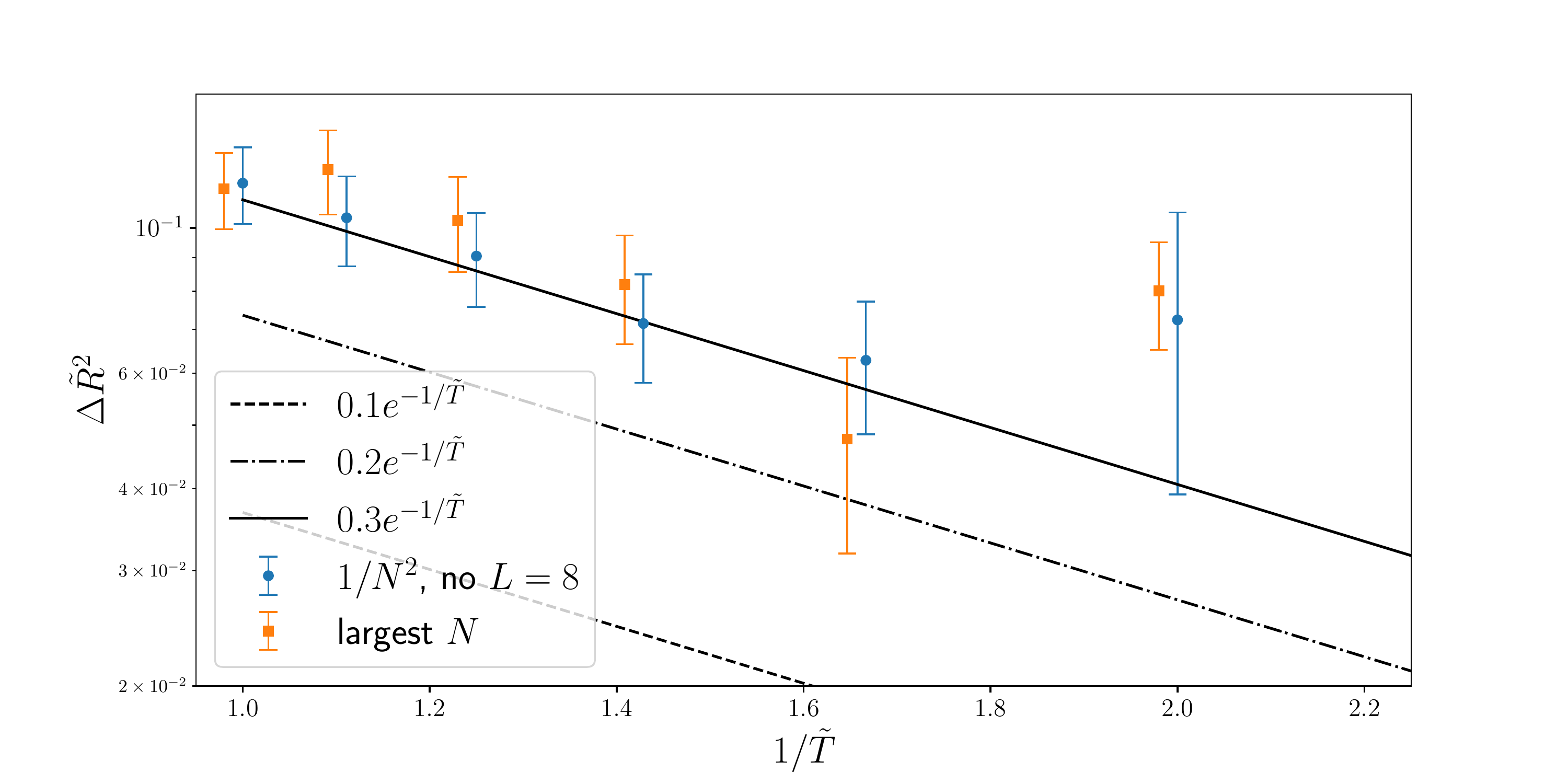}}
\scalebox{0.45}{
\includegraphics{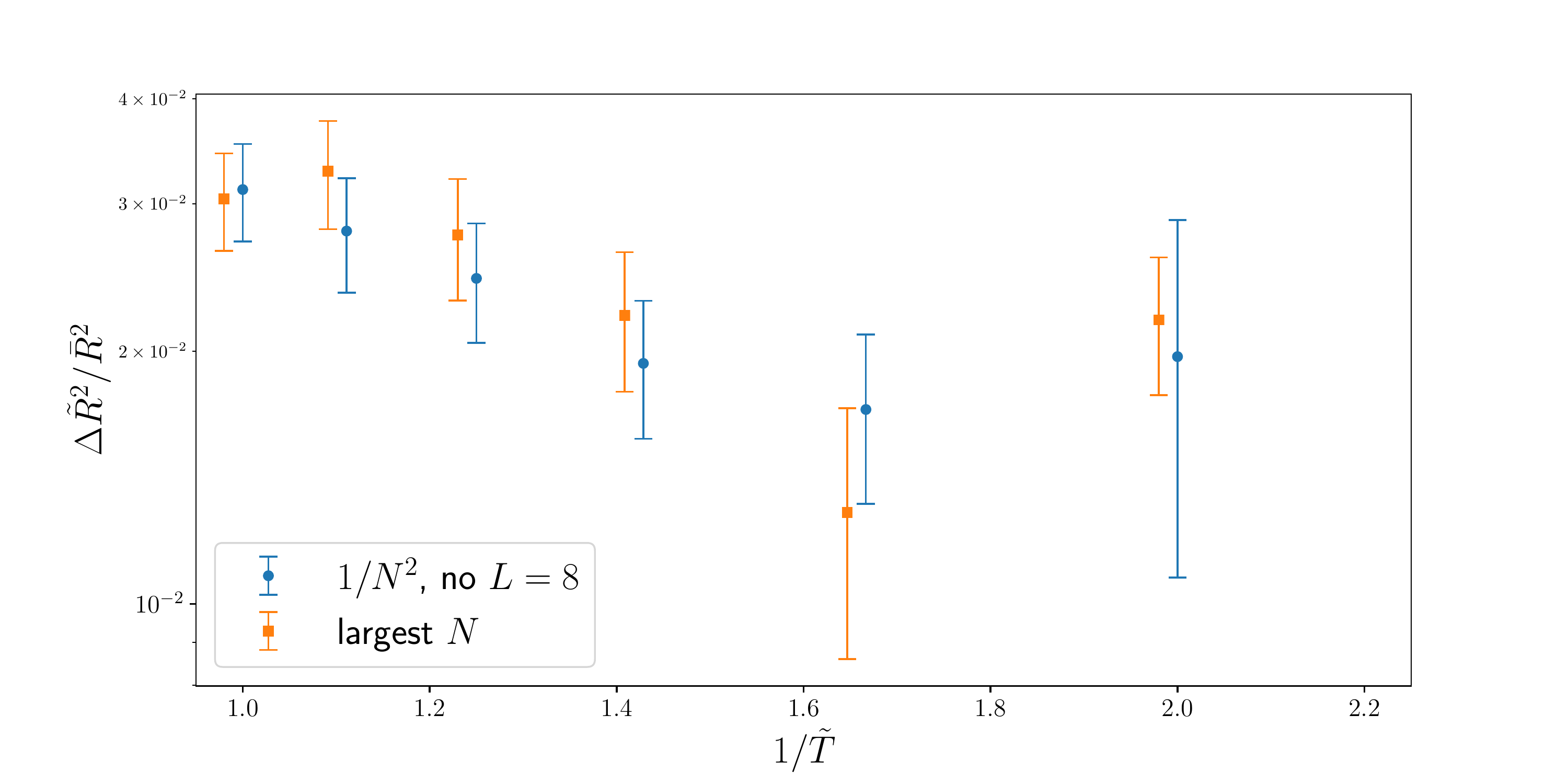}}
\end{center}
\caption{[Upper] $\Delta\tilde{R}^2$ as functions of $1/\tilde{T}$ together with some representative functional forms to guide the eye (as done for $\Delta\tilde{F}^2$, but with ten times smaller coefficients). The datasets shown are in the large-$N$ continuum limit or at our largest $N$ value. 
[Lower] $\Delta \tilde{R}^2/\bar{R}^2$ as functions of $1/\tilde{T}$, where $\bar{R}^2$ is the average value between the ungauged and the gauged theory. There is a clear decreasing trend as the temperature decreases, but the statistical uncertainties become too large at the smallest temperatures to say anything conclusive about the $\tilde{T}=0$ limit.
\label{fig:dR2-summary}}
\end{figure}

In the upper panel of Fig.~\ref{fig:dF2-summary} and Fig.~\ref{fig:dR2-summary}, we have plotted $\Delta \tilde{F}^2$ and $\Delta \tilde{R}^2$, respectively, as a function of the inverse temperature.
We have also plotted the relative difference $\Delta \tilde{F}^2/\bar{F}^2$, where $\bar{F}^2$ is just the average value of $\tilde{F}^2$ between the ungauged and the gauged theory (similarly for $\tilde{R}^2$).
Despite large error bars at $1/\tilde{T}\ge 2.0$, the trend that $\Delta\tilde{R}^2$ and $\Delta\tilde{F}^2$ go to zero as $\tilde{T}\to 0$ can be seen.
The relative difference is on the order of a few percent everywhere at $\tilde{T}\le 1.0$, which suggests typical matrix configurations important in the path integral are very close. 
Note that the relative difference of the energy, $\Delta \tilde{E}/\bar{E}$, is larger. 
However, it does not seem to be an immediate problem, because $\bar{E}$, the average internal energy between the ungauged and gauge theory, is zero at zero temperature 
due to the cancellation between bosonic and fermionic contributions.  
If we take the ratio of $\Delta \tilde{E}$ and another natural energy scale, e.g. the zero-point energy in the bosonic theory, it can be equally small.
Moreover, it is interesting to stress that, at very high temperatures, the ratios above are identical to that of the bosonic theory, which is $\Delta \tilde{E}/\bar{E}=\Delta \tilde{F}^2/\bar{F}^2 \sim 12\%$ (see Appendix~\ref{sec:high-T}), and that at our simulated temperatures we are already seeing a dramatic reduction, with ratios down to the 2\% level.

\begin{figure}[ht]
\begin{center}
\scalebox{0.45}{ \includegraphics{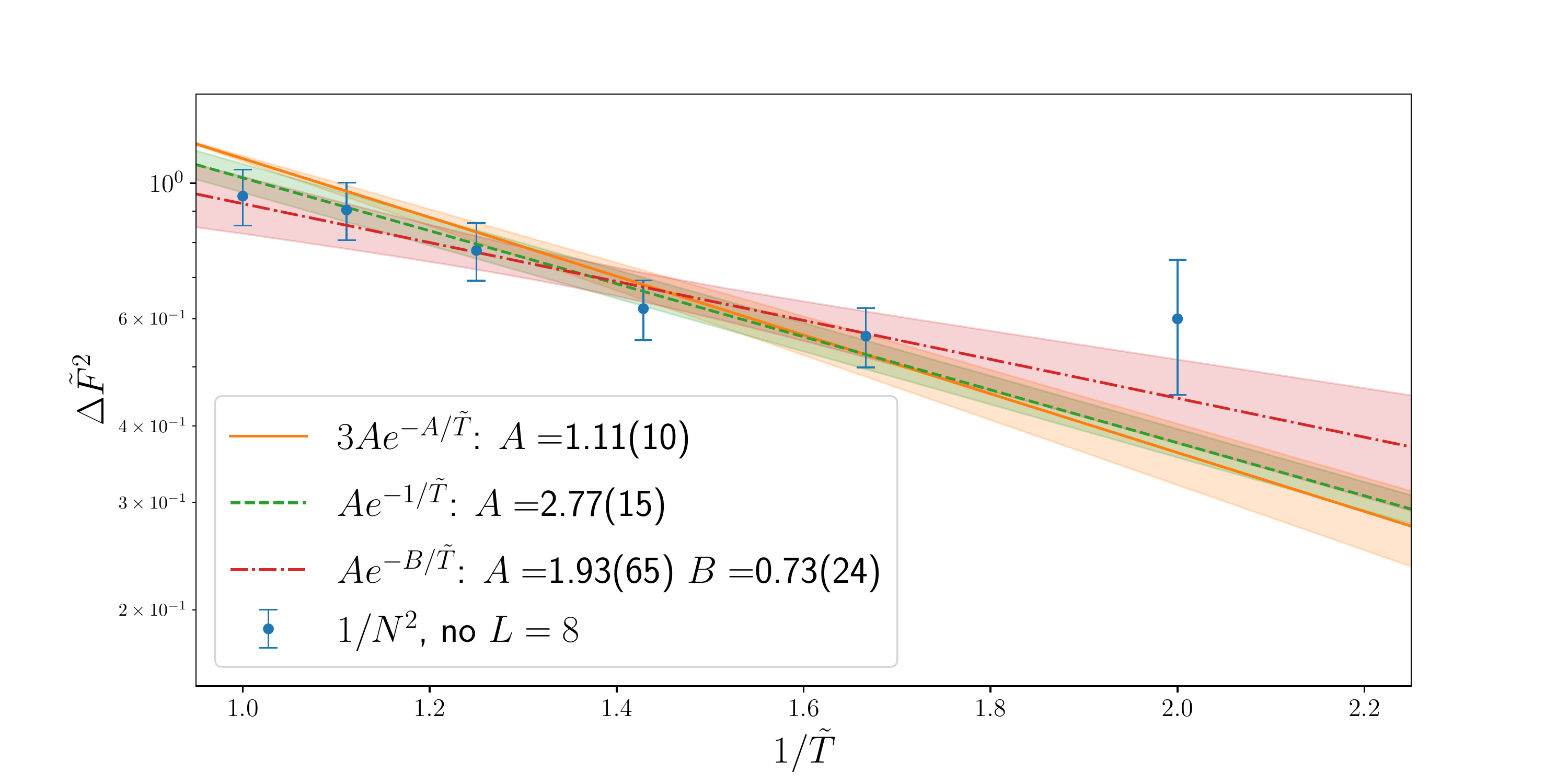}}
\scalebox{0.45}{ \includegraphics{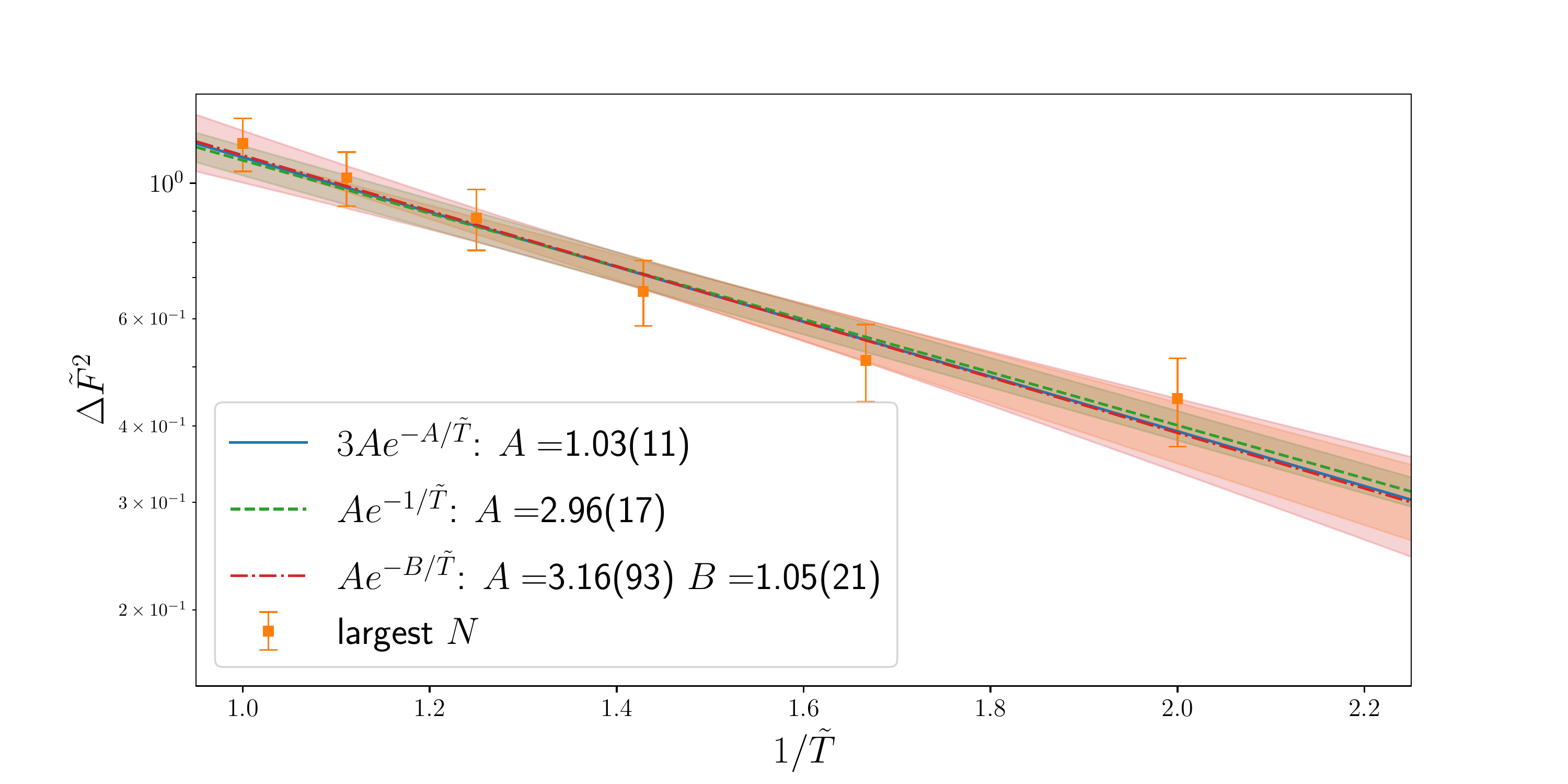}}
\end{center}
\caption{\label{fig:dF2-fit-quad}
[Upper] Different exponential fits are shown together with the lattice data obtained from a simultaneous continuum and large-$N$ extrapolation (reported in Tab.~\ref{tab:data-f2-r2-largeN}). [Lower] The same exponential fits are shown using a different dataset where the largest-$N$ results are used at each temperature (reported in Tab.~\ref{tab:data-f2-r2-largestN}).
In both cases, only five rightmost data points are included in the fits, corresponding to $\tilde{T} \le 0.9$. The fits are all compatible with the data and among each other, with the two-parameter fit having the largest uncertainties.
Although we don't find any theoretical reason that the overall coefficient is integer, numerically it appears to be rather close to 3.
}
\end{figure}

For $\Delta \tilde{F}^2$ we can attempt to do a fit to the data, similarly to what we did for the internal energy, while, unfortunately, it is hard to make any quantitative statement for $\Delta \tilde{R}^2$, where the error bars are larger. 
The exponential decay of $\Delta \tilde{F}^2$ is consistent with $e^{-\tilde{C}\adj/\tilde{T}}$ with $\tilde{C}\adj \simeq 1.0$, as we can see in Fig.~\ref{fig:dF2-fit-quad}.
The fits we performed on $\Delta \tilde{F}^2$ are similar to the ones we did for the internal energy in Eq.~\eqref{eq:fit-conjecture-fixD} to Eq.~\eqref{eq:fit-free} and they all have reduced $\chi^2$ close to one.
Fig.~\ref{fig:dF2-fit-quad} shows all three fits on the aforementioned two different datasets independently.
Although we don't find any theoretical reason that the overall coefficient is integer, numerically it appears to be rather close to 3, while it was closer to 2 for the internal energy.
The exponential decay of $\Delta \tilde{E}$ has the same scale: $\tilde{C}\adj\simeq1.0$.
We have seen this observable-independence in the bosonic case as well, and we remark that this is expected from the conjecture.

\clearpage

%%%%%%%%%%%%%%%
%%%%%%%%%%%%%%%
\section{Conclusion and Discussion}
\hspace{0.25in}
%%%%%%%%%%%%%%%
%%%%%%%%%%%%%%%
In this paper we have tested a recent conjecture by Maldacena and Milekhin\cite{ungauged} by studying the ungauged D0-brane matrix model with numerical lattice Monte Carlo simulations.
Our results shown in Sec.~\ref{sec:fermionic} appear to be consistent with the conjecture, given our statistical accuracy.
More detailed tests of this conjecture --- with higher precision, at lower temperatures, and for more observables --- are straightforward in principle and should be performed in the future to increase confidence in the conjecture.
While analytic methods, such as supersymmetric localization \cite{Pestun:2016zxk}, may work\footnote{
See Ref.~\cite{ungauged} for a supersymmetric modification of the ungauged theory.}
for certain problems, Monte Carlo simulation can be applied to more generic situations.
It is important to pursue various approaches which can lead to complementary results.
If the conjectured duality between gravity and the ungauged theory is correct, there are several interesting directions.
On the gravity side of the story, in addition to the issues raised already in Ref.~\cite{ungauged}, whether the ungauged theory fits into the supermembrane interpretation \cite{deWit:1988ig}, in which the gauge transformation is identified with the area-preserving diffeomorphism, would be an important question to address.
It would also be interesting to see how the basic results like D0-brane scattering \cite{Danielsson:1996uw,Kabat:1996cu,Becker:1997xw} may or may not be modified.
Another interesting question is whether such ungauging procedure can make sense for other theories which do not have dual gravity interpretations;
hopefully the ungauging procedure can solve some technical issues already mentioned in the introduction.

%%%%%%%%%%%%%%%
%%%%%%%%%%%%%%%
\section*{Acknowledgement}
\hspace{0.25in}
%%%%%%%%%%%%%%%
%%%%%%%%%%%%%%%
We would like to thank Juan Maldacena and Alexey Milekhin for valuable discussions at various stages of this work. 
M.~H. also thanks to Dio Anninos, Frank Ferrari and Hidehiko Shimada for insightful comments. 
The work of M.~H. is supported in part by the Grant-in-Aid of the Japanese Ministry of Education, Sciences and Technology, Sports and Culture (MEXT) for Scientific Research (No.~17K14285).
E.~R. is supported by a RIKEN Special Postdoctoral fellowship.
This research was supported in part by the International Centre for Theoretical Sciences (ICTS) during a visit (M.~H. and E.~B.) for the program --- Nonperturbative and Numerical Approaches to Quantum Gravity, String Theory and Holography (ICTS/NUMSTRINGS/2018/01).
This work is supported in part by the DFG and the NSFC through funds provided to the Sino-German CRC 110 ``Symmetries and the Emergence of Structure in QCD'' (E.~B.).
This work was performed under the auspices of the U.S. Department of Energy by Lawrence Livermore National Laboratory under contract~{DE-AC52-07NA27344}.
Computing was provided through the Lawrence Livermore National Laboratory (LLNL) Institutional Computing Grand Challenge program.

\newpage
%%%%%%%%%%%%%%%%%%%%%%%%%%%%%%%%%%%%%%%%%%%%%%

\newpage

\appendix

%%%%%%%%%%%%%%%
%%%%%%%%%%%%%%%
\section{Monte Carlo data analysis}\label{sec:numerics-detail}
\hspace{0.25in}
%%%%%%%%%%%%%%%
%%%%%%%%%%%%%%%

This appendix contains some technical details about the numerical analysis of our lattice Monte Carlo data.
We describe our data analysis workflow and we report various details about the extrapolations to the continuum limit and to the large-$N$ limit.
Summary tables for the observables measured in the ungauged bosonic matrix model and the ungauged full matrix model are reported in this section.

%%%%%%%%%%%%%%%
%%%%%%%%%%%%%%%
\subsection{Bosonic theory}
\hspace{0.25in}
%%%%%%%%%%%%%%%
%%%%%%%%%%%%%%%
We have studied the ungauged version of the bosonic matrix model by using the `naive' regularization of the continuum action:
\begin{eqnarray}
S_{b,ungauged}
&= &
\frac{N}{2a\lambda}\sum_{t,M}{\rm Tr}\left(X_M(t+a)-X_M(t)\right)^2
-
\frac{Na}{4\lambda}\sum_{t,M,N}{\rm Tr}[X_M(t),X_N(t)]^2.  
\nonumber\\
\end{eqnarray} 
The simulation parameters $\tilde{T}$, $N$ and $L$ are summarized in Tab.~\ref{tab:bosonic-ungauged} together with the accumulated Monte Carlo samples and the average value of two observables, the internal energy $\tilde{E}/N^2$ and the extent of space $\tilde{R}^2$.
The Monte Carlo samples are retained after 5000 samples of thermalization, and are binned in independent bins that are five times larger than the autocorrelation time of each observable, also reported in Tab.~\ref{tab:bosonic-ungauged}.
Different observable can have different autocorrelation functions and this is observed in the different number of independent bins that are available for $\tilde{E}/N^2$ and $\tilde{R}^2$: the latter observables tend to have less bins because it has longer autocorrelation times.

\begin{figure}
\begin{center}
\scalebox{0.5}{
\includegraphics{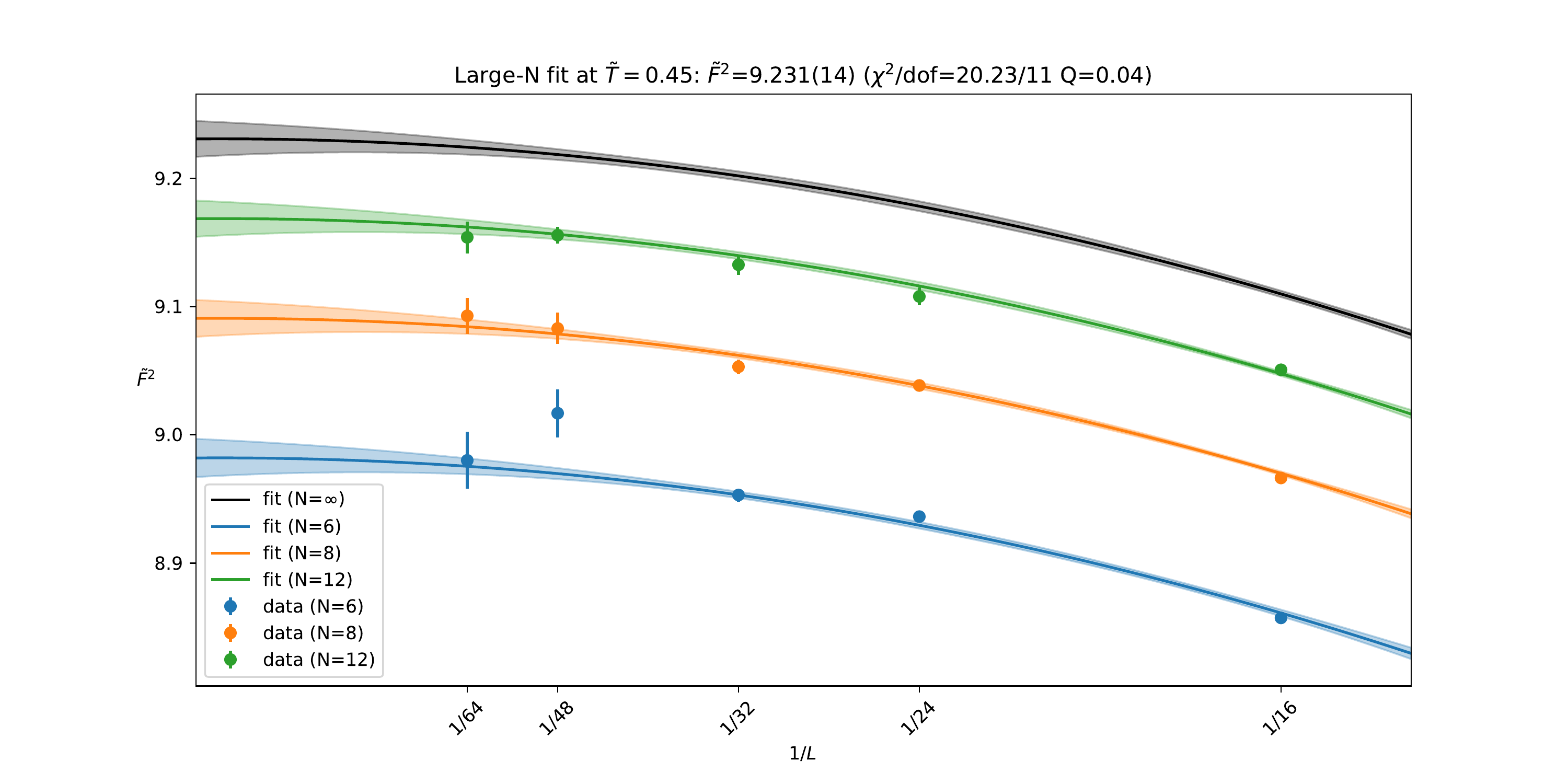}}
\end{center}
\caption{
Continuum and large-$N$ limit extrapolation of $\tilde{F}^2=\frac{4}{3}\tilde{E}/N^2$ for the bosonic ungauged matrix model at $\tilde{T}=0.45$. The title of the plot summarizes the large-$N$ continuum limit value of $\tilde{F}^2$, the $\chi^2$ with corresponding degrees of freedom, and the $Q$ value of the fit. 
}\label{fig:energy-bosonic-continuum-extrapolation}
\end{figure}

\begin{figure}
\begin{center}
\scalebox{0.5}{
\includegraphics{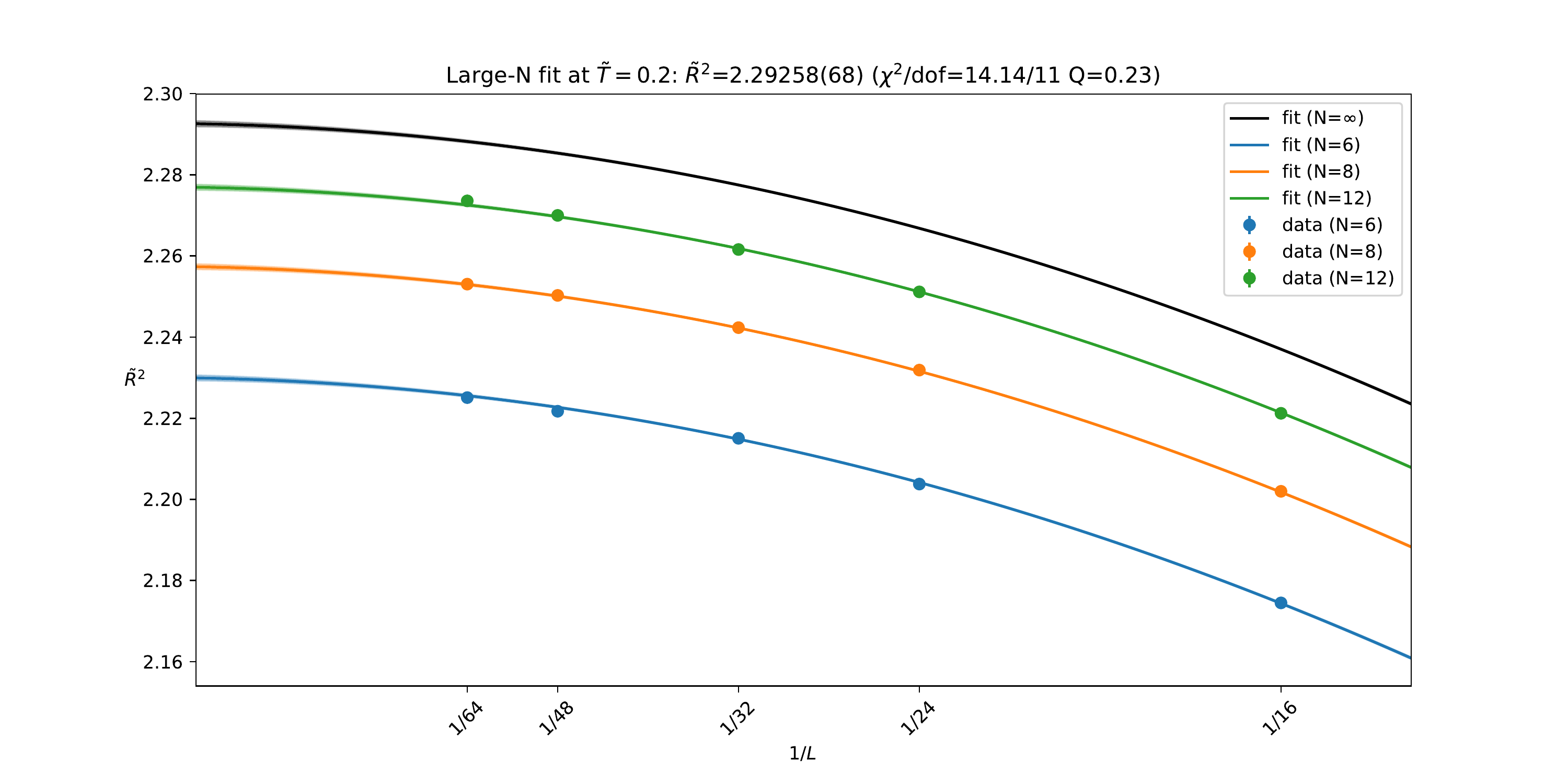}}
\end{center}
\caption{
Continuum and large-$N$ limit extrapolation of $\tilde{R}^2$ for the bosonic ungauged matrix model at $\tilde{T}=0.2$.
The title of the plot summarizes the large-$N$ continuum limit value of $\tilde{R}^2$, the $\chi^2$ with corresponding degrees of freedom, and the $Q$ value of the fit.
}\label{fig:r2-bosonic-continuum-extrapolation}
\end{figure}

For the simultaneous continuum and large-$N$ extrapolations of both our observables, we have used the fitting ansatz defined in Ref.~\cite{Berkowitz:2016jlq},
\begin{equation}
\mathcal{O}(\tilde{T},N,L) \; = \; \sum_{i,j\ge 0}o_{ij}N^{-2i}L^{-j} \, ,
\end{equation}
where $\mathcal{O}(\tilde{T},N,L)$ is the expectation value of an observable directly measured on the lattice at fixed $\tilde{T}$, $N$ and $L$.
We performed a 4-parameter fit with $o_{00}$, $o_{01}$, $o_{02}$ and $o_{10}$ while the other coefficients are set to zero.
The fits are done on the observables $\tilde{F}^2$ and $\tilde{R}^2$ and representative results are shown in Fig.~\ref{fig:energy-bosonic-continuum-extrapolation} and Fig.~\ref{fig:r2-bosonic-continuum-extrapolation}. 
The plots show both the fitted data points and the fitted curve---the different curves are constant-$N$ slices of the best-fit surface as a function of $1/L$, while the black curve is the large-$N$ limit.
The final results for the internal energy $\tilde{E}/N^2$ and the extent of space $\tilde{R}^2$ are summarized in Tab.~\ref{tab:energy-bosonic} and Tab.~\ref{tab:r2-bosonic}.

\begin{table}[ht]
\begin{center}
\begin{tabular}{|r|lr|}
  \hline
  $\tilde{T}$ & \multicolumn{1}{c}{$\tilde{E}/N^2$} & $\chi^2$/dof \\
  \hline
  \hline
  0.20 & 6.7143(36) & 1.58 \\
  0.25 & 6.7142(43) & 4.19 \\
  0.30 & 6.7376(58) & 2.23 \\
  0.35 & 6.7676(65) & 1.29 \\
  0.40 & 6.8310(93) & 1.90 \\
  0.45 & 6.923(10) & 1.84 \\
  0.50 & 7.052(14) & 1.52 \\
  0.55 & 7.217(19) & 2.37 \\
  0.60 & 7.353(25) & 1.60 \\
  \hline
\end{tabular}
\end{center}
\caption{Continuum large-$N$ values for the internal energy of the bosonic ungauged theory. Reduced $\chi^2$ values are also reported. Extrapolations at $\tilde{T}=0.25$ have a larger-then-usual $\chi^2/{\rm dof} > 4$. \label{tab:energy-bosonic}}
\end{table}

\begin{table}[ht]
\begin{center}
\begin{tabular}{|r|lr|}
    \hline
  $\tilde{T}$ & \multicolumn{1}{c}{$\tilde{R}^2$} & $\chi^2$/dof \\
    \hline
    \hline
0.20 & 2.29258(68) & 1.29 \\
0.25 & 2.29352(81) & 3.63 \\
0.30 & 2.2984(11) & 2.13 \\
0.35 & 2.3044(12) & 1.35 \\
0.40 & 2.3165(17) & 1.88 \\
0.45 & 2.3341(20) & 1.58 \\
0.50 & 2.3571(25) & 1.34 \\
0.55 & 2.3873(36) & 2.23 \\
0.60 & 2.4136(46) & 1.65 \\
\hline
\end{tabular}
\end{center}
\caption{Continuum large-$N$ values for $\tilde{R}^2$ of the bosonic ungauged theory. Reduced $\chi^2$ values are also reported. Extrapolations at $\tilde{T}=0.25$ have a larger-then-usual $\chi^2/{\rm dof} \sim 4$.\label{tab:r2-bosonic}}
\end{table}

%%%%%%%%%%%%%%% 
%%%%%%%%%%%%%%%
\subsection{Full theory with fermions}
\hspace{0.25in}
%%%%%%%%%%%%%%%
%%%%%%%%%%%%%%%
For the gauged theory, we take the data from Ref.~\cite{Berkowitz:2016jlq}.
The interested reader can find the tables in Appendix B of Ref.~\cite{Berkowitz:2016jlq}.
For the observable $\tilde{E}/N^2$, the large-$N$ and continuum values for the gauged theory were presented in Ref.~\cite{Berkowitz:2016jlq} and were fitted to 
\begin{equation}
  \label{eq:appendix-sugra}
  \tilde{E}/N^{2}(\tilde{T}) = a_0 \tilde{T}^{14/5}+a_1\tilde{T}^{23/5}+a_2\tilde{T}^{29/5} \quad .
\end{equation}
We perform the same fit on the data at $\tilde{T}\le 0.9$ and reproduce the results of Ref.~\cite{Berkowitz:2016jlq} within statistical accuracy:
\begin{align}
  \label{eq:parameters-sugra}
  a_0 & = 7.28(46) \\ \nonumber
  a_1 & = -9.3(2.0)  \\ \nonumber
  a_2 & = 5.3(1.7)
\end{align}
The full covariance matrix for the fitted parameters is reported in Tab.~\ref{tab:covariance-sugra} and it is used to obtain accurate error bars on $\tilde{E}/N^{2}(\tilde{T})$ for the same values of $\tilde{T}$ used in the ungauged theory.
\begin{table}
  \begin{center}
    \begin{tabular}{c|ccc}
      $\Sigma$ & $a_0$ & $a_1$ & $a_2$ \\ \hline
      $a_0$ & 0.21424042 & -0.93004255 & 0.77234741 \\
      $a_1$ & -0.93004255 &  4.17485676 & -3.52155514 \\
      $a_2$ &  0.77234741 & -3.52155514 &  2.99485795 \\
\end{tabular}
\end{center}
\caption{\label{tab:covariance-sugra} Covariance matrix for the parameters of the supergravity fit with string corrections in Eq.~\eqref{eq:appendix-sugra}.}
\end{table}

For the other observables, $\tilde{F}^2$ and $\tilde{R}^2$, we perform simultaneous large-$N$ and continuum fits using the data in  Ref.~\cite{Berkowitz:2016jlq}.
However, a functional dependence on the temperature can not be fitted for these observables, since there is no corresponding expectation from the gauge/gravity duality.

For the ungauged theory we perform new fits for all three observables, $\tilde{E}/N^2$, $\tilde{F}^2$ and $\tilde{R}^2$, to the data in Tab.~\ref{tab:bfss-ungauged}.

The analysis of the Monte Carlo data for the observables of the ungauged theory proceeds in a similar manner to the one done in the gauged theory~\cite{Berkowitz:2016jlq}:
\begin{itemize}
\item for each $N$, $L$ and $\tilde{T}$ value, we look at the Monte Carlo history of the observables and remove the configurations where we find instabilities due to the flat direction; this rarely happen because we simulate large enough $N$ at each $\tilde{T}$
\item for each $N$, $L$ and $\tilde{T}$ value, we remove $5000$ configurations from the beginning of the simulation to reduce thermalization effects
\item for each $N$, $L$ and $\tilde{T}$ value, we compute the integrated autocorrelation time $\tau$ using the Madras-Sokal algorithm and we make sure that the sample size is sufficient to include at least 10 times this autocorrelation. If this is not possible, we discard the whole ensemble because of limited statistics.
\item for each remaining $N$, $L$ and $\tilde{T}$ value, we compute the average and the standard deviation coming from binned measurements of size $5\times \tau$.
\item for each $\tilde{T}$, we extrapolate the results from the previous step to $\{N,L\}=\{\infty,\infty\}$ with a simultaneous large-$N$ and continuum limit.
\item for each $\tilde{T}$, we also extrapolate the results from the previous step to $L=\infty$ at fixed $N$, for comparison.
\end{itemize}

\begin{figure}
\begin{center}
\scalebox{0.45}{
\includegraphics{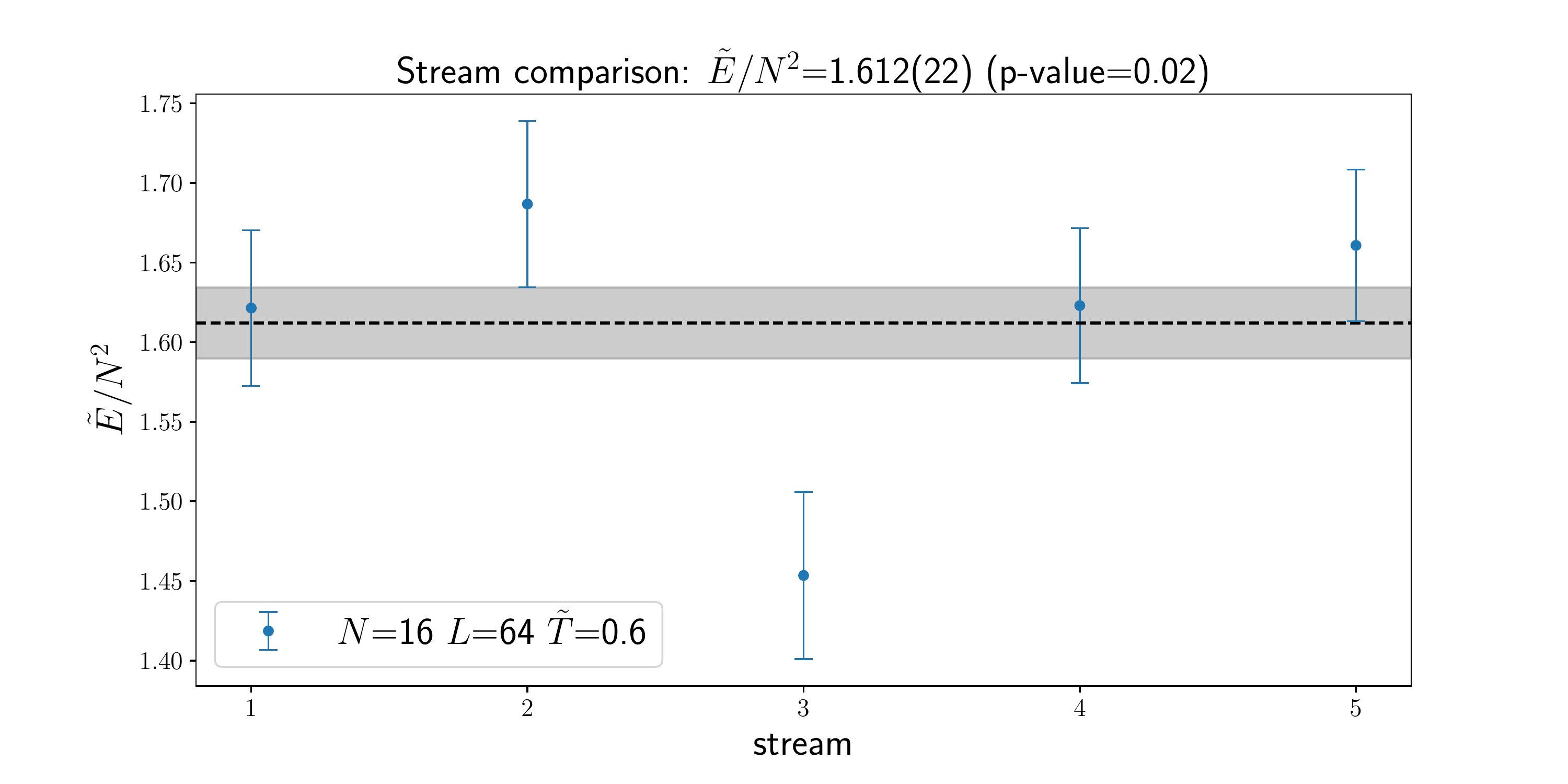}}
\scalebox{0.45}{
\includegraphics{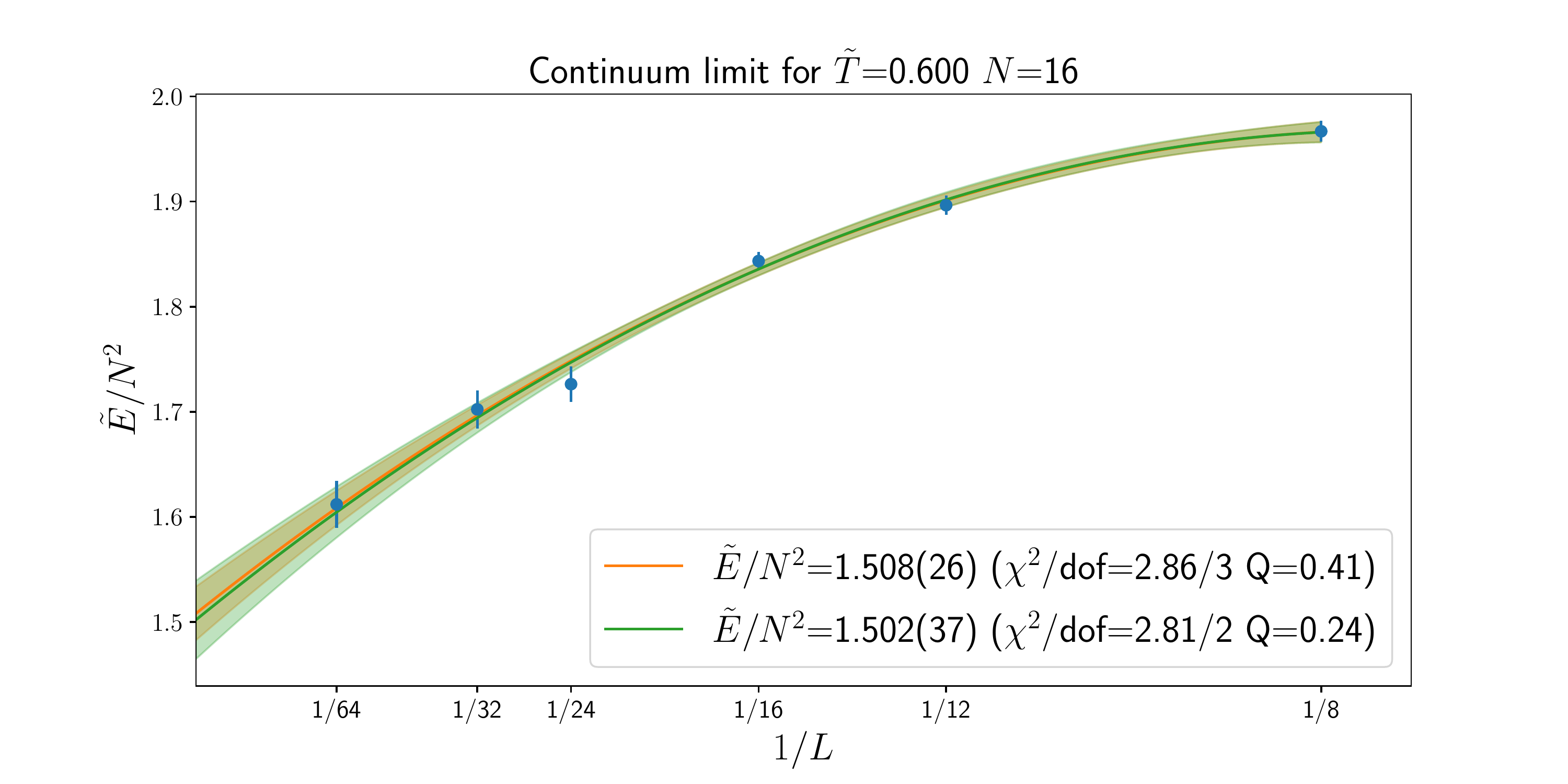}}
\end{center}
\caption{[Upper] Weighted average of five independent streams at $N=16$, $L=64$ and $\tilde{T}=0.6$. One of the streams is more than two standard deviations away from the weighted average over the different streams. [Lower] Continuum fit at $N=16$ and $\tilde{T}=0.6$ with and without the $L=64$ point. The results are indistinguishable. (The legend reports the internal energy value in the continuum and the $\chi^2$/dof and Q-value of the fit.)}
\label{fig:energy-N16L64T06}
\end{figure}

Similar challenges to the gauged theory are present in the ungauged theory, like the treatment of long autocorrelations in the Monte Carlo samples and the sensitivity of the continuum and large-$N$ extrapolations to the fitting ansatz and data set.
However, there are also some differences that we highlight in the following.

For several parameters $\{N,L,\tilde{T}\}$, we have simulated multiple streams of Markov Chain Monte Carlo.
This is a common procedure in order to increase the statistics with multiple independent copies of Monte Carlo samples (sometimes called `replicas').

The first step in the analysis of the multiple streams amounts to combining them as if they were separate independent `experiments', using a weighted average (or constant fit). 
Although this is a straightforward procedure, it can happen that some streams have central values that are not compatible with the weighted average within two standard deviations.
This is reflected by a poor `p-value' (or equivalently a large $\chi^2$ in this case).

An example of this happens at $N=16$, $L=64$ and $\tilde{T}=0.6$, where one of the five independent streams is $\sim 2.5$ standard deviations from the average of the streams when looking at the energy observable.
This is shown in the upper panel of Fig.~\ref{fig:energy-N16L64T06} and it is actually the only occurrence we encountered in the analysis.
We looked for possible sources of systematic errors in the numerical simulations of these particular streams, like a small acceptance rate, a limited number of samples, unusually large fluctuations, etc...
We did not find anything suspicious other than the possibility of long autocorrelations, much longer than the currently accumulated samples, which could result in underestimating the uncertainty.
Therefore, we keep this point at $N=16$, $L=64$ and $\tilde{T}=0.6$ in the analysis, and we interpret it as a statistical fluctuation.
We have also checked that a fit to the continuum at $N=16$ and $\tilde{T}=0.6$ with a quadratic function in $1/L$ is not dramatically influenced by including or removing this $L=64$ point, as can be seen in the lower panel of Fig.~\ref{fig:energy-N16L64T06}.

Another issue is that the $N$-dependence can be very mild and often the data is not sufficient to resolve the $1/N^2$ corrections unambiguously.
Let us focus on the energy observable where this is more dramatic.
Similarly to the gauged theory case, we use the fitting ansatz
\begin{equation}
\frac{\tilde{E}}{N^2} \; = \; \sum_{i,j\ge 0}e_{ij}N^{-2i}L^{-j} \, ,
\end{equation}
where we only include $\{e_{00},e_{01},e_{02},e_{10}\}$ terms, where $e_{00}$ is the continuum large-$N$ contribution, $e_{10}$ is the leading continuum large-$N$ correction, and the other coefficients characterize discretization artifacts and finite-$N$ corrections.
The $L^{-j}$ terms are needed to model the data at different lattice sizes $L$ up to the coarsest lattices ($L=8$). 
The dependence on $N$ is only modeled by the $N^{-2}$ term, but the corresponding coefficient $e_{10}$ is often not constrained by the data at $N=16$, 24 and 32.

\begin{figure}
\begin{center}
\scalebox{0.5}{
\includegraphics{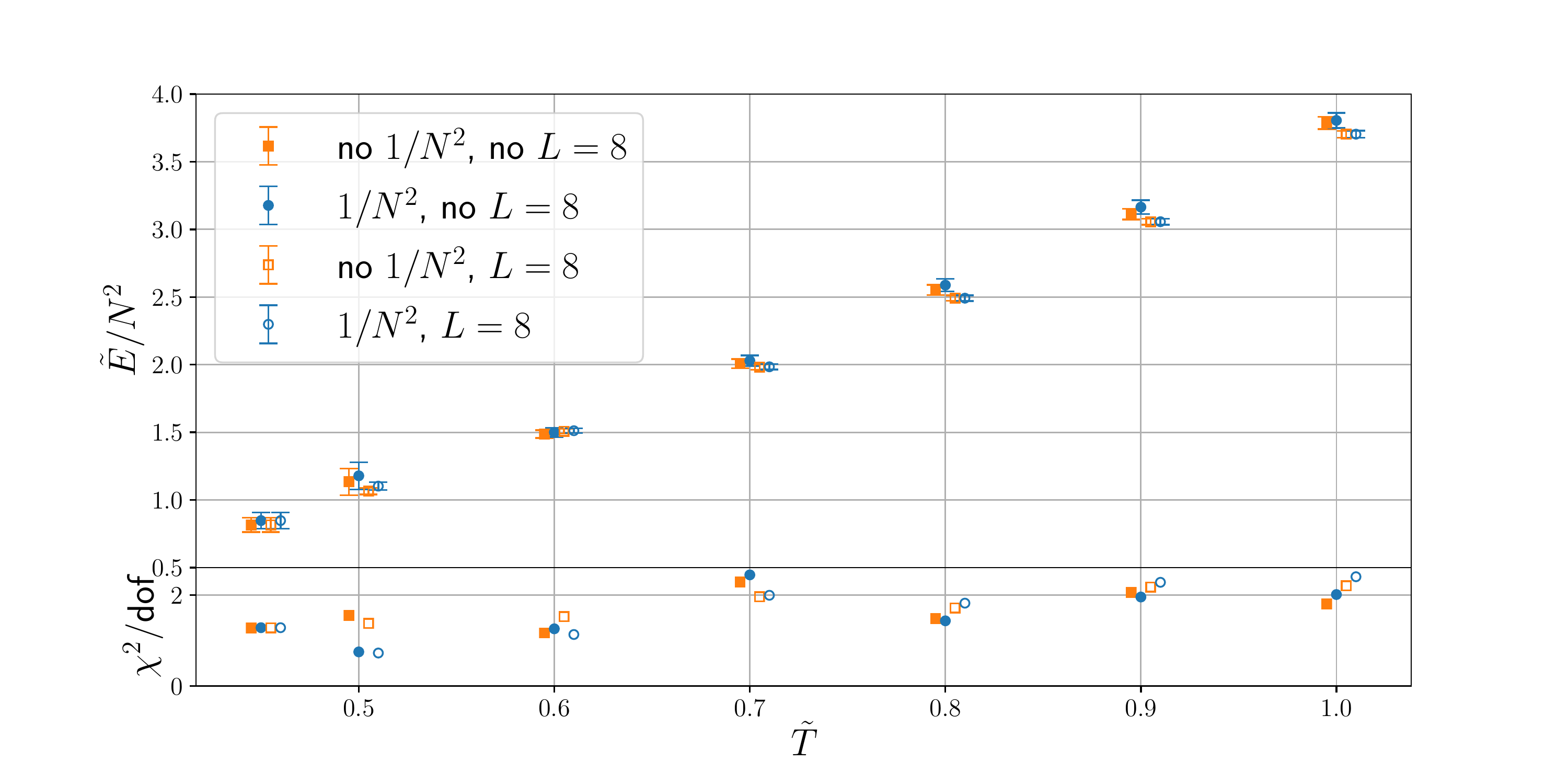}}
\end{center}
 \caption{The values of the energy in the ungauged theory (and the reduced $\chi^2$) as a function of the temperature from different simultaneous large-$N$ and continuum extrapolations: with and without $1/N^2$ corrections and with and without $L=8$ data points.\label{fig:energy-extrapolation-systematics}}
\end{figure}

When this is the case, a simple fit without $1/N^2$ corrections, or equivalently with an average of the datasets at different $N$ values, can be used to represent the data.
In other words, we can use results at fixed $N$ as if they were already in the $N=\infty$ limit.
Fig.~\ref{fig:energy-extrapolation-systematics} shows the results for the coefficient $e_{00}$, the energy in the large-$N$ and continuum limit $\tilde{E}$, as a function of temperature, for four different extrapolations.
It compares fits with and without the leading $1/N^2$ corrections when all the datapoints are included (all values of $L$ and $N$) and fits to data where $L=8$ points have been removed.
In all cases we note that the quality of the fit, measured by the reduced $\chi^2$ ($\chi^2$/dof), is always around or below 2, being a little worse for the higher temperatures.
At the same time, including or excluding $1/N^2$ corrections gives compatible results.
Removing points with the coarsest lattice spacing helps estimating additional systematic uncertainties of our extrapolation models.
For the fitting models considered above, there is a discrepancy of a bit more than one standard deviation when two different datasets are used at higher temperatures $\tilde{T}\ge 0.8$.
Our most conservative approach is to consider the results with the largest statistical errors, the ones obtained by fitting the datasets without $L=8$ and with the $1/N^2$ corrections. 
They are compatible within two standard deviations with the results from all the other fits.

\begin{figure}[ht]
\begin{center}
\scalebox{0.4}{
\includegraphics{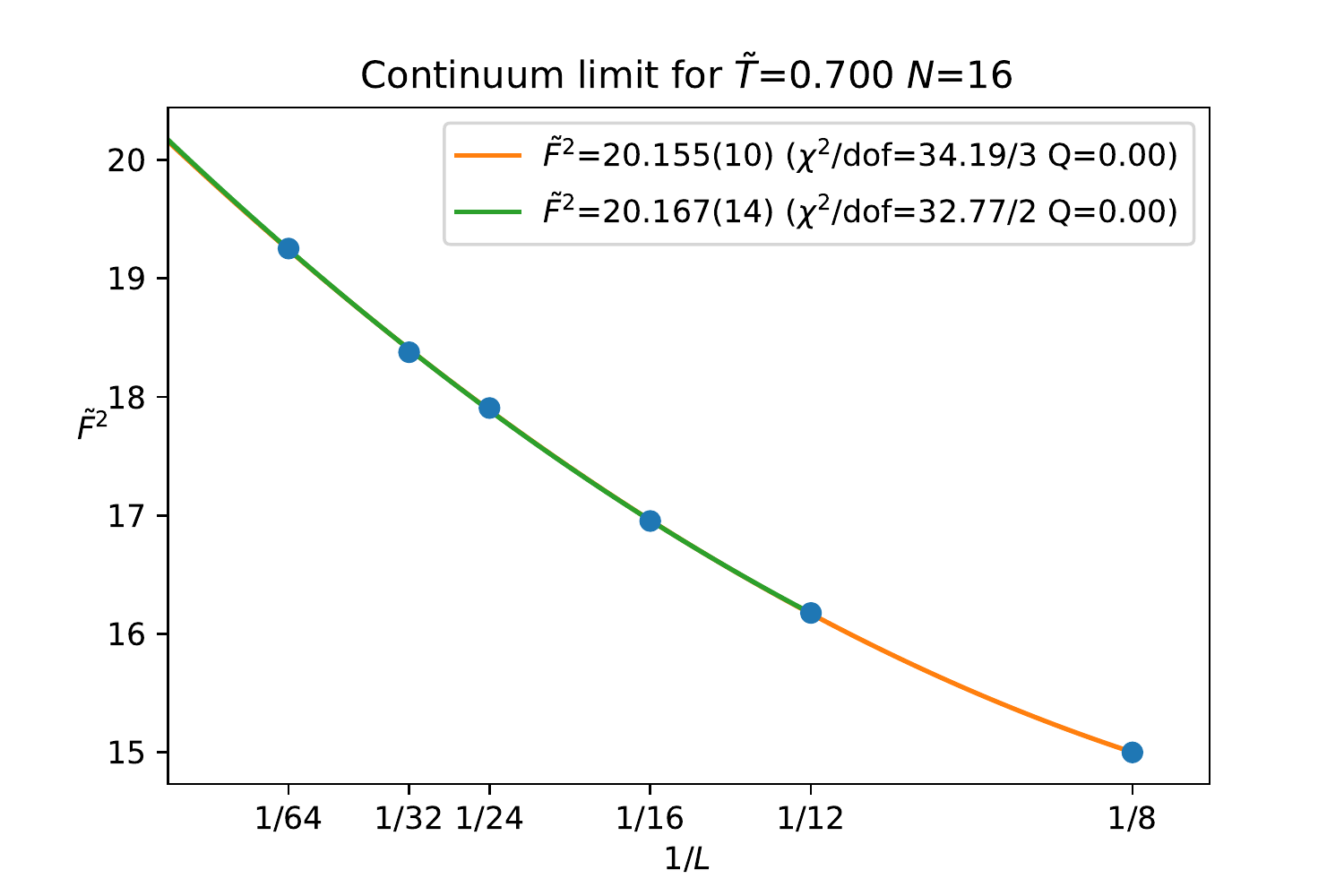}}
\scalebox{0.4}{
\includegraphics{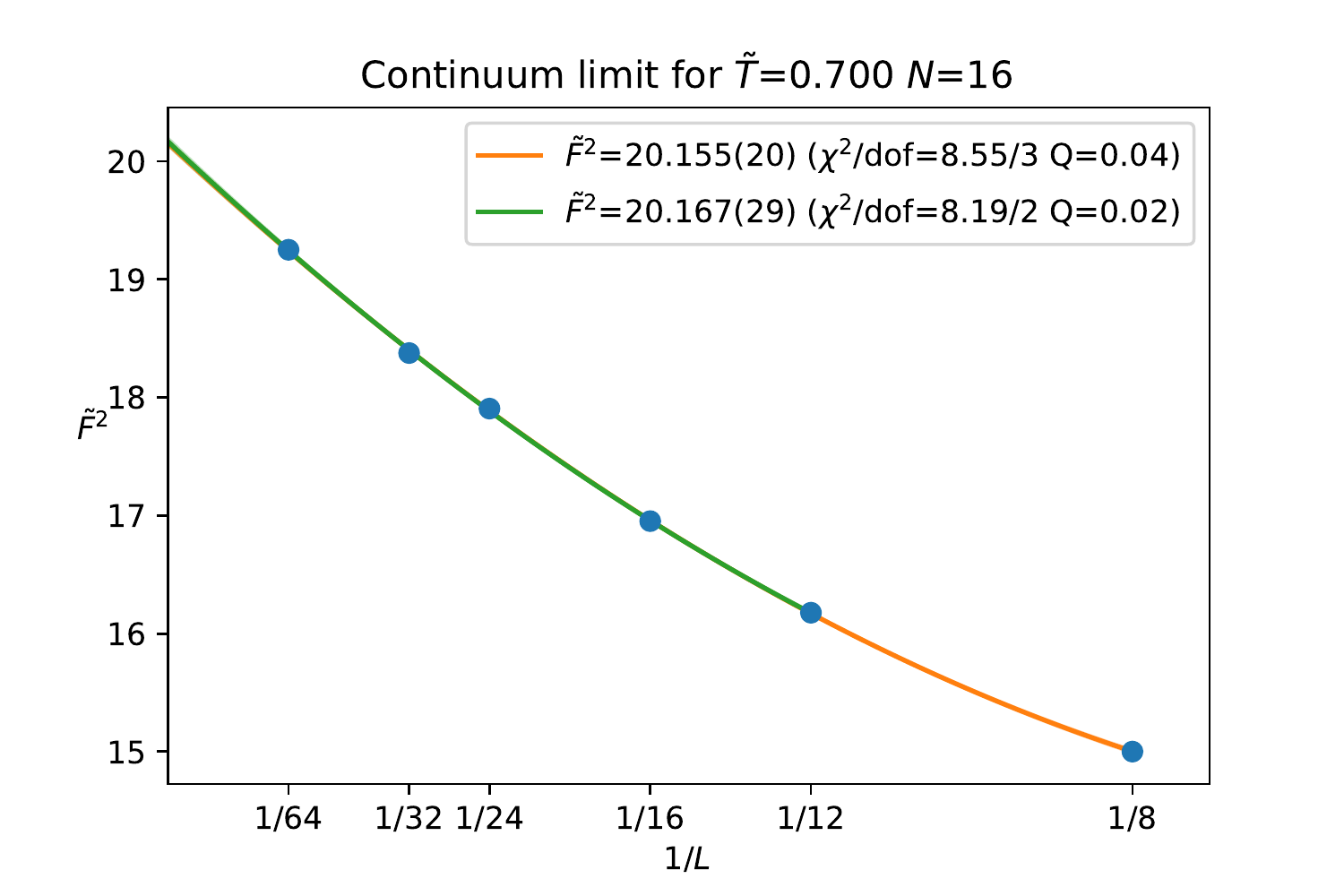}}
\scalebox{0.4}{
\includegraphics{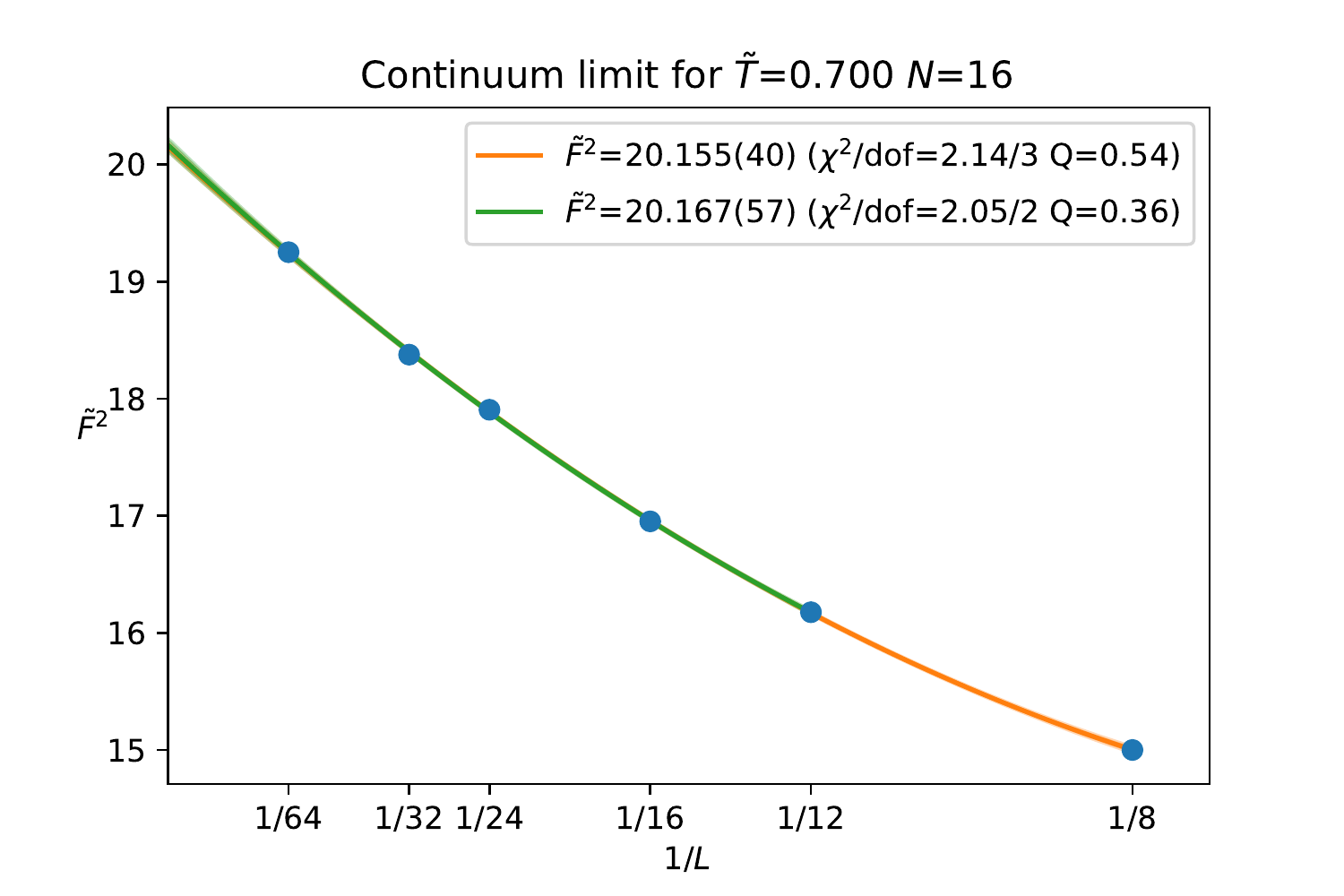}}
\end{center}
\caption{Continuum limit at fixed $N=16$ and $\tilde{T}=0.7$ for the ungauged $\tilde{F}^2$ with and without $L=8$ points. Each panel, starting from the top, shows fits to points with inflated error bars by a factor of 1, 2 and 4, respectively. The fits in each panel are consistent, showing that a quadratic extrapolation function works well, even for $L=8$, but the quality of the fits is bad. This indicates that our errors are underestimated.\label{fig:increase-errors}}
\end{figure}

We also noticed that observables like $\tilde{F}^2$ and $\tilde{R}^2$ have relative statistical errors that are orders of magnitude smaller than the ones for the $\tilde{E}/N^2$ observable.
This turns out to be an issue when doing least-square extrapolations because highly-precise data can artificially drive the reduced $\chi^2$ of the fit to values much larger than one: it does not mean that the functional form used in the fit is necessarily bad in reproducing the data, but only that the final errors on the fit parameters is under-estimated.
Of course, this latter scenario has to be tested, for example by manually increasing the uncertainty on the individual data points and tracking how the fit parameters change.
In Fig.~\ref{fig:increase-errors} we show the continuum extrapolation of $\tilde{F}^2$ data at $N=16$ and $\tilde{T}=0.7$ in the ungauged model.
The three panels show data points with statistical uncertainties increased by a factor of 1, 2 and 4 (clockwise), but they are still too small to be seen on the plots.
At the same time, each panel shows the continuum extrapolation with a second order polynomial in $1/L$ performed on the full data set and on the data set without the $L=8$ point.
The two fits in each panel are always compatible with each other and the functional form goes through the data points quite nicely, indicating that the fit can be used to reliably extrapolate the points to the continuum.
However, the reduced $\chi^2$ is always much larger than one, unless the individual point error bars are increased by a factor of 4.
The four-fold error bar increase is reflected by a corresponding four-fold increase in the fit parameters error bars.
We adopt this error-inflating procedure for $\tilde{F}^2$ and $\tilde{R}^2$ for both the ungauged and the gauged theory, but for the latter we only need a factor of 2 increase.

\begin{figure}
\begin{center}
\scalebox{0.5}{
\includegraphics{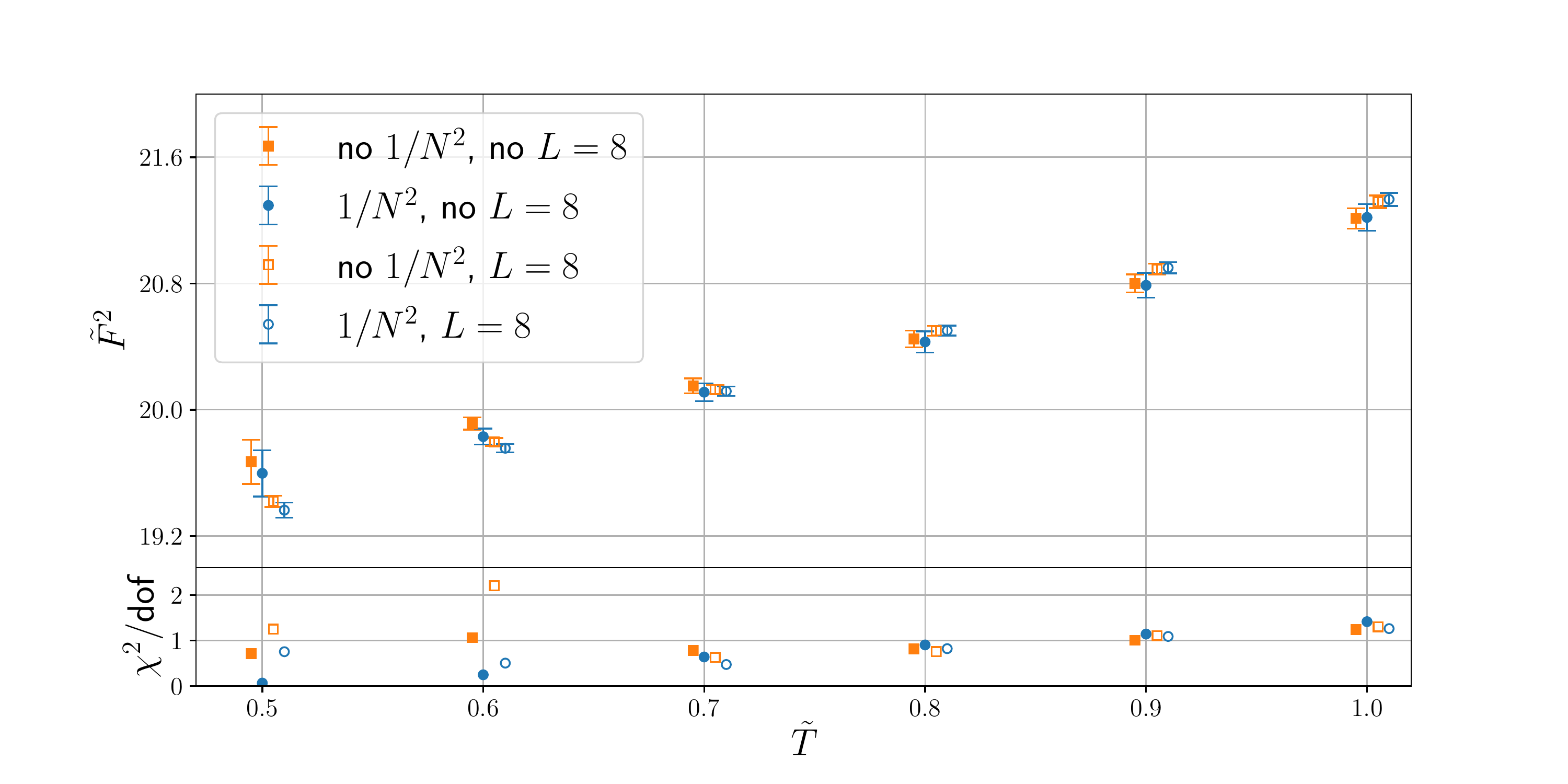}}
\end{center}
 \caption{The values of $\tilde{F}^2$ in the ungauged theory (and the reduced $\chi^2$) as a function of the temperature from different simultaneous large-$N$ and continuum extrapolations: with and without $1/N^2$ corrections and with and without $L=8$ data points. The error bars of the individual points at fixed $N$, $L$ and $\tilde{T}$ used in the extrapolations have been increased by a factor of four.\label{fig:F2-extrapolation-systematics-u}}
\end{figure}

\begin{figure}
\begin{center}
\scalebox{0.5}{
\includegraphics{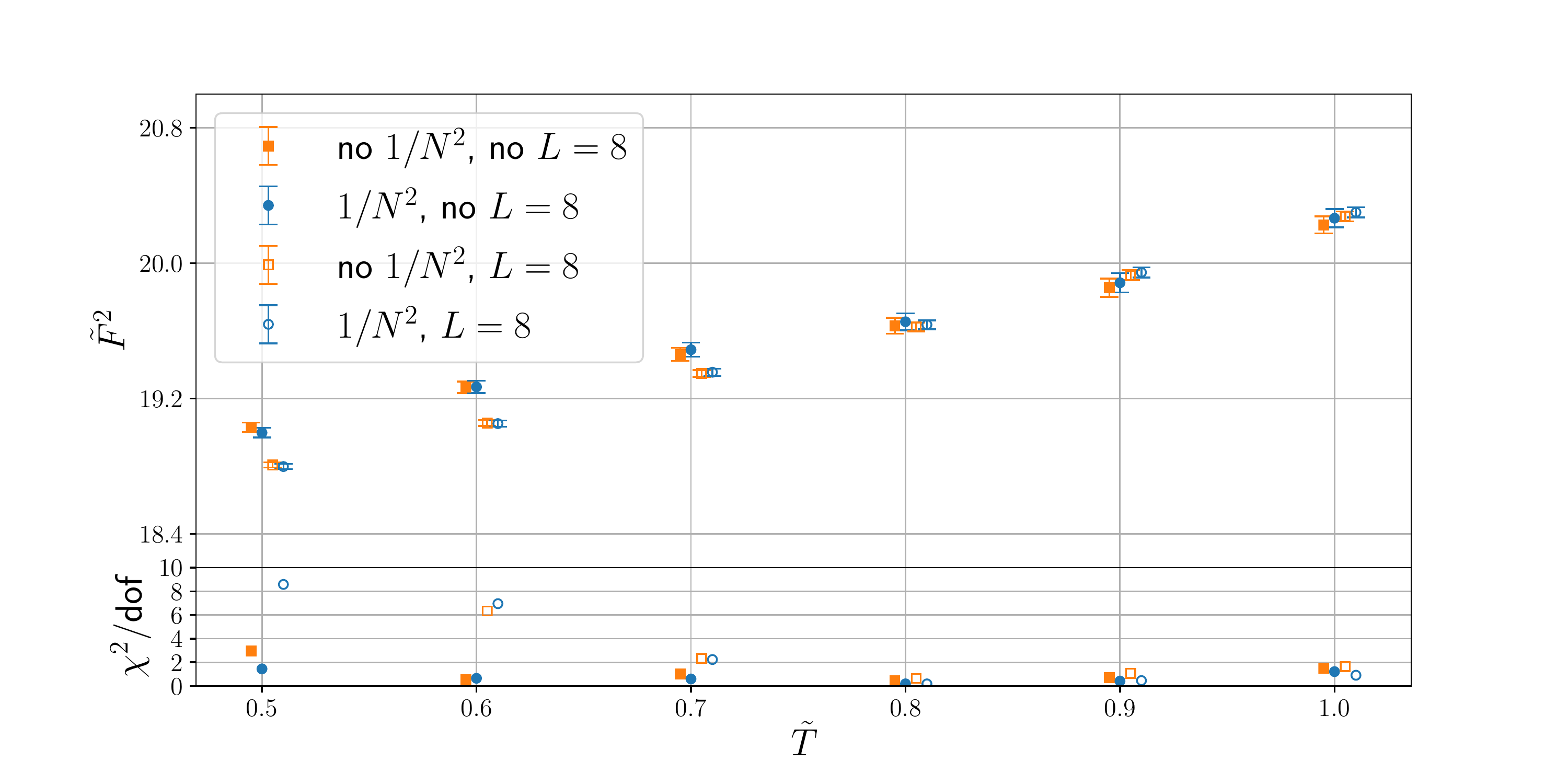}}
\end{center}
 \caption{The values of $\tilde{F}^2$ in the gauged theory (and the reduced $\chi^2$) as a function of the temperature from different simultaneous large-$N$ and continuum extrapolations: with and without $1/N^2$ corrections and with and without $L=8$ data points. The error bars of the individual points at fixed $N$, $L$ and $\tilde{T}$ used in the extrapolations have been increased by a factor of two.\label{fig:F2-extrapolation-systematics-g}}
\end{figure}

The simultaneous large-$N$ and continuum extrapolated results for $\tilde{F}^2$ at each $\tilde{T}$ are shown in Fig.~\ref{fig:F2-extrapolation-systematics-u} for the ungauged theory and in Fig.~\ref{fig:F2-extrapolation-systematics-g} for the gauged theory.
Fits to the datasets with and without $L=8$ are performed with and without $1/N^2$ corrections.
We notice that for the ungauged theories, the four different fits give compatible results, although fits with $L=8$ have a somewhat worse fit quality.
On the other hand, for the gauged theory, fits with $L=8$ give significantly different results at low temperatures $\tilde{T} \le 0.7$.
Given the worse quality of the fits in those cases, we therefore only consider fits without $L=8$ when we construct differences with the ungauged theory at each temperature.

\begin{figure}[ht]
\begin{center}
\scalebox{0.45}{
\includegraphics{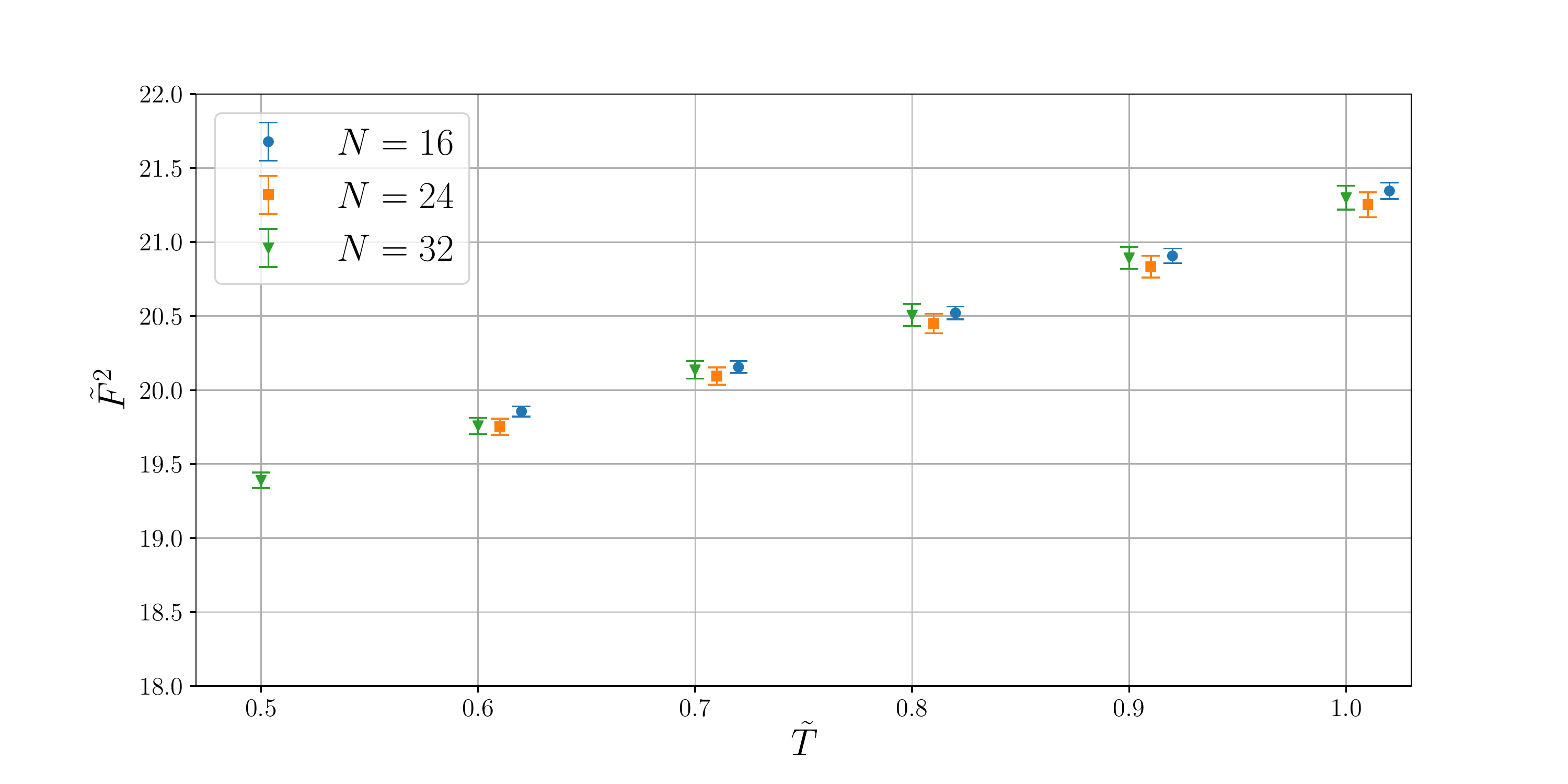}}
\scalebox{0.45}{
\includegraphics{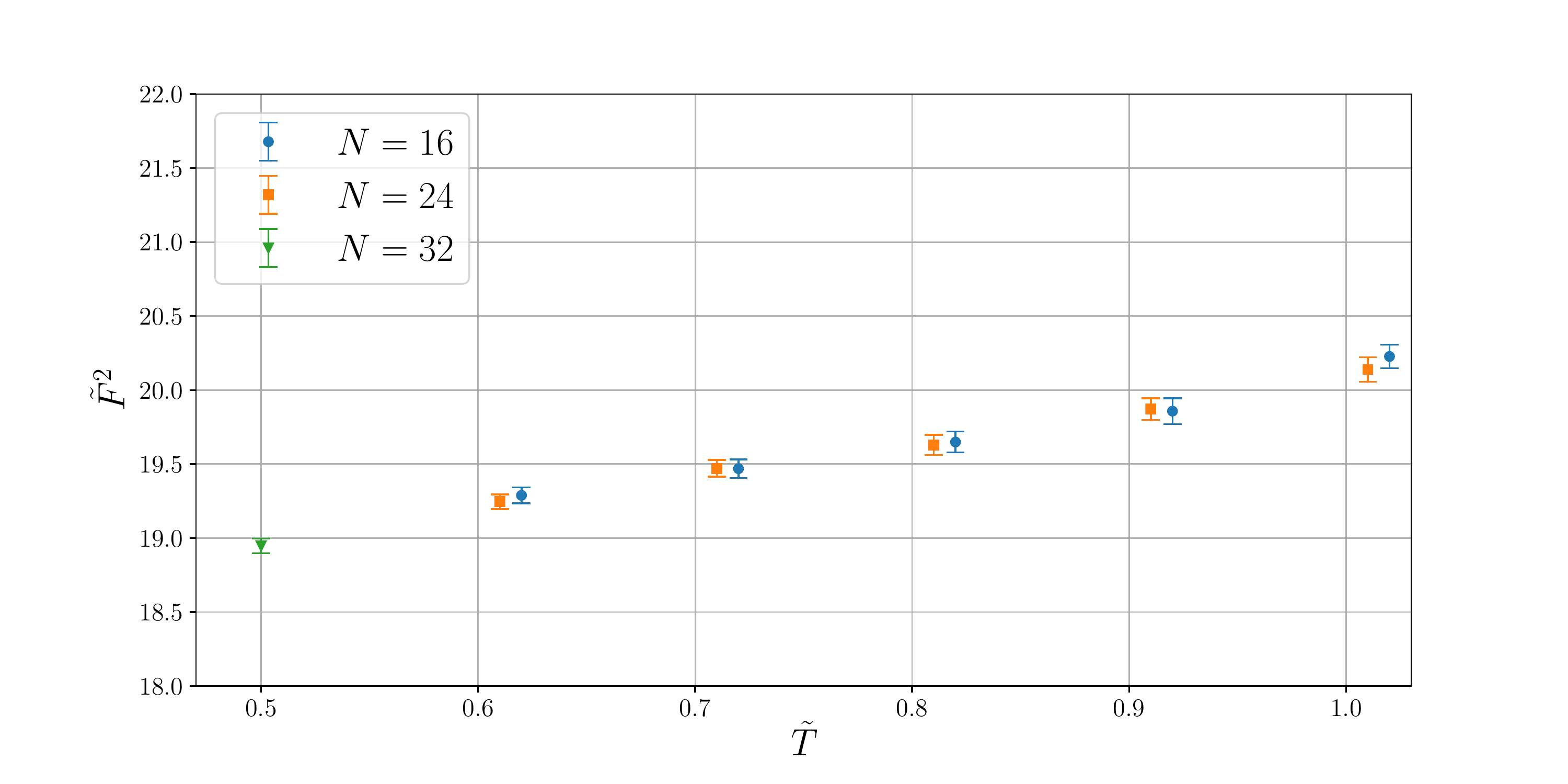}}
\end{center}
 \caption{[Upper] The values of $\tilde{F}^2$ in the continuum ungauged theory as a function of the temperature at three different values of $N$ (including all lattice spacings in the continuum fit). The error bars of the individual points at fixed $N$, $L$ and $\tilde{T}$ used in the continuum extrapolations have been increased by a factor of four. [Lower] Same as above, but for the gauged theory. The error bars of the individual points at fixed $N$, $L$ and $\tilde{T}$ used in the continuum extrapolations have been increased by a factor of two.\label{fig:F2-fixedN-u-g}}
\end{figure}

A different approach to the $N=\infty$ limit that has been used in past studies is to consider continuum results at fixed $N$, and if $N$ is large enough one can neglect $1/N^2$ corrections.
We plot the fixed-$N$ continuum results for $\tilde{F}^2$ in the ungauged and gauged theory at each temperature in Fig.~\ref{fig:F2-fixedN-u-g}.
When multiple values of $N$ are present at the same temperature in the continuum, we see negligible finite-$N$ corrections, as all points are statistically compatible.
This justifies the approach of considering continuum results for the largest $N$ at each temperature as a proxy for the $N=\infty$ result.

\begin{figure}
\begin{center}
\scalebox{0.45}{
\includegraphics{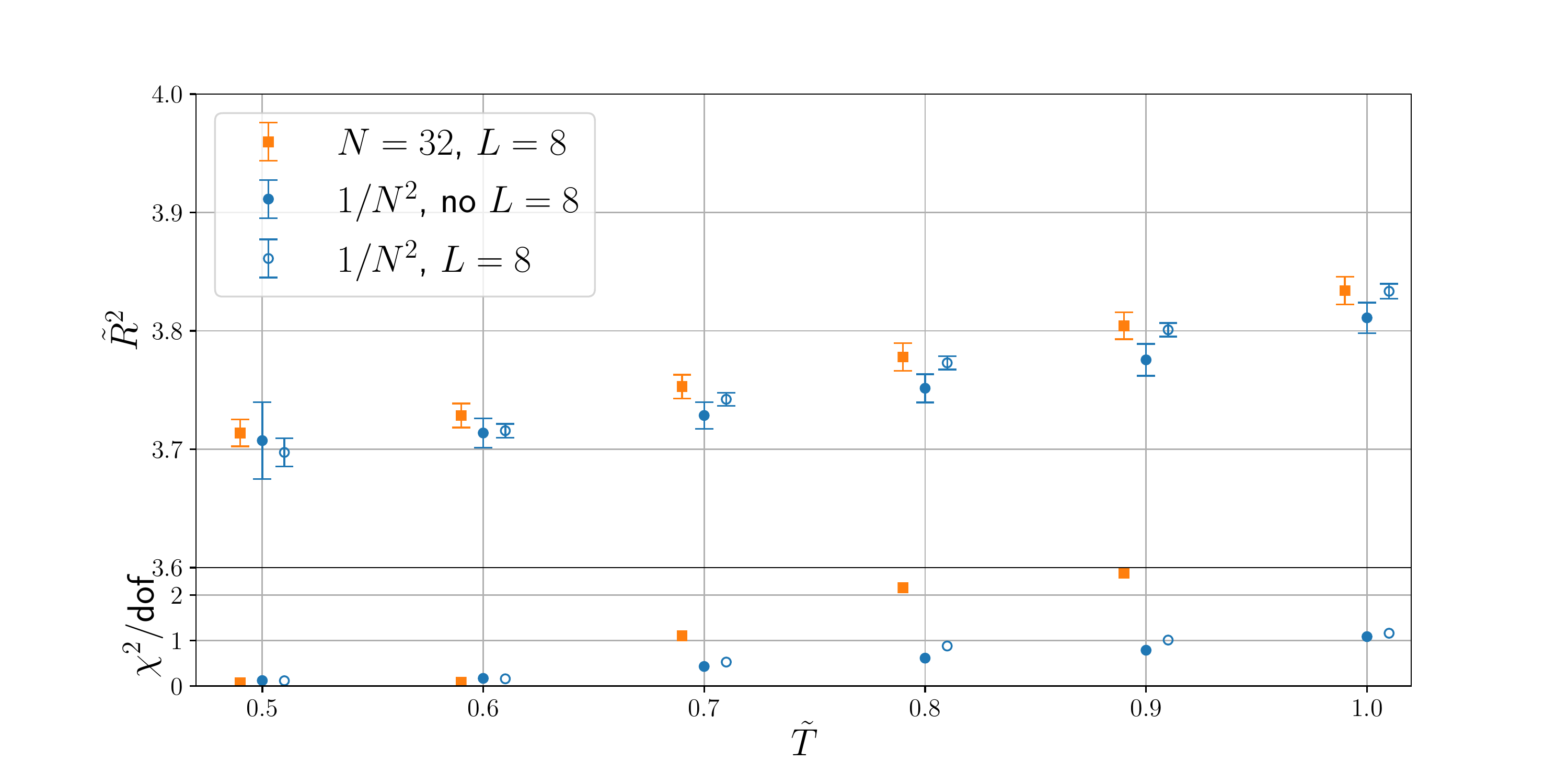}}
\scalebox{0.45}{
\includegraphics{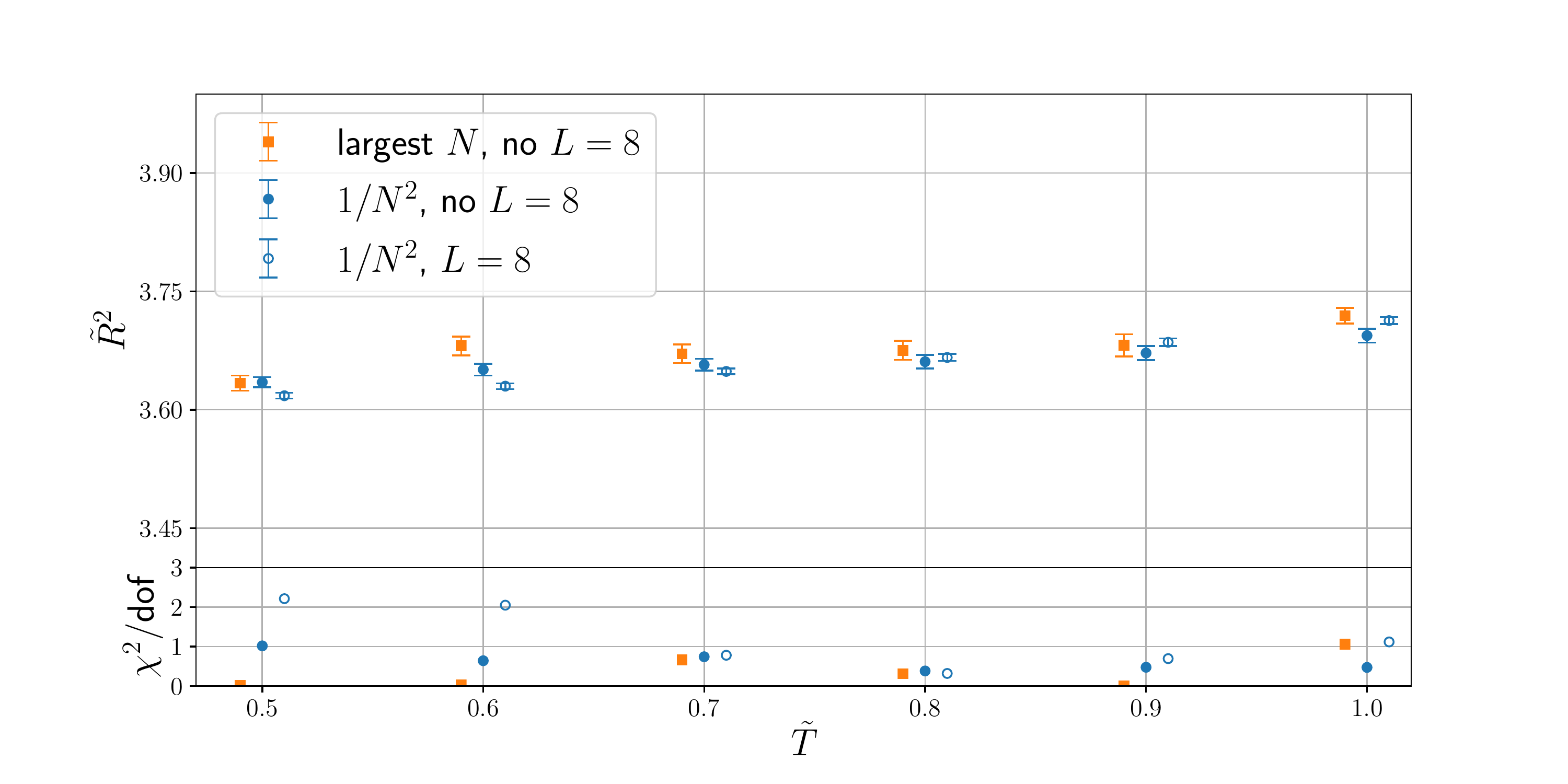}}
\end{center}
 \caption{[Upper] The values of $\tilde{R}^2$ in the ungauged theory as a function of the temperature for three different extrapolations: one continuum extrapolation at fixed $N=32$ and two simultaneous continuum large-$N$ extrapolations. The error bars of the individual points at fixed $N$, $L$ and $\tilde{T}$ used in the continuum extrapolations have been increased by a factor of four. [Lower] Same as above, but for the gauged theory. The error bars of the individual points at fixed $N$, $L$ and $\tilde{T}$ used in the continuum extrapolations have been increased by a factor of two.\label{fig:R2-largeN-u-g}}
\end{figure}

In the case of $\tilde{R}^2$, however, the $N$ dependence is larger.
$\tilde{R}^2$ is, by definition, very sensitive to the flat direction and, even when simulations do not show samples affected by the instability, the value of $\tilde{R}^2$ tends to be larger at smaller $N$.
We can see this trend in Fig.~\ref{fig:R2-largeN-u-g}, where for some temperatures, the largest value of $N$ can still be different than the value obtained by extrapolating at $N=\infty$, and has a systematic shift to larger values.
In order to assess the error one is making by not considering a full simultaneous large-$N$ and continuum limit, we keep two data sets in the analysis of the conjecture: the largest $N$ dataset and the dataset obtained by including $1/N^2$ corrections and without the coarsest lattice spacing $L=8$.

%%%%%%%%%%%%%%%
%%%%%%%%%%%%%%%
\section{High temperature behavior}\label{sec:high-T}
\hspace{0.25in}
%%%%%%%%%%%%%%%
%%%%%%%%%%%%%%%
At high temperature, the D0-brane matrix model and its bosonic analogue both reduce to the classical matrix model. 
The kinetic and potential energies are related by the virial theorem as $({\rm kinetic\ energy})=2\times({\rm potential\ energy})$, 
and hence the total energy $E$ satisfies $E=\frac{3}{2}\times ({\rm kinetic\ energy})=3\times({\rm potential\ energy})$. 
The kinetic energy in the classical theory simply counts the number of degrees of freedom: 
\begin{eqnarray}
({\rm kinetic\ energy})
=
\frac{T}{2}\times ({\rm \# d.o.f})
=
\frac{T}{2}\times
\left\{
\begin{array}{cc}
 8N^2 & {\rm (gauged)}\\
 9N^2& {\rm (ungauged)}
\end{array}
\right.
\end{eqnarray}
(Strictly speaking, we need to take into account the conservation laws coming from the constant shift of eigenvalues and $SO(9)$ rotations.
They affect subleading terms with respect to $1/N^2$.)
Hence $E/N^2=6T$ and $6.75T$ for gauged and ungauged theories, respectively. 
$E$ and $F^2$ are related by $E/N^2=\frac{3}{4\lambda}F^2$, because $({\rm potential\ energy})=\frac{N^2}{4\lambda}F^2$.  

\clearpage
%%%%%%%%%%%%%%%
%%%%%%%%%%%%%%%
\section{Summary tables}\label{sec:data-tables}
\hspace{0.25in}
%%%%%%%%%%%%%%%
%%%%%%%%%%%%%%%
\begin{longtable}{|cccr|lr|lr|}
\hline
    $\tilde{T}$ &  $N$ &  $L$ & \# config &  \multicolumn{1}{c}{$\tilde{E}/N^2$}  & \# bins & \multicolumn{1}{c}{$\tilde{R}^2$} & \# bins \\
\hline
\hline
\endhead
\hline
\endfoot
0.20 & 6 & 12 & 24596 &5.8270(43) &534 & 2.13894(77) & 599  \\
    &      & 16 & 50840 &6.0854(23) &2118 & 2.17449(40) & 2310  \\
    &      & 24 & 85000 &6.3092(13) &7727 & 2.20378(23) & 7083  \\
    &      & 32 & 85000 &6.3982(16) &5312 & 2.21505(29) & 4722  \\
    &      & 48 & 39861 &6.4566(38) &813 & 2.22172(73) & 724  \\
    &      & 64 & 85000 &6.4815(31) &1231 & 2.22508(59) & 1103  \\
    & 8 & 12 & 66905 &5.8999(15) &2158 & 2.16516(27) & 2389  \\
    &      & 16 & 85000 &6.1660(10) &5312 & 2.20199(18) & 5666  \\
    &      & 24 & 85000 &6.39324(98) &7727 & 2.23186(18) & 7083  \\
    &      & 32 & 85000 &6.4772(15) &3148 & 2.24232(29) & 2833  \\
    &      & 48 & 66611 &6.5416(28) &1024 & 2.25027(51) & 979  \\
    &      & 64 & 52683 &6.5652(34) &605 & 2.25305(64) & 548  \\
    & 12 & 12 & 85000 &5.9529(10) &2073 & 2.18406(18) & 2297  \\
    &      & 16 & 85000 &6.22076(76) &4250 & 2.22122(14) & 4722  \\
    &      & 24 & 85000 &6.44818(75) &5666 & 2.25114(14) & 5312  \\
    &      & 32 & 45800 &6.5327(12) &2290 & 2.26158(22) & 2081  \\
    &      & 48 & 50384 &6.5996(28) &445 & 2.26999(51) & 423  \\
    &      & 64 & 24317 &6.6262(44) &199 & 2.27357(81) & 191  \\
0.25 & 6 & 12 & 85000 &6.0387(23) &1976 & 2.16850(41) & 2179  \\
    &      & 16 & 85000 &6.2360(13) &9444 & 2.19493(23) & 9444  \\
    &      & 24 & 85000 &6.3900(17) &5666 & 2.21466(31) & 5312  \\
    &      & 32 & 85000 &6.4452(21) &3541 & 2.22130(39) & 3269  \\
    &      & 48 & 85000 &6.4784(34) &1440 & 2.22478(64) & 1180  \\
    &      & 64 & 85000 &6.4977(45) &787 & 2.22716(83) & 726  \\
    & 8 & 12 & 85000 &6.1215(12) &4473 & 2.19649(20) & 5312  \\
    &      & 16 & 85000 &6.3180(10) &7727 & 2.22268(19) & 8500  \\
    &      & 24 & 51774 &6.4675(16) &3698 & 2.24152(30) & 3235  \\
    &      & 32 & 85000 &6.5253(17) &3269 & 2.24862(32) & 2931  \\
    &      & 48 & 85000 &6.5666(46) &381 & 2.25350(88) & 351  \\
    &      & 64 & 28836 &6.5846(96) &175 & 2.2564(18) & 164  \\
    & 12 & 12 & 62589 &6.17842(97) &3129 & 2.21600(17) & 3477  \\
    &      & 16 & 85000 &6.36860(76) &7727 & 2.24114(14) & 6538  \\
    &      & 24 & 85000 &6.52305(92) &4722 & 2.26075(17) & 4250  \\
    &      & 32 & 85000 &6.5847(12) &2656 & 2.26868(23) & 2361  \\
    &      & 48 & 77113 &6.6270(30) &365 & 2.27376(56) & 345  \\
    &      & 64 & 85000 &6.6427(30) &376 & 2.27549(56) & 354  \\
0.30 & 6 & 12 & 85000 &6.1937(22) &2428 & 2.19067(39) & 2575  \\
    &      & 16 & 85000 &6.3384(15) &8500 & 2.20926(28) & 8500  \\
    &      & 24 & 85000 &6.4447(20) &4722 & 2.22239(38) & 4250  \\
    &      & 32 & 85000 &6.4805(22) &4250 & 2.22639(40) & 3863  \\
    &      & 48 & 85000 &6.5139(43) &988 & 2.23048(81) & 894  \\
    &      & 64 & 85000 &6.5425(59) &491 & 2.2355(12) & 418  \\
    & 8 & 12 & 85000 &6.2750(12) &6538 & 2.21832(21) & 7083  \\
    &      & 16 & 85000 &6.4168(13) &6538 & 2.23649(23) & 6538  \\
    &      & 24 & 85000 &6.5261(17) &3695 & 2.24987(32) & 3269  \\
    &      & 32 & 85000 &6.5653(24) &1931 & 2.25463(45) & 1734  \\
    &      & 48 & 40072 &6.5874(81) &147 & 2.2566(15) & 111  \\
    &      & 64 & 38729 &6.612(10) &102 & 2.2608(19) & 95  \\
    & 12 & 12 & 85000 &6.32828(90) &4722 & 2.23717(16) & 5000  \\
    &      & 16 & 85000 &6.47113(90) &6071 & 2.25556(17) & 5312  \\
    &      & 24 & 85000 &6.5828(12) &3541 & 2.26951(22) & 3148  \\
    &      & 32 & 85000 &6.6265(26) &787 & 2.27486(49) & 714  \\
    &      & 48 & 85000 &6.6542(34) &508 & 2.27814(61) & 449  \\
    &      & 64 & 51130 &6.6734(52) &214 & 2.28095(97) & 201  \\
0.35 & 6 & 12 & 85000 &6.3212(15) &8500 & 2.20988(28) & 8500  \\
    &      & 16 & 85000 &6.4272(18) &7083 & 2.22308(34) & 6538  \\
    &      & 24 & 85000 &6.5065(26) &3269 & 2.23250(49) & 2931  \\
    &      & 32 & 85000 &6.5345(26) &3541 & 2.23553(47) & 3269  \\
    &      & 48 & 85000 &6.5641(61) &551 & 2.2396(11) & 497  \\
    &      & 64 & 85000 &6.5705(84) &312 & 2.2398(15) & 295  \\
    & 8 & 12 & 85000 &6.3977(13) &7083 & 2.23658(24) & 7083  \\
    &      & 16 & 85000 &6.5051(14) &6071 & 2.24994(27) & 5312  \\
    &      & 24 & 85000 &6.5898(21) &3148 & 2.26037(39) & 2833  \\
    &      & 32 & 76307 &6.6223(31) &1467 & 2.26458(58) & 1315  \\
    &      & 48 & 85000 &6.6483(64) &344 & 2.2682(12) & 294  \\
    &      & 64 & 85000 &6.6517(87) &238 & 2.2678(16) & 220  \\
    & 12 & 12 & 85000 &6.45743(98) &5312 & 2.25672(18) & 5666  \\
    &      & 16 & 85000 &6.5639(10) &6071 & 2.26988(19) & 5666  \\
    &      & 24 & 85000 &6.6489(15) &2656 & 2.28026(28) & 2361  \\
    &      & 32 & 27210 &6.6862(57) &191 & 2.2852(10) & 179  \\
    &      & 48 & 85000 &6.7052(40) &314 & 2.28714(78) & 285  \\
    &      & 64 & 85000 &6.7004(47) &213 & 2.28613(84) & 197  \\
0.40 & 6 & 12 & 85000 &6.4455(18) &8500 & 2.23005(33) & 7727  \\
    &      & 16 & 85000 &6.5258(21) &6071 & 2.23931(39) & 5666  \\
    &      & 24 & 85000 &6.5925(26) &4047 & 2.24746(49) & 3695  \\
    &      & 32 & 85000 &6.6113(32) &2741 & 2.24965(61) & 2428  \\
    &      & 48 & 85000 &6.631(12) &176 & 2.2518(21) & 177  \\
    &      & 64 & 85000 &6.617(16) &79 & 2.2489(31) & 69  \\
    & 8 & 12 & 85000 &6.5209(15) &6538 & 2.25659(28) & 6538  \\
    &      & 16 & 85000 &6.6069(18) &5312 & 2.26709(33) & 4722  \\
    &      & 24 & 83064 &6.6750(26) &2307 & 2.27535(49) & 2076  \\
    &      & 32 & 65850 &6.6818(38) &1045 & 2.27510(71) & 940  \\
    &      & 48 & 85000 &6.7248(82) &256 & 2.2818(15) & 247  \\
    &      & 64 & 70000 &6.720(10) &139 & 2.2808(18) & 122  \\
    & 12 & 12 & 85000 &6.58132(99) &7083 & 2.27663(19) & 6071  \\
    &      & 16 & 85000 &6.6683(13) &4250 & 2.28738(24) & 3695  \\
    &      & 24 & 85000 &6.7298(19) &1808 & 2.29443(36) & 1603  \\
    &      & 32 & 85000 &6.7522(38) &464 & 2.29704(70) & 412  \\
    &      & 48 & 68007 &6.7766(57) &201 & 2.3004(11) & 183  \\
    &      & 64 & 39233 &6.7729(79) &169 & 2.2989(14) & 158  \\
0.45 & 6 & 12 & 85000 &6.5784(20) &7727 & 2.25264(38) & 7083  \\
    &      & 16 & 85000 &6.6429(26) &4722 & 2.25986(47) & 4473  \\
    &      & 24 & 85000 &6.7021(29) &3863 & 2.26737(53) & 3541  \\
    &      & 32 & 85000 &6.7148(38) &2023 & 2.26860(73) & 1808  \\
    &      & 48 & 85000 &6.763(14) &149 & 2.2759(28) & 128  \\
    &      & 64 & 85000 &6.735(17) &122 & 2.2717(29) & 111  \\
    & 8 & 12 & 73214 &6.6575(18) &5229 & 2.28000(33) & 5229  \\
    &      & 16 & 85000 &6.7247(21) &4250 & 2.28757(39) & 3863  \\
    &      & 24 & 85000 &6.7788(35) &1465 & 2.29389(66) & 1307  \\
    &      & 32 & 85000 &6.7898(43) &977 & 2.29506(81) & 867  \\
    &      & 48 & 85000 &6.8122(92) &196 & 2.2985(19) & 175  \\
    &      & 64 & 85000 &6.820(11) &195 & 2.2991(20) & 171  \\
    & 12 & 12 & 76609 &6.7148(12) &5472 & 2.29940(23) & 5107  \\
    &      & 16 & 85000 &6.7880(15) &3541 & 2.30827(28) & 3148  \\
    &      & 24 & 40473 &6.8309(51) &318 & 2.31295(97) & 304  \\
    &      & 32 & 44577 &6.8494(59) &234 & 2.3153(11) & 216  \\
    &      & 48 & 85000 &6.8667(48) &337 & 2.31724(96) & 299  \\
    &      & 64 & 44502 &6.8654(93) &112 & 2.3165(18) & 115  \\
0.50 & 6 & 12 & 85000 &6.7182(23) &6538 & 2.27746(43) & 6538  \\
    &      & 16 & 85000 &6.7750(28) &4473 & 2.28386(52) & 4047  \\
    &      & 24 & 85000 &6.8153(34) &3148 & 2.28835(63) & 2741  \\
    &      & 32 & 85000 &6.8468(70) &720 & 2.2929(13) & 611  \\
    &      & 48 & 85000 &6.838(16) &130 & 2.2897(30) & 121  \\
    &      & 64 & 85000 &6.816(20) &113 & 2.2866(37) & 95  \\
    & 8 & 12 & 85000 &6.8120(19) &5666 & 2.30717(34) & 5666  \\
    &      & 16 & 85000 &6.8612(19) &6071 & 2.31217(34) & 5666  \\
    &      & 24 & 85000 &6.8968(39) &1349 & 2.31603(74) & 1148  \\
    &      & 32 & 85000 &6.9149(54) &765 & 2.3178(10) & 714  \\
    &      & 48 & 85000 &6.9159(88) &379 & 2.3171(16) & 358  \\
    &      & 64 & 85000 &6.929(12) &167 & 2.3194(22) & 131  \\
    & 12 & 12 & 85000 &6.8651(14) &4473 & 2.32572(27) & 4047  \\
    &      & 16 & 85000 &6.9219(18) &3035 & 2.33235(34) & 2656  \\
    &      & 24 & 42702 &6.9685(57) &261 & 2.3376(11) & 201  \\
    &      & 32 & 85000 &6.9829(50) &379 & 2.33936(93) & 344  \\
    &      & 48 & 85000 &7.0066(94) &110 & 2.3425(18) & 88  \\
    &      & 64 & 47861 &7.005(16) &45 & 2.3424(26) & 48  \\
0.55 & 6 & 12 & 85000 &6.8901(26) &6538 & 2.30815(48) & 5666  \\
    &      & 16 & 85000 &6.9394(31) &4473 & 2.31350(57) & 4047  \\
    &      & 24 & 85000 &6.9693(41) &2428 & 2.31679(77) & 2179  \\
    &      & 32 & 85000 &6.9683(84) &531 & 2.3152(15) & 454  \\
    &      & 48 & 85000 &7.025(17) &140 & 2.3249(29) & 117  \\
    &      & 64 & 70000 &6.993(21) &113 & 2.3180(42) & 94  \\
    & 8 & 12 & 85000 &6.9725(20) &5312 & 2.33577(36) & 5000  \\
    &      & 16 & 85000 &7.0226(22) &5000 & 2.34151(40) & 4722  \\
    &      & 24 & 31639 &7.0631(85) &292 & 2.3468(15) & 224  \\
    &      & 32 & 85000 &7.0440(96) &303 & 2.3423(19) & 242  \\
    &      & 48 & 85000 &7.072(13) &152 & 2.3460(22) & 133  \\
    &      & 64 & 71689 &7.104(13) &99 & 2.3520(25) & 94  \\
    & 12 & 12 & 85000 &7.0380(16) &3863 & 2.35650(31) & 3269  \\
    &      & 16 & 85000 &7.0813(21) &2361 & 2.36118(39) & 2125  \\
    &      & 24 & 85000 &7.1198(51) &429 & 2.36562(92) & 397  \\
    &      & 32 & 85000 &7.1394(53) &382 & 2.3680(10) & 335  \\
0.60 & 6 & 12 & 70000 &7.0745(33) &4666 & 2.34197(60) & 4117  \\
    &      & 16 & 85000 &7.1130(32) &4722 & 2.34507(59) & 4250  \\
    &      & 24 & 85000 &7.1407(46) &2361 & 2.34801(85) & 2073  \\
    &      & 32 & 85000 &7.1465(89) &634 & 2.3486(17) & 562  \\
    &      & 48 & 85000 &7.165(23) &116 & 2.3504(42) & 102  \\
    &      & 64 & 85000 &7.104(26) &65 & 2.3390(44) & 62  \\
    & 8 & 12 & 85000 &7.1503(21) &6071 & 2.36776(38) & 5666  \\
    &      & 16 & 85000 &7.1931(26) &4250 & 2.37236(47) & 4047  \\
    &      & 24 & 70000 &7.2263(62) &648 & 2.3761(12) & 560  \\
    &      & 32 & 85000 &7.240(12) &190 & 2.3774(23) & 177  \\
    &      & 48 & 85000 &7.250(15) &106 & 2.3800(27) & 95  \\
    &      & 64 & 36379 &7.264(22) &80 & 2.3822(45) & 68  \\
    & 12 & 12 & 85000 &7.2190(18) &3541 & 2.38907(34) & 3148  \\
    &      & 16 & 79438 &7.2662(25) &2036 & 2.39460(47) & 1805  \\
    &      & 24 & 70000 &7.2892(57) &322 & 2.3966(11) & 284  \\
    &      & 32 & 85000 &7.2952(62) &315 & 2.3972(11) & 288  \\
\hline \hline
\caption{Summary table for the parameters and expectation values of observables in the bosonic ungauged matrix model.\label{tab:bosonic-ungauged}}
 \end{longtable}

\begin{longtable}{|cccr|lr|lr|lr|}
\hline
    $\tilde{T}$ &  $N$ &  $L$ & \# config &  \multicolumn{1}{c}{$\tilde{E}/N^2$}  & \# bins & \multicolumn{1}{c}{$\tilde{F}^2$} & \# bins  & \multicolumn{1}{c}{$\tilde{R}^2$} & \# bins \\
\hline
\hline
\endhead
\hline
\endfoot
0.45 & 24 & 16 & 5576 &1.212(24) &51 & 15.2494(94) & 69 & -- & --  \\
    &      & 24 & 36920 &1.168(14) &300 & 16.3390(56) & 277 & 3.4380(16) & 77  \\
    &      & 32 & 20176 &1.085(16) &288 & 17.0174(57) & 240 & -- & --  \\
    & 32 & 12 & 32608 &1.3127(54) &582 & 14.3985(41) & 184 & -- & --  \\
    &      & 16 & 16249 &1.266(10) &208 & 15.2289(48) & 162 & 3.3240(15) & 49  \\
    &      & 24 & 18229 &1.167(13) &189 & 16.3154(49) & 200 & -- & --  \\
    &      & 32 & 21489 &1.094(16) &191 & 16.9860(48) & 226 & -- & --  \\
0.50 & 16 & 24 & 12405 &1.291(27) &217 & -- & -- & -- & --  \\
    &      & 32 & 18000 &1.274(28) &285 & 17.369(17) & 81 & -- & --  \\
    & 24 & 16 & 38610 &1.4180(97) &508 & 15.6109(46) & 402 & 3.3611(15) & 72  \\
    &      & 24 & 77150 &1.335(11) &653 & 16.6676(40) & 637 & 3.45866(92) & 258  \\
    &      & 32 & 47098 &1.284(12) &682 & 17.3021(38) & 735 & 3.51756(88) & 261  \\
    & 32 & 8 & 81525 &1.5790(28) &1772 & 13.4749(24) & 627 & 3.1673(12) & 119  \\
    &      & 16 & 44800 &1.4435(62) &711 & 15.5880(27) & 640 & 3.34742(61) & 296  \\
    &      & 24 & 37995 &1.3376(99) &436 & 16.6513(37) & 391 & 3.44802(87) & 132  \\
    &      & 32 & 23365 &1.282(18) &156 & 17.2927(54) & 189 & 3.5079(11) & 106  \\
0.60 & 16 & 8 & 35489 &1.9669(98) &865 & 14.280(11) & 134 & -- & --  \\
    &      & 12 & 94960 &1.8965(92) &1396 & 15.5090(55) & 797 & 3.3577(16) & 218  \\
    &      & 16 & 76700 &1.8435(87) &2191 & 16.3184(41) & 1420 & 3.4363(11) & 511  \\
    &      & 24 & 38633 &1.726(17) &858 & 17.3404(70) & 529 & 3.5351(18) & 108  \\
    &      & 32 & 60839 &1.702(18) &965 & 17.8823(66) & 602 & 3.5822(16) & 224  \\
    &      & 64 & 78466 &1.612(22) &1318 & 18.8460(58) & 861 & 3.6672(14) & 288  \\
    & 24 & 8 & 49303 &1.9782(57) &1146 & 14.2312(39) & 573 & 3.2155(10) & 188  \\
    &      & 16 & 43540 &1.828(11) &649 & 16.2966(46) & 544 & 3.41313(96) & 235  \\
    &      & 24 & 95005 &1.750(11) &969 & 17.2897(41) & 819 & 3.50643(75) & 461  \\
    &      & 32 & 58712 &1.704(13) &863 & 17.8506(37) & 1030 & 3.55674(70) & 431  \\
    & 32 & 8 & 85250 &1.9894(33) &1813 & 14.2281(19) & 1522 & 3.20939(45) & 600  \\
    &      & 16 & 54720 &1.8519(66) &994 & 16.2872(29) & 729 & 3.40559(55) & 411  \\
    &      & 24 & 52896 &1.7386(98) &661 & 17.2848(36) & 539 & 3.49982(67) & 238  \\
    &      & 32 & 31240 &1.696(16) &240 & 17.8483(56) & 226 & 3.5506(10) & 127  \\
0.70 & 16 & 8 & 47580 &2.4294(99) &1106 & 15.0002(70) & 587 & 3.2961(18) & 214  \\
    &      & 12 & 104350 &2.3780(98) &1683 & 16.1766(54) & 1054 & 3.4069(12) & 366  \\
    &      & 16 & 84565 &2.3210(99) &2349 & 16.9525(43) & 1691 & 3.48264(91) & 735  \\
    &      & 24 & 76014 &2.199(14) &1900 & 17.9056(53) & 1134 & 3.5734(11) & 434  \\
    &      & 32 & 72060 &2.183(19) &1242 & 18.3763(66) & 783 & 3.6114(13) & 335  \\
    &      & 64 & 50755 &2.146(32) &820 & 19.2510(79) & 642 & 3.6877(17) & 271  \\
    & 24 & 8 & 53054 &2.4428(45) &2411 & 14.9840(25) & 1964 & 3.27989(51) & 1040  \\
    &      & 16 & 52690 &2.309(10) &958 & 16.9438(48) & 634 & 3.46690(86) & 339  \\
    &      & 24 & 115524 &2.221(11) &1242 & 17.8811(43) & 855 & 3.55417(73) & 502  \\
    &      & 32 & 60100 &2.195(15) &858 & 18.3685(42) & 985 & 3.59586(71) & 484  \\
    & 32 & 8 & 95795 &2.4466(49) &1182 & 14.9874(33) & 704 & 3.27650(64) & 334  \\
    &      & 16 & 59160 &2.3301(73) &969 & 16.9395(33) & 739 & 3.46137(56) & 402  \\
    &      & 24 & 56768 &2.207(10) &810 & 17.8784(34) & 777 & 3.54895(59) & 368  \\
    &      & 32 & 34512 &2.180(18) &305 & 18.3693(60) & 218 & 3.5913(10) & 130  \\
0.80 & 16 & 8 & 49060 &2.944(11) &1168 & 15.7228(76) & 605 & 3.3573(15) & 280  \\
    &      & 12 & 116950 &2.9178(77) &3654 & 16.8254(42) & 2165 & 3.45939(78) & 1044  \\
    &      & 16 & 91720 &2.844(16) &1132 & 17.5739(79) & 603 & 3.5325(16) & 186  \\
    &      & 24 & 83924 &2.728(15) &2046 & 18.4607(59) & 1198 & 3.6142(11) & 586  \\
    &      & 32 & 78707 &2.716(20) &1457 & 18.8790(69) & 874 & 3.6469(12) & 399  \\
    &      & 64 & 46001 &2.657(38) &842 & 19.6830(92) & 561 & 3.7176(17) & 261  \\
    & 24 & 8 & 58866 &2.9656(51) &2452 & 15.7091(28) & 2102 & 3.34472(49) & 1308  \\
    &      & 16 & 58865 &2.839(11) &1132 & 17.5723(52) & 588 & 3.52018(92) & 275  \\
    &      & 24 & 130168 &2.747(11) &1496 & 18.4516(45) & 936 & 3.60035(71) & 634  \\
    &      & 32 & 77420 &2.718(15) &1106 & 18.8793(48) & 806 & 3.63605(78) & 472  \\
    & 32 & 8 & 107005 &2.9648(52) &1321 & 15.7171(35) & 775 & 3.34288(58) & 451  \\
    &      & 16 & 67980 &2.8605(95) &819 & 17.5722(48) & 409 & 3.51571(77) & 263  \\
    &      & 24 & 57976 &2.735(11) &840 & 18.4545(42) & 616 & 3.59734(67) & 402  \\
    &      & 32 & 37064 &2.715(19) &366 & 18.8880(66) & 239 & 3.6330(10) & 134  \\
0.90 & 16 & 8 & 55060 &3.507(12) &1342 & 16.4221(84) & 605 & 3.4175(15) & 341  \\
    &      & 12 & 127950 &3.4864(83) &3877 & 17.4777(48) & 2030 & 3.51486(82) & 1066  \\
    &      & 16 & 99320 &3.418(17) &1360 & 18.1907(88) & 591 & 3.5830(15) & 241  \\
    &      & 24 & 74182 &3.276(19) &1513 & 19.0386(86) & 662 & 3.6604(15) & 331  \\
    &      & 32 & 78983 &3.279(22) &1579 & 19.3894(79) & 858 & 3.6851(12) & 467  \\
    &      & 64 & 53548 &3.214(42) &823 & 20.136(10) & 474 & 3.7493(20) & 191  \\
    & 24 & 8 & 90740 &3.5347(60) &2387 & 16.4153(36) & 1463 & 3.40846(60) & 840  \\
    &      & 16 & 65155 &3.400(12) &1229 & 18.1921(60) & 509 & 3.57291(99) & 285  \\
    &      & 24 & 140560 &3.319(12) &1802 & 19.0197(49) & 924 & 3.64708(74) & 560  \\
    &      & 32 & 84856 &3.288(16) &1178 & 19.4000(53) & 778 & 3.67736(80) & 493  \\
    & 32 & 8 & 109695 &3.5319(55) &1443 & 16.4258(39) & 731 & 3.40731(60) & 470  \\
    &      & 16 & 72560 &3.4269(83) &1295 & 18.1992(40) & 797 & 3.57038(62) & 490  \\
    &      & 24 & 62280 &3.299(12) &958 & 19.0222(48) & 571 & 3.64451(71) & 347  \\
    &      & 32 & 38120 &3.292(20) &443 & 19.4050(74) & 210 & 3.6744(12) & 131  \\
1.00 & 16 & 8 & 57300 &4.107(12) &1469 & 17.1166(88) & 629 & 3.4780(15) & 364  \\
    &      & 12 & 135400 &4.087(11) &2763 & 18.1234(69) & 1128 & 3.5692(11) & 586  \\
    &      & 16 & 106760 &4.018(17) &1570 & 18.8107(95) & 606 & 3.6343(15) & 290  \\
    &      & 24 & 82194 &3.891(19) &2004 & 19.6005(87) & 790 & 3.7041(13) & 439  \\
    &      & 32 & 92646 &3.911(21) &1971 & 19.9184(83) & 899 & 3.7263(12) & 482  \\
    &      & 64 & 59154 &3.804(44) &909 & 20.639(11) & 538 & 3.7887(17) & 309  \\
    & 24 & 8 & 95304 &4.1359(66) &2382 & 17.1173(39) & 1512 & 3.47184(61) & 843  \\
    &      & 16 & 68150 &4.001(13) &1310 & 18.8177(64) & 524 & 3.62646(98) & 291  \\
    &      & 24 & 148856 &3.929(12) &2011 & 19.5917(53) & 942 & 3.69417(75) & 588  \\
    &      & 32 & 72952 &3.912(20) &1013 & 19.9319(64) & 623 & 3.71970(93) & 412  \\
    & 32 & 8 & 112250 &4.1357(59) &1580 & 17.1272(42) & 763 & 3.47081(61) & 514  \\
    &      & 16 & 75472 &4.017(10) &1179 & 18.8226(52) & 486 & 3.62353(74) & 315  \\
    &      & 24 & 72000 &3.900(13) &1043 & 19.6012(51) & 517 & 3.69296(78) & 302  \\
    &      & 32 & 42832 &3.921(27) &297 & 19.9422(70) & 278 & 3.7185(10) & 169  \\
\hline \hline
\caption{Summary table for the parameters and expectation values of observables in the full ungauged BFSS matrix model. When the expectation value of an observable is not shown, it means that the autocorrelation time estimation was not robust enough given the accumulated samples.\label{tab:bfss-ungauged}}
 \end{longtable}

\end{document}